\documentclass[aos,preprint]{imsart}
\usepackage{xr}
\usepackage{amsthm,amsmath,natbib}
\usepackage{algpseudocode, algorithm, algorithmicx}
\usepackage[colorlinks,citecolor=blue,urlcolor=blue]{hyperref}
\usepackage{graphicx,color}
\usepackage{verbatim}
\usepackage{multirow}

\newtheorem{thm}{Theorem}[section] 
\newtheorem{lemma}{Lemma}[section] 
\newtheorem{cor}{Corollary}[section]

\newtheorem{definition}{Definition}[section]

\newcommand{\bed}{\begin{definition}}
\newcommand{\eed}{\end{definition}}
\newcommand{\beq}{\begin{equation}}
\newcommand{\eeq}{\end{equation}}

\newcommand{\eps}{\epsilon}

\newcommand{\bitem}{\begin{itemize}}
\newcommand{\eitem}{\end{itemize}}

\newcommand{\goto}{\rightarrow}

\newcommand{\margmax}{\mathrm{argmax}}

\newcommand{\beqn}{\begin{equation}}
\newcommand{\eeqn}{\end{equation}}
\newcommand{\balign}{\begin{align}}
\newcommand{\ealign}{\end{align}}

\newcommand{\cH}{\mathcal{H}}

\newcommand{\sgn}{\mathrm{sgn}}
\newcommand{\diag}{\mathrm{diag}}

\newcommand{\heta}{\hat{\eta}}
\newcommand{\hamm}{\mathrm{Hamm}}

\newcommand{\bel}{\begin{eqnarray}\label}
\newcommand{\eel}{\end{eqnarray}}

\newcommand{\bes}{\begin{eqnarray*}}
\newcommand{\ees}{\end{eqnarray*}}

\def\sgn{\hbox{\rm sgn}}

\newcommand{\sW}{W^{\hat{S}_p(t_p(q))}}
\usepackage{amssymb}
\begin{document}

\begin{frontmatter}

\title{Influential Features PCA for high dimensional clustering}
\runtitle{Influential Features PCA}

\begin{aug}
\author{\fnms{Jiashun} \snm{Jin}\thanksref{t1}\thanksref{t2}\ead[label=e1]{jiashun@stat.cmu.edu}}
\and
\author{\fnms{Wanjie} \snm{Wang}\thanksref{t1}\thanksref{t3}\ead[label=e2]{wanjiew@wharton.upenn.edu}}
\thankstext{t1}{Supported in part by NSF grants  DMS-1208315 and DMS-1513414.}
\runauthor{J. Jin and W. Wang}

\affiliation{Carnegie Mellon University\thanksmark{t2}  and University of 
Pennsylvania\thanksmark{t3}}

\address{J. Jin\\
Department of Statistics\\
Carnegie Mellon University\\
Pittsburgh, Pennsylvania, 15213\\
USA\\
\printead{e1}}

\address{W. Wang\\
Department of Statistics\\
The Wharton School\\
University of Pennsylvania\\
Philadelphia, Pennsylvania, 19104\\
USA\\
\printead{e2}}
\end{aug}

\begin{abstract}
We consider a clustering problem where we observe feature vectors
$X_i \in R^p$,  $i = 1, 2, \ldots, n$,  from $K$ possible classes.  The class labels are unknown and the main interest is to estimate them.  We are primarily interested in the modern regime of $p \gg n$, where classical clustering methods
face challenges.

We propose  Influential Features PCA (IF-PCA) as a new clustering procedure.
In IF-PCA,  we select a small fraction of features with the largest Kolmogorov-Smirnov (KS) scores,   obtain the first $(K-1)$ left singular vectors of the post-selection normalized data matrix,  and then estimate the labels by applying the classical $k$-means procedure to these singular vectors.  In this procedure, the only tuning parameter is the threshold in the feature selection step.  We set the threshold in a data-driven fashion by adapting the recent notion of Higher Criticism. As a result,  IF-PCA is a tuning-free clustering method.

We apply IF-PCA  to $10$ gene microarray data sets. The method has competitive
performance in clustering.  Especially, in three of the
data sets, the error rates  of IF-PCA are only $29\%$ or less of the error rates by other methods. We have also rediscovered  a phenomenon on empirical null by \cite{Efron} on microarray data.

With delicate analysis, especially post-selection eigen-analysis,  we derive tight probability bounds on the Kolmogorov-Smirnov statistics and
show that IF-PCA yields clustering consistency in a broad context.  The clustering problem is connected
to the problems of  sparse PCA and low-rank matrix recovery, but it
is different in important ways. We reveal an interesting phase transition phenomenon associated with
these problems and identify the range of interest for each.
\end{abstract}



\end{frontmatter}

\section{Introduction}  \label{sec:intro}
\setcounter{equation}{0}
 Consider a clustering problem where we have feature  vectors  $X_i \in R^p$, $i = 1, 2, \ldots, n$,   from $K$ possible classes. For simplicity,  we assume $K$ is small and is known to us. The class labels
$y_1$,  $y_2$,  $\ldots$, $y_n$ 
take values from $\{1, 2, \ldots, K\}$, but are unfortunately unknown to us, and
the main interest is to estimate them.

Our study is largely motivated by clustering using gene microarray data. In a typical setting,
we have patients from several different  classes (e.g., normal, diseased), and
for each patient, we have measurements (gene expression levels) on the same   set of  genes. The class labels of the patients are unknown and it is of interest to use the   expression  data  to predict them.

 Table \ref{table:data} lists $10$ gene microarray data sets (arranged alphabetically).  Data sets $1$, $3$, $4$, $7$, $8$, and $9$ were analyzed and cleaned in   \cite{Dettling}, Data set $5$ is from \cite{lungcancer},  Data sets $2$, $6$, $10$ were analyzed and grouped into two classes in   \cite{Yousefi},  among which Data  set $10$ was cleaned  by us in the same way as by   \cite{Dettling}.  All the data sets can be found at
\verb+www.stat.cmu.edu/~jiashun/Research/software/GenomicsData+.  The data sets are analyzed in Section \ref{subsec:realdata}, after our  approach is  fully introduced.

In these data sets, the true labels are given but (of course) we do not use them  for clustering;  the true labels are thought of as the `ground truth' and are only used for comparing the error rates of different methods.

\begin{table}[hb!]
\centering
\caption{Gene microarray data sets investigated in this paper. Note that $K$ is small and $p \gg n$ ($p$: number of genes; $n$:    number of subjects).}
\scalebox{0.9}{ 
\begin{tabular}{|l|l|l|l||l|c|c|}
\hline
$\#$  & Data Name & Abbreviation & Source  & $K$ &  $n$  & $p$  \\
\hline
1 & Brain &  Brn & Pomeroy (02)   & 5    &  42 &  5597  \\
2 & Breast Cancer & Brst & Wang et al. (05) &    2  & 276    & 22215   \\
3 & Colon Cancer &Cln & Alon et al. (99)  & 2  &  62  &2000 \\
4 & Leukemia		  	&  Leuk & Golub et al. (99) 	& 2  &   72 	 & 3571\\
5 & Lung Cancer(1)  		& Lung1 & Gordon et al. (02) 	&  2 &  181	 & 12533\\
6 & Lung Cancer(2)   & Lung2 & Bhattacharjee et al. (01) & 2 & 203 & 12600  \\
7 & Lymphoma &  Lymp & Alizadeh et al. (00) & 3   &  62 & 4026  \\
8 & Prostate Cancer & Prst& Singh et al. (02) &2   &  102 & 6033 \\
9 & SRBCT & SRB & Kahn (01) &  4    &  63  & 2308 \\
10 & SuCancer & Su & Su et al (01) & 2 & 174 & 7909  \\
\hline
\end{tabular}
} 
\label{table:data}
\end{table}

View each $X_i$ as the sum of a `signal component' and a `noise component':
\begin{equation} \label{model1}
X_i = E[X_i] + Z_i, \qquad Z_i  \equiv X_i  - E[X_i].
\end{equation}
For any  numbers $a_1, a_2, \ldots, a_p$,  let  $\diag(a_1, a_2, \ldots, a_p)$ be the $p \times p$ diagonal matrix where the $i$-th diagonal entry is $a_i$, $1 \leq i \leq p$.
We assume
\begin{equation} \label{model2}
Z_i  \stackrel{iid}{\sim} N(0, \Sigma),  \qquad \mbox{where} \qquad \Sigma = \mathrm{diag}\bigl(\sigma^2(1), \sigma^2(2), \ldots, \sigma^2(p)\bigr),
\end{equation}
and the vector $\sigma = (\sigma(1), \sigma(2), \ldots, \sigma(p))'$ is unknown to us.
Assumption \eqref{model2}  is only for simplicity:  our method to be introduced below is not tied to such an assumption, and works well
with most of the data sets in Table \ref{table:data};  see Sections  \ref{subsec:IFPCA} and \ref{subsec:realdata} for more discussions.

Denote the overall mean vector by
$\bar{\mu} =  \frac{1}{n}\sum_{i = 1}^n E[X_i]$.  
For $K$ different vectors $\mu_1, \mu_2, \ldots, \mu_K \in R^p$,  we model $E[X_i]$ by ($y_i$ are class labels) 
\begin{equation} \label{model3}
E[X_i]  = \bar{\mu} + \mu_k, \qquad  \mbox{if and only if}  \qquad   y_i = k.
\end{equation}
For $1 \leq k \leq K$, let $\delta_k$ be the fraction of samples in Class $k$. Note that 
\begin{equation} \label{sumofmu}
\delta_1 \mu_1 + \delta_2 \mu_2+ \ldots  +  \delta_K \mu_K = 0,  
\end{equation}
so $\mu_1, \mu_2, \ldots, \mu_K$ are linearly dependent. However, it is natural to assume
\begin{equation} \label{model4}
\mbox{$\mu_1, \mu_2, \ldots, \mu_{K-1}$ are linearly independent}.
\end{equation}
\bed
We call feature $j$ a useless feature (for clustering)  if  $\mu_1(j)  = \mu_2(j) = \ldots  =  \mu_K(j) = 0$, and a useful feature otherwise.
\eed
We call  $\mu_k$  the {\it contrast mean vector} of Class $k$, $1 \leq k \leq K$.
In many applications, the contrast mean vectors are sparse in the sense that only a small fraction of the features are useful.  Examples include but are not limited to gene microarray data:   it is widely believed that only a small fraction of genes are differentially expressed, so the contrast mean vectors are sparse.

We are primarily interested in the modern regime of $p \gg n$.
In such a  regime,   classical methods (e.g.,  $k$-means,  
hierarchical clustering,  Principal Component Analysis (PCA) \citep{HTF})   are  either computationally challenging or
ineffective. Our primary interest   is to develop new methods  that are appropriate for this regime.

\subsection{Influential Features PCA (IF-PCA)}  \label{subsec:IFPCA}  
Denote the data matrix by:
\[
X   =   [ X_1, X_2, \ldots, X_n]'.
\]
We propose IF-PCA as a new spectral clustering method.  Conceptually,  IF-PCA contains an IF part  and a PCA part. In the IF part, we select features by exploiting the sparsity of the contrast mean vectors,  where we remove many columns of  $X$ leaving only those we think are influential for clustering (and so the name of  Influential Features). 
In the PCA part,  we apply the classical PCA to the post-selection data matrix.\footnote{Such a two-stage clustering idea (i.e., feature selection followed by post-selection clustering) is not completely new and can be found in \cite{CH} for example. Of course, their procedure  is very different from ours.}

We normalize each column of $X$ and denote the resultant matrix by $W$:
\[
W(i,j) = [X_i(j) - \bar{X}(j)]/\hat{\sigma}(j), \qquad 1 \leq i \leq n,   \; 1 \leq j \leq p, 
\]
where $\bar{X}(j)  = \frac{1}{n} \sum_{i = 1}^n X_i(j)$  and $\hat{\sigma}(j) = [\frac{1}{n-1} \sum_{i = 1}^n (X_{i}(j) - \bar{X}(j))^2]^{1/2}$ are the empirical mean and standard deviation associated with feature $j$,     respectively.  
Write
\[
W = [W_1, W_2, \ldots, W_n]'. 
\]  
For any  $1 \leq j \leq p$,  denote the empirical CDF associated with feature $j$ by   
\[
F_{n, j}(t)  =   \frac{1}{n} \sum_{i = 1}^n 1\{W_i(j)   \leq t\}.  
\] 
IF-PCA  contains two `IF' steps and  two  `PCA' steps   as follows.  
\begin{center} 
Input: data matrix $X$, number of classes $K$, and  parameter $t$.  \\
Output: predicted $n \times 1$ label vector $\hat{y}_{t}^{IF}  =  (\hat{y}_{t,1}^{IF}, \hat{y}_{t,2}^{IF}, \ldots, \hat{y}_{t,n}^{IF})$. 
\end{center} 
\begin{itemize} 
\item IF-1.  For each $1 \leq j \leq p$,  compute a Kolmogorov-Smirnov (KS)  statistic by
\begin{equation} \label{DefineKS}
\psi_{n,j} =  \sqrt{n} \cdot \sup_{-\infty < t < \infty} | F_{n, j}(t) -  \Phi(t) |,  \qquad (\mbox{$\Phi$: CDF of $N(0,1)$)}.
\end{equation}
\item IF-2.   Following the suggestions by   \cite{Efron}, we renormalize by 
\begin{equation} \label{KSthreshold}
\psi_{n,j}^*  =  [\psi_{n,j}  -  \mbox{mean of all $p$  KS-scores}]/\mbox{SD of all $p$  KS-scores}.\footnote{Alternatively, we can normalize the KS-scores with sample median and Median Absolute Deviation (MAD); see Section \ref{subsec:IFPCAvar} for more discussion.} 
\end{equation}
\item PCA-1.   Fix a threshold $t > 0$.  For short, let $W^{(t)}$ be the matrix formed by restricting the columns of $W$ 
to the set of retained indices $\hat{S}_p(t)$, where 
\begin{equation} \label{DefinehatSt}
\hat{S}_p(t) =  \{1 \leq j \leq p:  \psi^*_{n,j}  \geq t\}.
\end{equation}
Let $\hat{U}^{(t)} \in R^{n, K-1}$ be the  matrix consisting the first $K - 1$ (unit-norm) left singular vectors  of $W^{(t)}$.\footnote{For a  matrix $M \in R^{n,m}$, the $k$-th left (right) singular vector is the 
eigenvector associated with the $k$-th largest eigenvalue of the matrix $MM'$  (of the matrix $M'M$).}    Define a matrix  $\hat{U}_*^{(t)} \in R^{n, K-1}$  by truncating $\hat{U}^{(t)}$ entry-wise with threshold $T_p = \log(p)/\sqrt{n}$.\footnote{That is,  
$\hat{U}_*^{(t)}(i,k) = \hat{U}(i,k)  1\{|\hat{U}(i,k)| \leq T_p \} + T_p \sgn(\hat{U}(i,k)) 1\{|\hat{U}(i,k)| >  T_p \}$, $1 \leq i \leq n, 1 \leq k \leq K-1$. We usually take $T_p = \log(p)/\sqrt{n}$ as above, but $\log(p)$ can be replaced by any sequence that tends to $\infty$ as $p \goto \infty$. The truncation is mostly for theoretical analysis in Section \ref{sec:main} and is not used in numerical study (real or simulated data).}
\item PCA-2.  Cluster by applying the classical $k$-means  to $\hat{U}_*^{(t)}$   assuming there are $\leq K$ classes.  Let 
$\hat{y}^{IF}_t$ be the predicted label vector. 
\end{itemize} 
In the procedure,  $t$ is the only tuning parameter. In Section \ref{subsec:HCT}, we propose a data-driven approach to choosing $t$, so the method becomes tuning-free. 
Step 2 is largely for gene microarray data, and is not necessary if Models (\ref{model1})-(\ref{model2}) hold.

In Table \ref{table:Difft}, we use the Lung Cancer(1) data to illustrate  how IF-PCA performs with different choices of $t$. The results show that  with $t$ properly set, the number of clustering errors of IF-PCA can be as low as $4$.  In comparison, classical PCA (column $2$ of Table \ref{table:Difft}; where $t = .000$ so we do not perform feature selection) 
has $22$ clustering errors.  

\begin{figure}[ht!]
\begin{center}
\includegraphics[height = 1.8 in, width = 4.8 in]{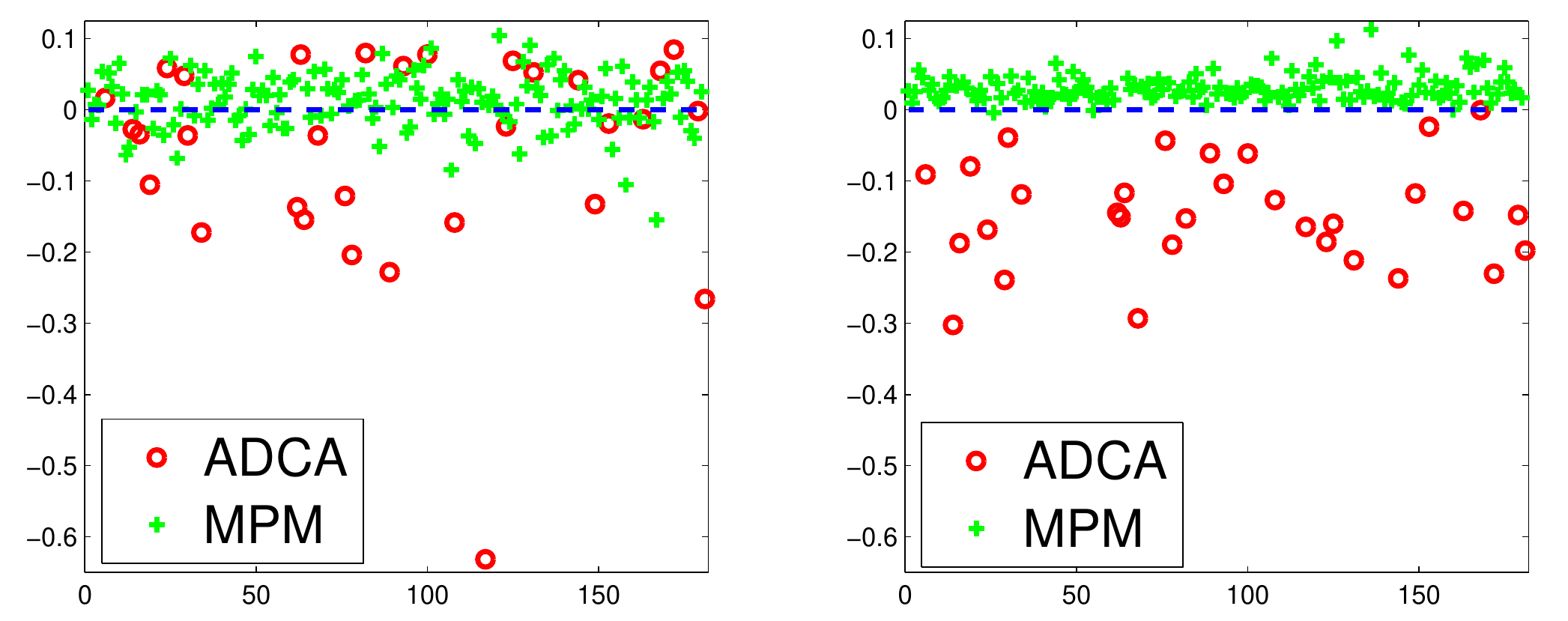}
\end{center}
\caption{Comparison of $\hat{U}^{(t)}$ for $t = .000$ (left; no feature selection) and $t = 1.057$ (right; $t$ is set by Higher Criticism in a data-driven fashion); note $\hat{U}^{(t)}$ is an $n \times 1$ vector since $K = 2$.  $y$-axis: entries of  $\hat{U}^{(t)}$, $x$-axis: sample indices.
Plots are based on Lung Cancer(1) data, where  ADCA and MPM represent two different classes.}
\label{fig:lung}
\end{figure}

\begin{table}[ht!]
\caption{Clustering errors and $\#$ of selected features for different choices of  $t$ (Lung 
Cancer(1) data). Columns  highlighted correspond to the sweet spot of the threshold choice.}  
\scalebox{0.9}{ 
\begin{tabular}{rrrrrrrrrr} 
Threshold $t$ & .000 & .608 & .828 & {\bf .938} & {\bf 1.048}  & {\bf 1.158}  & 1.268 & 1.378 & 1.488  \\
\# of selected features & 12533 & 5758 & 1057 & {\bf 484} & {\bf 261} & {\bf 129}  & 63 & 21 & 2\\
Clustering errors & 22 & 22 & 24 &{\bf 4}  & {\bf 5} & {\bf 7} &38  & 39 & 33 
\end{tabular}  
}
\label{table:Difft} 
\end{table} 
In Figure \ref{fig:lung}, we compare IF-PCA with classical PCA by investigating $\hat{U}^{(t)}$ defined in Step 3  for two choices of $t$: (a) $t = .000$ so $\hat{U}^{(t)}$ is   the first  singular vector of pre-selection data matrix $W$, and (b) a data-driven threshold choice by Higher Criticism to be introduced in Section \ref{subsec:HCT}. 
For (b), the entries of $\hat{U}^{(t)}$  can be clearly divided into two groups, yielding almost error-free clustering results. Such a clear separation does not exist for   (a).   
These results suggest that  IF-PCA may  significantly improve classical PCA.

Two important questions arise:
\begin{itemize}
\item In (\ref{KSthreshold}), we use a modified KS statistic  for feature selection. What is the rationale behind the use of KS statistics and the modification?
\item The clustering errors critically depend on the threshold $t$. How to set $t$ in a data-driven fashion?
\end{itemize}
In Section \ref{subsec:KSrationale},
we address the first question. In Section \ref{subsec:HCT}, we propose a data-driven threshold choice by the recent notion of Higher Criticism.

\subsection{KS statistic, normality assumption,  and Efron's empirical null} \label{subsec:KSrationale}
The goal in Steps 1-2  is to find an easy-to-implement method to rank the features.
The focus of Step 1 is on a data matrix satisfying Models (\ref{model1})-(\ref{model4}),
and the focus of Step 2 is to  adjust  Step 1  in a way so to work well with microarray data. We consider two steps separately.

Consider the first step.  The interest is to test for each fixed $j$, $1 \leq j \leq p$, whether feature $j$ is useless or useful. Since we have no prior information about the class labels, the problem can be reformulated as that of  testing whether
all $n$ samples associated with the $j$-th feature are iid Gaussian
\begin{equation} \label{testnull}
H_{0,j}: \qquad X_i(j)  \stackrel{iid}{\sim}  N(\bar{\mu}(j), \sigma^2(j)),  \qquad i = 1, 2, \ldots, n,
\end{equation}
or they are iid from a $K$-component heterogenous Gaussian mixture:
\begin{equation} \label{testalt}
H_{1, j}:  \qquad   X_i(j) \stackrel{iid}{\sim}  \sum_{k = 1}^K \delta_k  N(\bar{\mu}(j) + \mu_k(j), \sigma^2(j)), \qquad i = 1, 2, \ldots, n,
\end{equation}
where $\delta_k > 0$ is the prior probability that $X_i(j)$ comes from Class $k$, $1 \leq k \leq K$.  Note that
$\bar{\mu}(j)$, $\sigma(j)$,  and $\bigl((\delta_1, \mu_1(j)), \ldots, (\delta_K, \mu_K(j))\bigr)$ are unknown.
\begin{figure}[htb!]
\begin{center}
\includegraphics[height = 1.4 in, width = 4 in]{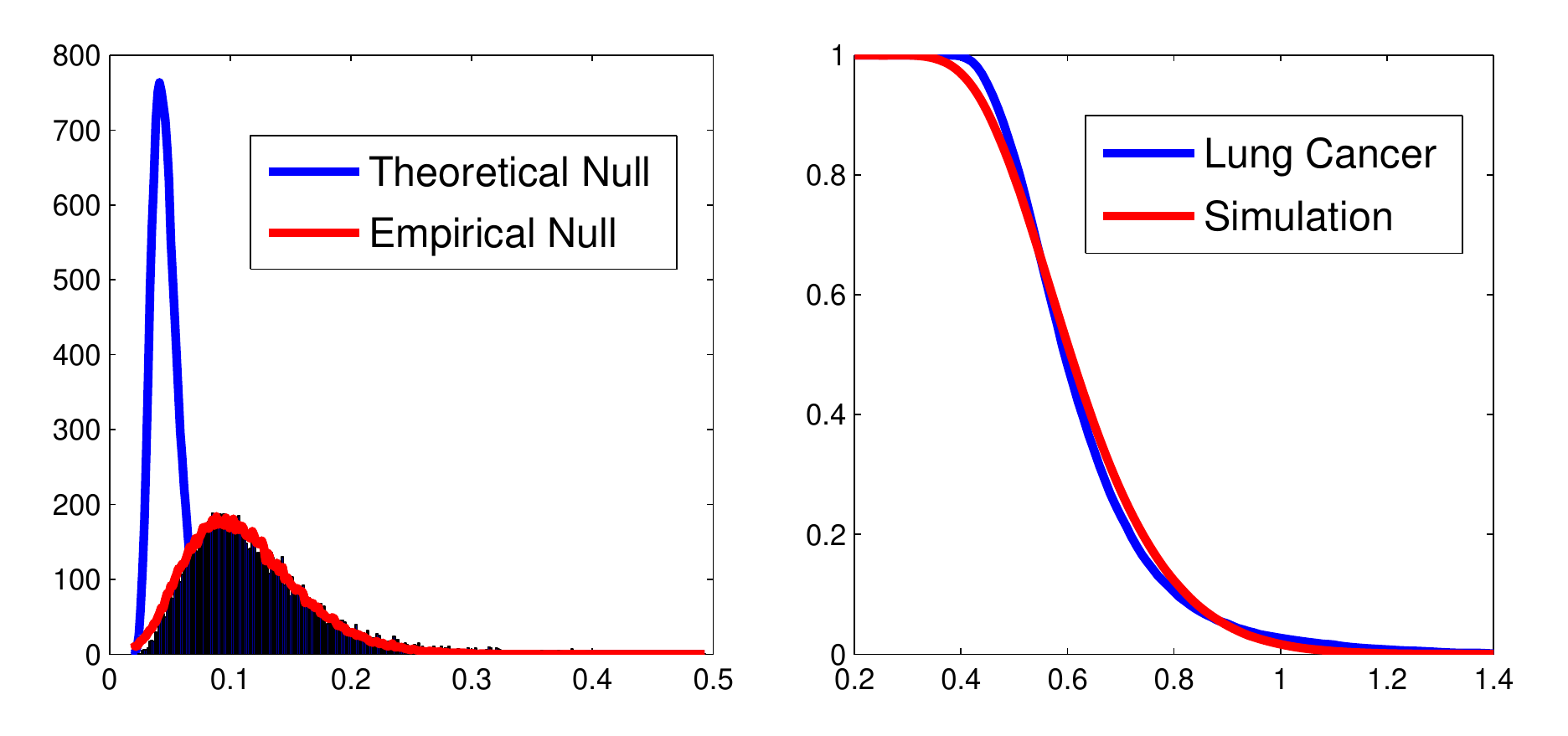}
\end{center}
\caption{Left: The histogram of KS-scores of the Lung Cancer(1) data. The two lines in blue and red
denote  the theoretical null and empirical null densities, respectively. Right: empirical survival function of the adjusted KS-scores based on Lung Cancer(1) data (red) and the survival function of theoretical null (blue).}
\label{fig:ks}	
\end{figure}
The above is a well-known difficult testing problem.  For example, in such a setting,   the classical Likelihood Ratio Test (LRT) is known to be not   well-behaved (e.g.,  \cite{ChenJH}).    

Our proposal is to use the Kolmogorov-Smirnov (KS) test, which measures the 
maximum difference between the empirical CDF for the normalized data and the CDF of  $N(0,1)$. 
The KS test is  a well-known goodness-of-fit test  (e.g.,  \cite{Wellner86}). 
 In the idealized Gaussian Model (\ref{testnull})-(\ref{testalt}),   the KS test is asymptotically equivalent to the optimal moment-based tests (e.g., see Section \ref{sec:main}),  but its success is not tied to a specific model for the alternative hypothesis, and is more robust against occasional outliers. Also, Efron's null correction (below) is more successful if we use KS instead of moment-based tests for feature ranking.  
This is our rationale for Step 1.

We now discuss our rationale for Step 2. We discover an interesting phenomenon which we illustrate with Figure \ref{fig:ks} (Lung Cancer(1) data).     Ideally, if the normality assumption (\ref{model2}) is valid for this data set, then the 
density function of the KS statistic for Model (\ref{testnull}) (the blue curve in left panel; obtained by simulations) should fit well with the 
histogram of the KS-scores based on the Lung Cancer(1) data. 
Unfortunately, this is not the case, and there is a substantial discrepancy in fitting.  
On the other hand, if we translate and rescale the blue curve so that  it has the same mean and standard deviation as the KS-scores associated with Lung Cancer(1) data, then the new curve  (red curve;  left panel of Figure \ref{fig:ks}) fits well with the histogram.\footnote{If we replace sample mean and standard deviation by sample median and MAD, respectively, then it gives rises to the normalization in  the second footnote of Section \ref{subsec:IFPCA}.}

A related phenomenon was discussed in \cite{Efron}, only considering Studentized $t$-statistics in a different setting.  As in  \cite{Efron},  we call the density functions associated with two curves (blue and red)  the  
 {\it theoretical null} and the {\it empirical null}, respectively. The phenomenon is then:  
 the theoretical null has a poor fit with the histogram of the KS-scores of the real data, but the empirical null may have a good fit.   
 
  In the right panel of Figure \ref{fig:ks}, we view this from a slightly different perspective, and show that  the survival function associated with the adjusted KS-scores (i.e., $\psi_{n, j}^*$)   of the real data fits well with the theoretical null.    
 
The above  observations explain the rationale for Step 2. Also, they suggest that IF-PCA does not critically depend on the normality assumption and works well  for microarray data.  This is further validated in Section \ref{subsec:realdata}.

{\bf Remark}.  \cite{Efron} suggests several possible reasons (e.g., dependence between different samples, dependence between the genes) for the discrepancy between the theoretical null and empirical null, but what has really caused such a discrepancy is not fully understood.  Whether Efron's empirical null is useful in other application areas or  other data types (and if so, to what extent)  is also  an open problem, and to understand it we need a good grasp on  the mechanism by which the data sets of interest  are  generated.

\subsection{Threshold choice by Higher Criticism} \label{subsec:HCT}
The performance of IF-PCA critically depends on the threshold $t$, and it is of interest
to set $t$ in a data-driven fashion.
We approach this by  the recent notion of Higher Criticism.

Higher Criticism (HC) was first introduced in \cite{DJ04}  
as a method for large-scale multiple testing. In \cite{DJ08},  HC was also found to be
useful to set a threshold for  feature selection in the context of classification.
HC is also useful in many other settings. See  \cite{DJ13, JK}  for reviews on HC.

To adapt HC for  threshold choice in IF-PCA, we must modify the procedure carefully, since the purpose is very different from those in previous literature. The approach contains three simple steps as follows.
\begin{itemize}
\item  For  $1 \leq j \leq p$,  calculate a $P$-value
$\pi_j  = 1 - F_0(\psi_{n,j})$, where   $F_0$ is the distribution of  $\psi_{n,j}$  under the null (i.e., feature $j$ is useless).
\item Sort all $P$-values in the ascending order $\pi_{(1)} < \pi_{(2)} < \ldots < \pi_{(p)}$.
\item Define the Higher Criticism score by
\begin{equation} \label{HCfunction}
HC_{p,j}  =
\sqrt{p}  ( j/p  -  \pi_{(j)} )  / \sqrt{ \max\{ \sqrt{n} ( j/p  -  \pi_{(j)}) , 0 \}  + j/p}.
\end{equation}
Let $\hat{j}$ be the index such that $\hat{j}=\margmax_{\{1\leq j\leq p/2, \pi_{(j)}>\log(p)/p\}} \{HC_{p,j} \}$.
The HC threshold $t_p^{HC}$ for IF-PCA is then the $\hat{j}$-th largest KS-scores.
\end{itemize} 
Combining HCT with IF-PCA gives a tuning-free clustering procedure IF-HCT-PCA, or IF-PCA for short if there is no confusion. See Table \ref{tab:code}. 
\begin{table}[htb!]
\caption{Pseudocode for IF-HCT-PCA (for microarray data; threshold set by Higher Criticism)} 
\scalebox{0.83}{ 
{\begin{tabular}{ll}\\ \hline
& \underline{Input}: data matrix $X$, number of classes $K$.      \underline{Output}: class label vector $\hat{y}^{IF}_{HC} $. \\
1. &Rank features: Let $\psi_{n,j}$  be the KS-scores as in (\ref{DefineKS})  and $F_0$ be the CDF of  $\psi_{n,j}$ under null,  $1 \leq j \leq p$.   \\
2. & Normalize KS-scores:    $\psi^*_n = (\psi_n - mean(\psi_n))/SD(\psi_n)$.    \\
3. &Threshold choice by HCT: Calculate $P$-values by  $\pi_j = 1 - F_0(\psi_{n,j}^*)$, $1 \leq j \leq p$  and sort them by   \\ 
     & $\pi_{(1)} <  \pi_{(2)}  < \ldots < \pi_{(p)}$.   Define $HC_{p, j} = \sqrt{p}  ( j/p  -  \pi_{(j)} )  / \sqrt{ \max\{ \sqrt{n} ( j/p  -  \pi_{(j)}) , 0 \}  + j/p}$, and let  \\
     &  $\hat{j} = \margmax_{\{ j: \pi_{(j)} > \log(p)/p, j < p/2\} } \{HC_{p, j}\}$.
    HC threshold  $t_p^{HC}$  is the $\hat{j}$-largest KS-score.  \\ 
4. & Post-selection PCA: Define post-selection data matrix  $W^{(HC)}$ (i.e., sub-matrix of $W$ consists of all   \\ 
& column $j$ of $W$ with $\psi^*_{n,j} > t_p^{HC}$).  Let $U \in R^{n, K-1}$    
 be the matrix of the first $(K - 1)$ left singular  \\
 &  vectors of  $W^{(HC)}$. Cluster by $\hat{y}^{IF}_{HC} = kmeans(U, K)$. \\
 \hline
\end{tabular}}
} 
\label{tab:code} 
\end{table}

\begin{figure}[ht!]
\begin{center}
\includegraphics[height = 1 in, width = 5 in]{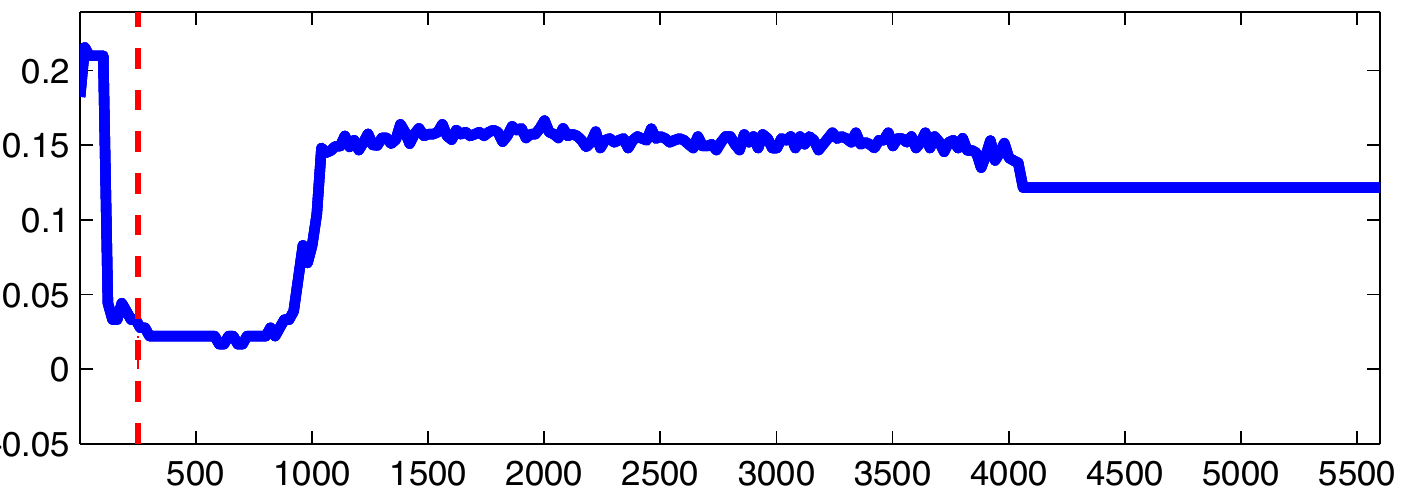}
\end{center}
\caption{Error rates by IF-PCA  (y-axis) with different number of selected features $k$ ($x$-axis) (Lung Cancer(1) data).    HCT corresponds to $251$ selected features  (dashed vertical line). }
\label{fig:Errorrate2}
\end{figure}

For illustration, we again employ the Lung Cancer(1) data. In this data set,  $\hat{j} = 251$,  $t_p^{HC}  = 1.0573$, and HC selects $251$ genes with the largest KS-scores.  
In Figure \ref{fig:Errorrate2},   we plot the error rates of IF-PCA applied to the $k$  features of $W$ with the 
largest KS-scores, where $k$ ranges from $1$ to $p/2$ (for different $k$, we are using the same ranking for all $p$ genes).  The figure shows that there is a `sweet spot' for $k$ where the error rates are the lowest. 
HCT corresponds to $\hat{j} = 251$ and $251$ is in this sweet spot. 
This suggests that HCT gives a reasonable threshold choice, at least for some real data sets. 

{\bf Remark}. When we apply HC to microarray data,  we follow the discussions  in Section \ref{subsec:KSrationale} and take $F_0$ to be the distribution of $\psi_{n,j}$  under the null  but with the mean and variance adjusted to match those of the KS-scores. In the definition,  we require $\pi_{(\hat{j})} > \log(p)/p$, as 
$HC_{p,j}$ may be ill-behaved for very small $j$ (e.g.,  \cite{DJ04}).

The rationale for HCT can also be explained theoretically.  For illustration, consider the case where $K = 2$ so we only have two classes.  
Fixing a threshold $t > 0$,  let $\hat{U}^{(t)}$ be  the first  left singular vector  of $W^{(t)}$ as in Section \ref{subsec:IFPCA}.   In a companion paper   \citep{JinWang2}, we show that when the signals are rare and weak, then for $t$ in the  range of interest,
\begin{equation} \label{t1}
\hat{U}^{(t)}  \propto    \widetilde{snr}(t)  \cdot  U     +  z + rem,
\end{equation}
where $U$ is an $n \times 1$  non-stochastic vector  with only two distinct entries (each determines one of two classes),
$\widetilde{snr}(t)$  is a non-stochastic function of $t$,   $z \sim N(0, I_n)$, and $rem$ is the remainder term (the entries of which are asymptotically of much smaller magnitude  than that of $z$ or $\widetilde{snr}(t) \cdot U$).  Therefore, performance of IF-PCA is best when we maximize $\widetilde{snr}(t)$ (though this is unobservable). We call such a threshold the {\it Ideal Threshold}:
$t_p^{ideal} = \mathrm{argmin}_{t > 0} \{\widetilde{snr}(t)\}$. 

Let $\bar{F}_p(t)$ be the survival function of $\psi_{n,j}$ under the null  (not dependent on $j$),  and let $\hat{G}_p(t)  =   \frac{1}{p} \sum_{j = 1}^p  1\{\psi_{n, j} \geq t\}$ be the empirical survival function. Introduce $HC_p(t) = \sqrt{p} [\hat{G}_p(t) - \bar{F}_p(t)] / \sqrt{ \hat{G}_p(t) + \sqrt{n }[ \max\{ \hat{G}_p(t)  -  \bar{F}_p(t), 0 \}]}$, and let $\psi_{(1)} >  \psi_{(2)}   > \ldots > \psi_{(p)}$ be the sorted values of $\psi_{n,j}$. 
Recall that $\pi_{(k)}$ is the $k$-th smallest $P$-value. By definitions, we have   
$\hat{G}_p(t) |_{t  = \psi_{(k)}} = k/p$ and $\bar{F}_p(t)|_{t = \psi_{(k)}} = \pi_{(k)}$. As a result, we have $HC_p(t) \bigl|_{t = \psi_{(k)}} =    [k/p - \pi_{(k)}] / \sqrt{k/p +  \sqrt{n} \max\{k/p - \pi_{(k)}, 0\}}$, where the right hand side is the form of HC introduced in \eqref{HCfunction}.  Note that  $HC_p(t)$ is a function which is only discontinuous at  $t = \psi_{(k)}$,  $1 \leq k \leq p$,  and between two adjacent discontinuous points, the function is monotone.  
Combining this with the definition of $t_p^{HC}$,     
$t_p^{HC} = \margmax_{t} \{HC_p(t)\}$.

Now, as $p \goto \infty$, some regularity appears, and $\hat{G}_p(t)$ converges to a non-stochastic counterpart, denoted by $\bar{G}_p(t)$, which can be viewed as the survival function associated with the marginal density of $\psi_{n,j}$.  Introduce $IdealHC(t) = \sqrt{p} [\bar{G}_p(t) -  \bar{F}_p(t)] / \sqrt{ \bar{G}_p(t) + \sqrt{n }[ \max\{ \bar{G}_p(t) - \bar{F}_p(t), 0 \}]}$ as the ideal counterpart  of $HC_p(t)$.  It is seen that  
$HC_p(t) \approx IdealHC(t)$ for $t$ in the range of interest,  and so $t_p^{HC} \approx t_p^{idealHC}$, where the latter is defined as the non-stochastic threshold $t$ that maximizes $IdealHC(t)$. 
 
In \cite{JinWang2}, we show that under a broad class of  rare and weak signal models, the leading term of the Taylor expansion of $\widetilde{snr}(t)$ is proportional to that of  $IdealHC(t)$  for $t$ in the range of interest, and so $t_p^{idealHC} \approx t_p^{ideal}$.   Combining this with the discussions above,  we have 
$t_p^{HC}  \approx t_p^{idealHC} \approx t_p^{ideal}$, which explains the rationale for HCT.

The above relationships are justified in   \cite{JinWang2}. The proofs are rather long ($70$ manuscript pages in Annals of Statistics format),  so we will report them in a separate paper.
The ideas above are similar to that in  \cite{DJ08} but the focus there is on classification and our focus is on clustering; our version of HC is also very different from theirs. 
 
\subsection{Applications to gene microarray data}
\label{subsec:realdata}
We compare IF-HCT-PCA  with four other clustering methods (applied to the normalized data matrix $W$ directly, without feature selection):
(1) SpectralGem \citep{Lee} which is the same as classical PCA introduced earlier,  (2) classical $k$-means, (3) hierarchical   clustering  \citep{HTF}, and 
(4) $k$-means$++$ \citep{kmeans+}.   In theory, $k$-means is NP hard, but heuristic algorithms are available; we use the built-in $k$-means package
in Matlab with the parameter `replicates' equal to $30$, so that the algorithm randomly samples initial cluster centroid positions $30$ times (in the last step of either classical PCA  or IF-HCT-PCA,  $k$-means is also used, where  the number of `replicates' is also  $30$).    The $k$-means$++$ \citep{kmeans+} is a recent modification of $k$-means. It improves the performance of $k$-means in some numerical studies, though the problem remains NP hard in theory.
For hierarchical clustering, we use `complete' as the linkage function; other choices give more or less the same results. 
In IF-HCT-PCA, the $P$-values associated with the KS-scores are computed using simulated KS-scores under the null with
$2 \times 10^3 \times p$ independent replications; see Section \ref{subsec:HCT} for remarks on $F_0$. In Table \ref{tab:code}, we repeat the main steps of IF-HCT-PCA for clarification, by presenting the pseudocode.

\begin{table}[htb!]
\caption{Comparison of clustering error rates by different methods for the $10$ gene microarray data sets introduced in Table \ref{table:data}. Column $5$: numbers in the brackets are the standard deviations (SD); SD for all other methods are negligible so are not reported.  Last column: see \eqref{Definer}.}
\scalebox{0.9}{
\begin{tabular}{|l| l |c|ccccc|r|}
\hline
$\#$ & Data set & $K$ & kmeans &kmeans++& Hier & SpecGem & IF-HCT-PCA & $r$\\
\hline
1 & Brain &5 & .286 &.427(.09) & .524 & .143 & .262 & 1.83\\
2 & Breast Cancer &2 & .442 &.430(.05) & .500 & .438 & .406 & .94\\
3 & Colon Cancer &2 & .443 & .460(.07) & .387 & .484 & .403 & 1.04\\
{\bf 4} & {\bf Leukemia } & {\bf 2}  & {\bf .278}  &{\bf .257(.09)}  & {\bf .278}  & {\bf .292}  & {\bf .069}  & {\bf .27}  \\
{\bf 5} & {\bf Lung Cancer(1)}  & {\bf 2}  & {\bf .116}  & {\bf .196(.09)}  & {\bf .177}  & {\bf .122}  & {\bf .033}  & {\bf .29}  \\
6 & Lung Cancer(2) & 2 & .436 &.439(.00)& .301 & .434 & .217 & .72 \\
{\bf 7}  &{\bf  Lymphoma } & {\bf 3} & {\bf .387}  & {\bf .317(.13)}  & {\bf .468}  & {\bf .226}  & {\bf .065}  & {\bf .29}  \\
8 & Prostate Cancer & 2 & .422 & .432(.01) & .480 & .422 & .382 & .91\\
9 & SRBCT & 4 & .556 &.524(.06) & .540 & .508 & .444 & .87\\
10 & SuCancer & 2 & .477 & .459(.05) & .448 & .489 & .333 & .74 \\
\hline
\end{tabular}
} 
\label{table:Error}
\end{table}

We applied all $5$ methods to each of the $10$ gene microarray data sets in Table~\ref{table:data}.
The results are reported in Table~\ref{table:Error}. Since all methods except hierarchical clustering have algorithmic randomness (they depend on built-in $k$-means package in Matlab which uses a random start), we report the mean error rate based on $30$ independent replications. The standard deviation of all methods is very small ($<.0001$) except for $k$-means$++$, so we only report the standard deviation of $k$-means$++$.
In the last column of Table \ref{table:Error},
\begin{equation} \label{Definer}
r  =  \frac{\mbox{error rate of IF-HCT-PCA}}{\mbox{minimum of the error rates of the other 4 methods}}.
\end{equation}
We find that $r < 1$ for all data sets except for two. In particular, $r \leq .29$ for three of the data sets, marking a substantial improvement,  and $r \leq .87$ for three other data sets, marking a moderate improvement.
The $r$-values also suggest an interesting point:  for `easier' data sets, IF-PCA tends to have more improvements over the other $4$ methods.

We make several remarks. First, for the Brain data set, unexpectedly,
IF-PCA underperforms classical PCA, but still outperforms other methods.  Among our data sets, 
the Brain data seem to be an `outlier'. 
Possible reasons include
 (a) useful features are not sparse, and (b) the sample size is very small ($n = 42$) so the
useful features are individually very weak. When (a)-(b) happen, it is almost impossible to successfully separate the useful features from useless ones, and it is preferable to use classical PCA.  Such a scenario may be found in \cite{JinWang3}; 
see for example Figure 1 (left) and related context therein.

Second, for Colon Cancer, all methods behave unsatisfactorily, and IF-PCA slightly underperforms  hierarchical clustering ($r=1.04$).  The data set is known to be a difficult one even for  classification (where class labels of training samples are known  \citep{DJ08}). For such a difficult data set, it is hard for IF-PCA to significantly outperform other methods.  

Last, for the SuCancer data set, the KS-scores are significantly skewed to the right. Therefore, instead of using the normalization \eqref{KSthreshold}, we normalize $\psi_{n,j}$ such that the mean and standard deviation for the lower $50\%$ of KS-scores match those for the lower $50\%$ of the simulated KS-scores under the  null; 
compare this with Section \ref{subsec:HCT} for remarks on  $P$-value calculations. 

\subsection{Three variants of IF-HCT-PCA} 
\label{subsec:IFPCAvar}
First,   in IF-HCT-PCA, we  normalize the KS-scores with the sample mean and sample standard deviation as in (\ref{KSthreshold}). Alternatively, we may  normalize the KS-scores by $\psi_{n,j}^*=[\psi_{n,j} - \mbox{median of all $KS$-scores}]/[\mbox{MAD of all $KS$-scores}]$  (MAD: Median Absolute Deviation), while other steps of IF-HCT-PCA are kept intact. Denote the resultant variant by IF-HCT-PCA-med (med: median). 
Second, recall that  IF-HCT-PCA has two stages: in the first one, we select features with a threshold determined by HC;   
in the second one, we apply PCA to the post-selection data matrix.  Alternatively,  in the second stage,  we may apply classical $k$-means or hierarchical clustering  to the post-selection data instead  (the first stage is intact). Denote these two alternatives by IF-HCT-kmeans and IF-HCT-hier, respectively. 

\begin{table}[htb!]
\caption{Clustering error rates of  IF-HCT-PCA,  IF-HCT-PCA-med, IF-HCT-kmeans, and IF-HCT-hier.}
\scalebox{.91}{
\begin{tabular}{lcccccccccc}
 &   Brn & Brst & Cln & Leuk & Lung1  & Lung2 & Lymp & Prst & SRB & Su \\
\hline
IF-HCT-PCA&  .262 &.406 &.403 & .069  & .033   &.217 & .065  &.382 &.444 &.333 \\
IF-HCT-PCA-med &  .333 & .424 & .436 & .014 & .017 & .217 & .097 & .382 & .206 & .333 \\
IF-HCT-kmeans &  .191 &.380 &.403 & .028  &  .033  &.217 & .032  &.382 &.401 &.328  \\
IF-HCT-hier & .476 &.351 &.371 &  .250  &  .177  & .227 & .355 &.412 &.603 &.500 \\
\end{tabular}
}
\label{table:IF-kmeans} 
\end{table} 
Table \ref{table:IF-kmeans} compares IF-HCT-PCA with the three variants (in IF-HCT-kmeans, the `replicate' parameter in k-means is taken to be $30$ as before), where the first three methods have similar performances, while the last one performs comparably less satisfactorily.  Not surprisingly,  these methods generally outperform their classical counterparts (i.e., classical PCA, classical k-means, and hierarchical clustering; see Table \ref{table:Error}).  

We remark that, for post-selection clustering, it is frequently preferable to use PCA than $k$-means. First,  $k$-means could be much slower than PCA, especially when the number of selected features in the IF step is large. Second,      the $k$-means algorithm we use in Matlab  is only a heuristic approximation of   the theoretical $k$-means (which is NP-hard), so it is not always easy to justify the performance of $k$-means algorithm theoretically.


\subsection{Connection to sparse PCA}  \label{subsec:connection}
The study is closely related to the recent interest on sparse PCA  (\cite{Ery1,  Wainwright, Johnstone, Jung,     Lei, Ma, ZHT}),  but is different in important ways. Consider the normalized data matrix $W = [W_1, W_2, \ldots, W_n]'$ for example.  
In our model, recall that $\mu_1, \mu_2, \ldots, \mu_K$ are the $K$ sparse contrast mean vectors and the noise covariance matrix  $\Sigma$ is diagonal,  we have  
\[
W \approx M \Sigma^{-1/2} + Z,  \qquad \mbox{where $Z \in R^{n,p}$ has $iid$ $N(0,1)$ entries}, 
\]
and $M \in R^{n,p}$ is the matrix where the $i$-th row is $\mu_k'$ if and only if $i \in \mbox{Class $k$}$. 
This is a setting that is frequently considered in the sparse PCA literature.  

However, we must note that the main focus of sparse PCA is to recover the supports of $\mu_1, \mu_2, \ldots, \mu_K$, while the main focus here is subject clustering. 
We recognize that, the two problems---support recovery and subject clustering---are essentially two different problems, and addressing one successfully does not necessarily address  the other successfully.  For illustration, consider two scenarios.
\begin{itemize}
\item If useful features are very sparse but each   is sufficiently strong, it is easy to identify the support of the useful features, but due to the extreme sparsity, it may be still impossible to have consistent clustering.
\item If most of the useful features are very weak with only a few of them very strong,
the latter will be easy to identify and may yield consistent clustering, still,  it may be impossible to satisfactorily recover the supports of $\mu_1, \mu_2, \ldots, \mu_K$,  as  most of the useful features are very weak.
\end{itemize} 
In  a forthcoming manuscript \cite{JinWang3}, we investigate the connections and differences between two problems more closely, and elaborate the above points with details. 
 
With that being said,  from a practical viewpoint,  one may still wonder how sparse PCA may help in subject clustering. 
A straight-forward clustering approach that exploits the sparse PCA ideas is the following:   
\begin{itemize} 
\item Estimate the first $(K-1)$ right singular vectors  of the matrix $M \Sigma^{-1/2}$  using the sparse PCA algorithm as in 
\cite[Equation (3.7)]{ZHT} (say). Denote the estimates by $\hat{\nu}^{sp}_1, \hat{\nu}_2^{sp}, \ldots, \hat{\nu}_{K-1}^{sp}$.  
\item Cluster by applying classical $k$-means to the $n \times K-1$ matrix $[W \hat{\nu}_1^{sp}, W \hat{\nu}_2^{sp}$, $\ldots, W \hat{\nu}_{K-1}^{sp}]$, assuming there are $\leq K$ classes.  
\end{itemize}   
For short, we call this approach Clu-sPCA. 
One problem here is that, Clu-sPCA is {\it not} tuning-free, as most existing sparse PCA algorithms have one or more tuning parameters.  How to set the tuning parameters in subject clustering is a challenging problem: for example,  since 
the class labels are unknown, using conventional cross validations (as we may use in classification where class labels of the training set are known) might not help.

\begin{table}[ht!]
\caption{Clustering error rates for IF-HCT-PCA   and Clu-sPCA. The tuning parameter of 
Clu-sPCA is chosen ideally to minimize the errors (IF-HCT-PCA is tuning-free).  Only SDs that are larger than 0.01 are reported (in brackets).  }
\scalebox{0.9}{
\begin{tabular}{lcccccccccc}
&  Brn & Brst & Cln & Leuk & Lung1  & Lung2 & Lymp & Prst & SRB & Su  \\
\hline
IF-HCT-PCA & .262 &.406 &.403 &{\bf  .069} &  {\bf .033}  & {\bf .217}   &  {\bf .065}    &.382 &.444 &{\bf .333} \\
Clu-sPCA &     .263 &.438 &.435 & {\bf .292}  &  {\bf .110} &  {\bf .433}  & {\bf .190(.01)} &.422 &.428 & {\bf .437} \\
\end{tabular}
}
\label{table:IF-sPCA} 
\end{table} 
In Table \ref{table:IF-sPCA}, we compare IF-HCT-PCA and Clu-sPCA using the $10$ data sets in Table \ref{table:data}. 
Note that in Clu-sPCA, the tuning parameter in the sparse PCA step  \cite[Equation (3.7)]{ZHT}  is {\it ideally chosen} to minimize the clustering errors, using the true class labels. The results are based on $30$ independent repetitions.  
Compared to Clu-sPCA, IF-HCT-PCA outperforms  for half of the data sets (bold face),    and has similar performances  
for the remaining half.

The above results support  our philosophy:  
the problem  of subject clustering and the problem of support recovery are related but  different, and success in one does not   automatically lead to the success in the other.

\subsection{Summary and contributions}  \label{subsec:summary}
Our contribution is three-fold: feature selection by the KS statistic, post-selection PCA for high dimensional clustering, and threshold choice by the recent idea of Higher Criticism.

In the first fold, we rediscover a phenomenon  found earlier by   \cite{Efron} for microarray study, but the focus there is on $t$-statistic or $F$-statistic, and the focus here is on the KS statistic.
We establish tight probability bounds on the KS statistic when the data is Gaussian or Gaussian mixtures where the means and variances are unknown; see Section \ref{subsec:KStail}.
While tight tail probability bounds have been available for decades in the case where
the data are $iid$ from $N(0,1)$, the current case is much more challenging.  Our results follow the work by   \cite{Siegmund} and  \cite{Loader} on  the  local Poisson approximation of boundary crossing probability, and are useful for pinning down the thresholds in KS screening.

In the second fold, we propose to use IF-PCA for clustering and have successfully applied it to gene microarray data. The method compares favorably with other methods, which suggests that both the IF step and the post-selection PCA step are effective. We also establish a theoretical framework where we investigate the clustering consistency carefully; see Section \ref{sec:main}. The analysis it entails is sophisticated and involves delicate post-selection eigen-analysis (i.e., eigen-analysis on the post-selection data matrix).   We also gain useful insight that
the success of feature selection depends on the feature-wise weighted third moment of the samples, while
the success of PCA depends more on the feature-wise weighted second moment. Our study is closely related to the SpectralGem approach by \cite{Lee}, but our focus is on KS screening,  post-selection PCA, and clustering with microarray data is different.

In the third fold, we propose to set the threshold by Higher Criticism.
We find an intimate relationship between the HC functional and the signal-to-noise ratio associated with post-selection
eigen-analysis.  As mentioned in Section \ref{subsec:HCT}, the full analysis on the HC threshold choice is difficult and long, so for reasons of space, we do not include it in this paper.

Our findings support the philosophy by \cite{50years}, that for real data analysis, we prefer to use simple models and methods that allow sophisticated theoretical analysis than complicated and computationally intensive methods (as an increasing trend in some other scientific communities).

\subsection{Content and notations}
\label{subsec:content}
Section \ref{sec:main} contains the main theoretical
results, where we show IF-PCA is consistent in clustering under some regularity conditions.  Section \ref{sec:Simul} contains the numerical studies, and
Section \ref{sec:Discu} discusses connection to other work and
addresses some future research. Secondary theorems and lemmas are proved in the appendix. 
In this paper, $L_p$ denotes a generic multi-$\log(p)$ term (see Section \ref{subsec:mainthm}). 
For a vector $\xi$, $\|\xi\|$ denotes the $\ell^2$-norm. For a real matrix  $A$,  
$\|A\|$ denotes the matrix spectral norm, $\|A\|_F$ denotes the matrix Frobenius norm,   and $s_{\min}(A)$ denotes the smallest nonzero singular value.

\section{Main results}  \label{sec:main}
\setcounter{equation}{0}
Section \ref{subsec:ACM} introduces our asymptotic model,  Section \ref{subsec:condi}  discusses  the main regularity conditions and  related notations. 
Section \ref{subsec:mainthm} presents the main theorem, and Section \ref{subsec:corollary} presents two corollaries, together with a  phase transition phenomenon.
Section \ref{subsec:KStail} discusses the
tail probability of the KS statistic, which is the key for the IF step.  Section \ref{subsec:postselect} studies post-selection eigen-analysis which is the key for the PCA step.  The main theorems and corollaries are proved in Section \ref{subsec:mainpf}.

To be utterly clear, the IF-PCA procedure we study in this section is the one presented in Table \ref{tab:code2}, where the threshold $t > 0$ is given. 
\begin{table}[ht!]
\caption{Pseudocode for IF-PCA (for a given threshold $t > 0$)} 
\scalebox{0.85}{ 
{\begin{tabular}{ll}\\ \hline
& \underline{Input}: data matrix $X$, number of classes $K$, threshold $t > 0$.       \underline{Output}: class label vector $\hat{y}^{IF}_t $. \\
1. &Rank features: Let $\psi_{n,j}$, $1 \leq j \leq p$, be the KS-scores as in (\ref{DefineKS}).   \\ 
2.  & Post-selection PCA: Define post-selection data matrix  $W^{(t)}$ (i.e, sub-matrix of $W$ consists of  all   \\   & column $j$ with $\psi_{n,j} > t$). Let $U \in R^{n, K-1}$ be   the matrix of the first $(K - 1)$ left singular vectors  \\ 
& of  $W^{(t)}$. Cluster by $\hat{y}^{IF}_t = kmeans(U, K)$. \\
 \hline
\end{tabular}}
} 
\label{tab:code2} 
\end{table}

\subsection{The Asymptotic Clustering Model}
\label{subsec:ACM}  
The model we consider is  (\ref{model1}), (\ref{model2}), (\ref{model3}) and (\ref{model4}), where the data matrix is  $X =  [X_1, X_2, \ldots, X_n]'$, with $X_i \sim N(\bar{\mu} + \mu_k, \Sigma)$ if and only if 
$i \in \mbox{Class $k$}$, $1 \leq k \leq K$,  and $\Sigma = \diag(\sigma_1^2, \sigma_2^2, \ldots, \sigma_p^2)$;  $K$ is the number of classes, $\bar{\mu}$ is the overall mean vector,  $\mu_1, \mu_2, \ldots, \mu_K$ are contrast mean vectors which satisfy   (\ref{model4}).  

We use $p$ as the driving asymptotic parameter, and let other parameters be tied to $p$ through fixed parameters.   Fixing $\theta \in (0,1)$, we let
\begin{equation} \label{modeln}
n = n_p = p^{\theta},
\end{equation}
so that as $p \goto \infty$,  $p \gg n \gg 1$.\footnote{For simplicity, we drop the subscript of $n_p$ as long as there is no confusion.}
Let $M \in R^{K, p}$ be the matrix
\beq \label{DefineM}
M =   [m_1, m_2, \ldots, m_K]', \qquad \mbox{where \;  $m_k = \Sigma^{-1/2} \mu_k$}.
\eeq
Denote the set of useful features by
\begin{equation} \label{DefineS}
S_p = S_p(M) = \{1 \leq j \leq p: \mbox{$m_k(j)  \neq 0$ for some $1 \leq k \leq K$}\},
\end{equation}
and let 
$s_p = s_p(M) = |S_p(M)|$ 
be the number of useful features. Fixing $\vartheta  \in (0,1)$,   we let
\begin{equation} \label{models}
s_p = p^{1 - \vartheta}.
\end{equation}
Throughout this paper, the number of classes $K$ is  fixed, as $p$ changes.
\bed
We call model (\ref{model1}), (\ref{model2}), (\ref{model3}), and (\ref{model4}) the Asymptotic Clustering Model if (\ref{modeln}) and (\ref{models}) hold, and denote it by $ACM(\vartheta, \theta)$.
\eed

It is more convenient to work with the normalized data matrix $W = [W_1, W_2, \ldots, W_n]'$, where,  as before,  
$W_i(j) =  [X_i(j) - \bar{X}(j)]/\hat{\sigma}(j)$, 
and $\bar{X}(j)$ and $\hat{\sigma}(j)$ are the empirical mean and standard deviation associated with the feature $j$, $1 \leq j \leq p$, $1 \leq i \leq n$. 
Introduce
$\hat{\Sigma} = \diag(\hat{\sigma}^2(1), \hat{\sigma}^2(2), \ldots, \hat{\sigma}^2(p))$ and 
$\widetilde{\Sigma} = E[\hat{\Sigma}]$. 
Note that $\hat{\sigma}^2(j)$ is an unbiased estimator for  $\sigma^2(j)$  when feature $j$ is useless but is not necessarily so when feature $j$ is useful. As a result, $\hat{\Sigma}$ is `closer' to $\widetilde{\Sigma}$ than to $\Sigma$; this causes (unavoidable) complications in notations.   Denote for short
\begin{equation} \label{DefineA}
\Lambda  = \Sigma^{1/2}\widetilde{\Sigma}^{-1/2} .
\end{equation}
This is a $p \times p$ diagonal matrix where most of the diagonals are $1$, and all other diagonals are close to $1$ (under mild conditions).
Let  ${\bf 1}_n$ be the $n \times 1$  vector of ones and $e_k \in R^K$ be the $k$-th standard basis vector of $R^K$, $1 \leq k \leq K$. Let
$L \in R^{n, K}$ be the matrix where the $i$-th row is $e_k'$ if and only if Sample $i \in \mbox{Class $k$}$. Recall the definition of $M$ in \eqref{DefineM}. With these notations, we can write 
\begin{equation} \label{modelmatrix1}
W  = [L M   + Z \Sigma^{-1/2}] \Lambda       +    R,  \qquad  \mbox{$Z \Sigma^{-1/2}$ has $iid$  $N(0,1)$ entries},    
\end{equation}
where $R$ stands for the remainder term
\begin{equation} \label{DefinetildeZ}
R = {\bf 1}_n (\bar{\mu} - \bar{X})' \hat{\Sigma}^{-1/2} + [L M\Sigma^{1/2} + Z](\hat{\Sigma}^{-1/2} - \widetilde{\Sigma}^{-1/2}).  
\end{equation}
Recall that $rank(LM) = K-1$ and $\Lambda$ is nearly the identity matrix.

\subsection{Regularity conditions and  related notations} \label{subsec:condi}
We use $C > 0$ as a generic constant, which may change from occurrence to occurrence, but does not depend on $p$. Recall that $\delta_k$ is the fraction of samples in Class $k$, and $\sigma^2(j)$ is the $j$-th diagonal of $\Sigma$. 
The following regularity conditions are mild:
\begin{equation} \label{Condition1}
\min_{1 \leq k \leq K} \{\delta_k\} \geq C, \qquad \mbox{and}
\qquad \max_{1 \leq j \leq p}  \{\sigma(j) +  \sigma^{-1}(j)\}   \leq C.
\end{equation}

Introduce the following two $p \times 1$ vectors $\kappa = (\kappa(1), \kappa(2), \ldots, \kappa(p))'$ and $\tau = (\tau(1), \tau(2),  \ldots, \tau(p))'$ by
\begin{align} 
\kappa(j) &= \kappa(j; M, p, n) =  \bigl( \sum_{k = 1}^K  \delta_k m_k^2(j) \bigr)^{1/2},    \label{Definekappa}   \\ 
\tau(j)  &= \tau(j; M, p, n)   =  (6\sqrt{2\pi})^{-1}  \cdot \sqrt{n} \cdot   \bigl| \sum_{k = 1}^K \delta_k m_k^3(j) \bigr|.   \label{Definetau}
\end{align}
Note that $\kappa(j)$ and $\tau(j)$ are related to the weighted second and third moments of the $j$-th column of $M$, respectively;  $\tau$ and $\kappa$ play a key role in the success of feature selection and post-selection PCA, respectively. 
In the case that $\tau(j)$'s are all small, the success of our method relies on higher moments of the columns of $M$; see Section~\ref{subsec:KStail} for more discussions. 
Introduce 
\[
\eps(M)  = \max_{1 \leq k \leq K, j\in S_p(M)} \{|m_k(j)|\},   \qquad   \tau_{min} = \min_{j\in S_p(M)} \{ \tau(j) \}.
\]
We are primarily interested in the range where the feature strengths are rare and weak, so we assume as $p \goto \infty$, 
\begin{equation} \label{Condition2}
\eps(M) \goto 0.\footnote{This condition is used in the post-selection eigen-analysis. Recall that $W^{(t)}$ is the short-hand notation for the post-selection normalized data matrix associated with threshold $t$. As  $W^{(t)}$ is the sum of a low-rank matrix and a noise matrix,   $(W^{(t)})'(W^{(t)})$ equals to  the sum of four terms, two of them are ``cross terms".  In eigen-analysis of $(W^{(t)})'(W^{(t)})$,  we need condition (\ref{Condition2})  to control the cross terms.}
\end{equation}
In Section \ref{subsec:KStail},  we shall see that $\tau(j)$ can be viewed as the Signal-to-Noise Ratio (SNR) associated with the $j$-th feature and $\tau_{min}$ is the minimum SNR of all useful features.
The most interesting range for $\tau(j)$ is $\tau(j)  \geq  O(\sqrt{\log(p)})$.
In fact, if $\tau(j)$s are  of a much smaller order,  then the useful features and the useless features are merely inseparable.
In light of this, we fix a constant $r > 0$ and assume
\begin{equation} \label{Condition3}
 \tau_{min}  \geq  a_0  \cdot  \sqrt{2 r   \log(p)},  \qquad  \mbox{where $a_0 = \sqrt{(\pi - 2)/(4\pi)}$}.\footnote{Throughout this paper, $a_0$ denotes the constant  $\sqrt{(\pi-2)/(4\pi)}$. The constant   comes from the analysis of the tail behavior of the KS statistic; see Theorems \ref{thm:KSnull}-\ref{thm:KSalt}.}  
\end{equation}
By the way $\tau(j)$ is defined,   the interesting range for non-zero $m_k(j)$ is
$|m_k(j)| \geq O\big((\log(p)/n)^{1/6}\big)$. 
We also need some technical conditions which can be largely relaxed with more complicated analysis:\footnote{Condition (\ref{KSaltCondition}) is only needed for Theorem \ref{thm:KSalt} on the tail behavior of the KS statistic associated with a useful feature.  The conditions ensure singular cases will not happen so the weighted third moment (captured by $\tau(j)$) is the leading term in the Taylor expansion. For more discussions, see the remark in Section \ref{subsec:KStail}.}
\begin{equation}  \label{KSaltCondition}
\max_{j\in S_p(M)}\Big\{ \frac{\sqrt{n}}{\tau(j)}\sum_{k=1}^K \delta_k m_k^4(j)\Big\} \leq Cp^{-\delta}, \;   \min_{\{(j,k): m_k(j)\neq 0]\}}  \{|m_k(j)|\} \geq C(\frac{\log(p)}{n})^{1/2},  
\end{equation}
for some $\delta > 0$.  As the most interesting range of $|m_k(j)|$ is $O((\log(p)/n)^{1/6})$, these conditions are mild.

Similarly, for the threshold $t$ in (\ref{DefinehatSt})  we use for the KS-scores,  the interesting range  is $t = O(\sqrt{\log(p)})$.
In light of this, we are primarily interested in threshold of the form
\begin{equation} \label{Definetp}
t_p(q) = a_0  \cdot  \sqrt{2 q \log(p)},  \qquad \mbox{where $q > 0$ is a constant}.
\end{equation}

We now define a quantity $err_p$, which is the clustering error rate of IF-PCA in our main results. Define
\[
\rho_1(L, M) = \rho_1(L, M; p,n) =  \frac{s_p \|\kappa\|_{\infty}^2 }{\|\kappa\|^2}. 
\]
Introduce
two $K \times K$ matrices $A$ and $\Omega$ (where $A$ is diagonal)  by
\[
A(k,k) = \sqrt{\delta_k} \|m_k\|,  \qquad  \Omega(k, \ell) = m_k'  \Lambda^2 m_{\ell}/(\|m_k\| \cdot \|m_{\ell}\|),
\qquad  1 \leq k, \ell  \leq K;
\]
recall that $\Lambda$ is `nearly' the identity matrix.
Note that $\|A \Omega A \| \leq  \|\kappa\|^2$, and that
when $\|m_1\|,\cdots, \|m_K\|$ have comparable magnitudes, 
all the eigenvalues of $A\Omega A$ have the same magnitude. 
In light of this, let $s_{\min}(A\Omega A)$ be the minimum singular value of $A$ and introduce the ratio
\[
\rho_2(L,M)  = \rho_2(L,M; p, n) = \|\kappa\|^2 / s_{\min}(A \Omega A).
\]
Define
\[ 
err_p = \rho_2(L, M)\biggl[   \frac{1 + \sqrt{\frac{p^{1-\vartheta\wedge q}}{n}}}{\|\kappa\|} + p^{-\frac{(\sqrt{r}-\sqrt{q})_+^2}{2K}}    
 +  \sqrt{p^{\vartheta - 1} + \frac{p^{(\vartheta-q)_+}}{n}}  \sqrt{\rho_1(L,M)} \biggr]. 
\] 
This quantity $err_p$ combines the `bias' term associated with the useful features that we have missed in feature selection and the `variance' term associated with retained features; see Lemmas \ref{lemma:Urem} and \ref{lemma:Zrem} for details.  
Throughout this paper, we assume that there is a constant $C > 0$ such that
\begin{equation} \label{Condition4}
err_p  \leq p^{-C}.
\end{equation}

{\bf Remark}. Note that  $\rho_1(L, M) \geq 1$ and $\rho_2(L,M) \geq 1$.  A relatively small $\rho_1(L, M)$ means that $\tau(j)$ are more or less in the same magnitude, and a relatively small $\rho_2(L,M)$ means that the $(K-1)$ nonzero eigenvalues of $L M \Lambda^2 M' L'$ have comparable magnitudes.
Our hope is that neither of these two ratios is unduly large.

\subsection{Main theorem: clustering consistency by IF-PCA}
\label{subsec:mainthm}
Recall $\psi_{n,j}$ is the KS statistic. For any threshold $t > 0$, denote the set of retained features by
\[
\hat{S}_p(t)   = \{1 \leq j \leq p:  \psi_{n, j} \geq t\}. 
\]
For any $n \times p$ matrix $W$,   let $W^{\hat{S}_p(t)}$ be 
the matrix formed by replacing all columns of $W$ with the index $j \notin \hat{S}_p(t)$ by the vector of zeros (note the slight difference compared with $W^{(t)}$ in Section \ref{subsec:IFPCA}).  Denote the $n \times (K-1)$ matrix of the first $(K-1)$  left singular vectors of $W^{\hat{S}_p(t_p(q))}$ by
\[
\hat{U}^{(t_p(q))} = \hat{U}(W^{\hat{S}_p(t_p(q))}) = [\heta_1, \heta_2,\cdots, \heta_{K-1}], \quad \mbox{where $\heta_k = \heta_k(W^{\hat{S}_p(t_p(q))})$}.
\]
Recall that $W = [LM + Z \Sigma^{-1/2}] \Lambda  + R$ and let $LM\Lambda  = U D V'$ be the Singular Value Decomposition (SVD) of $LM\Lambda$ such that $D \in R^{K-1, K-1}$ is a diagonal matrix with the diagonals being singular values arranged descendingly, $U \in R^{n, K-1}$ satisfies $U'U = I_{K-1}$, and $V \in R^{p, K-1}$ satisfies $V'V = I_{K-1}$. Then $U$ is the non-stochastic counterpart of $\hat{U}^{(t_p(q))}$. 
We hope that  
the linear space spanned by columns of $\hat{U}^{(t_p(q))}$ is ``close" to that spanned by columns of $U$. 

\bed 
$L_p > 0$ denotes a   multi-$\log(p)$ term that may vary from occurrence to occurrence but satisfies $L_p p^{-\delta} \goto 0$ and $L_p p^{\delta} \goto \infty$,  $\forall \delta > 0$. 
\eed 
For any $K \geq 1$,  let 
\begin{equation}\label{eqn:defineHk}
{\cal H}_K = \{\mbox{All  $K \times K$ orthogonal matrices}\}.  
\end{equation}
The following theorem is proved in Section~\ref{subsec:mainpf}, which shows that the singular vectors IF-PCA obtains span a low-dimensional subspace that is ``very close" to its counterpart in the ideal case where there is no noise. 
\begin{thm} \label{thm:Frob}
Fix $(\vartheta, \theta) \in (0,1)^2$, and consider $ACM(\vartheta, \theta)$.  Suppose the regularity conditions (\ref{Condition1}), (\ref{Condition2}), (\ref{Condition3}), (\ref{KSaltCondition}) and (\ref{Condition4}) hold, and the threshold in IF-PCA is set as $t=t_p(q)$ as in \eqref{Definetp}.
Then there is a matrix $H$ in ${\cal H}_{K-1}$ such that as $p \goto \infty$, with probability at least $1 - o(p^{-2})$,
$\|\hat{U}^{(t_p(q))}  - UH \|_F \leq L_p  err_p$.
\end{thm}
Recall that in IF-PCA, once $\hat{U}^{(t_p(q))}$ is obtained,  we estimate the class labels  by truncating   $\hat{U}^{(t_p(q))}$ entry-wise (see the PCA-1 step and the footnote   in Section \ref{subsec:IFPCA}) and then cluster by  applying the classical $k$-means. Also,  the estimated class labels are denoted  by $\hat{y}_{t_p(q)}^{IF} = (\hat{y}_{t_p(q),1}^{IF}, \hat{y}_{t_p(q),2}^{IF},  \hat{y}_{t_p(q), n}^{IF})'$.  
We measure the clustering errors by the Hamming distance
\[
\hamm_p^*(\hat{y}_{t_p(q)}^{IF}, y) = \min_{\pi}  \big\{  \sum_{i = 1}^{n}  P(\hat{y}_{t_p(q), i}^{IF} \neq \pi(y_i)) \big\},
\]
where $\pi$ is any permutation in $\{1, 2, \ldots, K\}$.
The following theorem is our main result, which gives an upper bound for the Hamming   errors of IF-PCA.  
\begin{thm} \label{thm:Clu}
Fix $(\vartheta, \theta) \in (0,1)^2$, and consider $ACM(\vartheta, \theta)$.  Suppose the regularity conditions (\ref{Condition1}), (\ref{Condition2}), (\ref{Condition3}), (\ref{KSaltCondition}) and (\ref{Condition4}) hold, and let $t_p=t_p(q)$ as in \eqref{Definetp} and $T_p=\log(p)/\sqrt{n}$ in IF-PCA.
As $p \goto \infty$,
\[
n^{-1}  \hamm_p^*(\hat{y}_{t_p(q)}^{IF}, y)    \leq L_p  err_p.
\]
\end{thm}
The theorem can be proved by Theorem \ref{thm:Frob} and an adaption of  \cite[Theorem 2.2]{SCORE}. In fact, 
by Lemma \ref{lemma:LM} below, the absolute values of all entries of $U$ are bounded by $C/\sqrt{n}$ from above. By the choice of $T_p$ and definitions, the truncated matrix $\hat{U}_*^{(t_p(q))}$ satisfies $\|\hat{U}_*^{(t_p(q))} - UH\|_F \leq \|\hat{U}^{(t_p(q))}  - UH\|_F$.  Using this and Theorem \ref{thm:Frob}, the proof of Theorem \ref{thm:Clu} is basically an exercise of classical theory on $k$-means algorithm. For this reason, we skip the proof.

\subsection{Two corollaries and a phase transition phenomenon} \label{subsec:corollary}
Corollary \ref{corollary1} can be viewed as a simplified version of Theorem \ref{thm:Frob}, so we omit the proof; recall that $L_p$  denotes  a generic multi-$\log(p)$ term. 
\begin{cor}  \label{corollary1}
Suppose conditions of Theorem \ref{thm:Frob} hold, and suppose
 $\max\{\rho_1(L,M), \rho_2(L,M)\}\leq L_p$ as $p\to\infty$. Then there is a matrix $H$ in ${\cal H}_{K-1}$ such that as $p \goto \infty$, with probability at least $1 - o(p^{-2})$,
\begin{align*}
\| &\hat{U}^{(t_p(q))}  - U H\|_F \leq  L_p p^{-[(\sqrt{r}-\sqrt{q})_+]^2/(2K)} \cr
& + L_p (\|\kappa\|^{-1}p^{(1-\vartheta)/2} + 1) \left\{
\begin{array}{lr}
p^{-\theta/2+[(\vartheta-q)_+]/2}, & \text{if } (1-\vartheta) > \theta,\\
p^{-(1-\vartheta)/2+ [(1-\theta-q)_+]/2}, & \text{if } (1-\vartheta) \leq \theta.
\end{array}\right.
\end{align*}
\end{cor}
By assumption \eqref{Condition3}, the interesting range for a nonzero $m_k(j)$ is $|m_k(j)|\asymp L_p n^{-1/6}$. It follows that $\| \kappa \|\asymp L_p p^{(1-\vartheta)/2}n^{-1/6}$ and $\|\kappa\|^{-1}p^{(1-\vartheta)/2} \to\infty$. In this range, we have the following corollary, which is proved in Section \ref{subsec:mainpf}.
\begin{cor}  \label{corollary2}
Suppose conditions of Corollary~\ref{corollary1} hold, and $\|\kappa\|= L_pp^{(1-\vartheta)/2}n^{-1/6}$. Then as $p\to\infty$, the following holds:
\begin{itemize}
\item[(a)] If $(1-\vartheta)<\theta/3$, for any $r>0$, whatever $q$ is chosen, the upper bound of $\min_{H\in\cH_{K-1}}\|\hat{U}^{(t_p(q))} - U H\|_F$ in Corollary~\ref{corollary1} goes to infinity.
\item[(b)] If $\theta/3<(1-\vartheta)<1-2\theta/3$, for any $r>\vartheta-2\theta/3$, there exists $q\in(0,r)$ such that $\min_{H\in\cH_{K-1}}\|\hat{U}^{(t_p(q))} - U H\|_F\to 0$ with probability at least $1-o(p^{-2})$.
In particular,  if $(1-\vartheta)\leq \theta$ and $r>(\sqrt{K(1-\vartheta)-K\theta/3} + \sqrt{1-\theta})^2$, by taking $q=1-\theta$,
\[
\min_{H\in\cH_{K-1}}\|\hat{U}^{(t_p(q))} - U H\|_F\leq L_p n^{1/6}s_p^{-1/2};
\]
if $(1-\vartheta)>\theta$ and $r>(\sqrt{2K\theta/3} + \sqrt{\vartheta})^2$, by taking $q=\vartheta$,
\[
\min_{H\in\cH_{K-1}}\|\hat{U}^{(t_p(q))} - U H\|_F\leq L_p n^{-1/3}.
\]
\item[(c)] If $(1-\vartheta)>1-2\theta/3$, for any $r>0$, by taking $q=0$, $\min_{H\in\cH_{K-1}}\|\hat{U}^{(t_p(q))} - U H\|_F\to 0$ with probability at least $1-o(p^{-2})$.
\end{itemize}
\end{cor}

To interpret Corollary~\ref{corollary2}, we take a special case where $K = 2$, all diagonals of $\Sigma$ are bounded from above and below by a constant,  and all nonzero features $\mu_k(j)$ have comparable magnitudes;  that is,  there is a positive number $u_0$  that may depend on $(n,p)$ and a constant $C > 0$ such that
\begin{equation} \label{constraintadd}
u_0  \leq  |\mu_k(j)| \leq C u_0, \qquad \mbox{for any $(k,j)$ such that $\mu_k(j) \neq 0$}.
\end{equation}
In our parametrization, $s_p =p^{1 - \vartheta}$, $n = p^{\theta}$,  and $u_0\asymp \tau_{min}^{1/3}/n^{1/6} \asymp (\log(p)/n)^{1/6}$ since $K = 2$. 
Cases (a)-(c) in  Corollary~\ref{corollary2}  translate to (a) $1\ll s_p\ll n^{1/3}$, (b) $n^{1/3} \ll s_p \ll p / n^{2/3}$, and (c) $s_p \gg p /n^{2/3}$, respectively. 

The primary interest in this paper is Case (b). In this case,
Corollary \ref{corollary2} says that both feature selection and post-selection PCA can be successful, provided that $u_0=c_0(\log(p)/n)^{1/6}$ for an appropriately large constant $c_0$. 
Case (a) addresses the case of very sparse signals, and Corollary~\ref{corollary2} says that we need stronger signals than that of $u_0\asymp (\log(p)/n)^{1/6}$ for IF-PCA to be successful.  Case (c) addresses the case where signals are relatively dense, and  PCA is successful without feature selection (i.e., taking $q=0$). 

We have been focused on the case $u_0 = L_p n^{-1/6}$ as our primary interest is on clustering by IF-PCA.  For a more complete picture, we model $u_0$ by $u_0 = L_p p^{-\alpha}$;  we let the exponent $\alpha$ vary and investigate what is the critical order for $u_0$ for some different problems and different methods. 
   In this case,  it is seen that $u_0\sim n^{-1/6}$ is the critical order for the success of feature selection (see Section~\ref{subsec:KStail}), $u_0\sim \sqrt{p/(ns)}$ is the critical order for the success of Classical PCA and $u_0\sim 1/\sqrt{s}$ is the critical order for IF-PCA in an idealized situation where the Screen step finds exactly all the useful features. 
These suggest an interesting phase transition phenomenon for IF-PCA. 
\begin{itemize}
\item  {\it Feature selection is trivial but clustering is impossible}.  $1 \ll s \ll n^{1/3}$ and $n^{-1/6} \ll  u_0  \leq 1/\sqrt{s}$.  Individually, useful features are sufficiently strong, so it is trivial to recover the support of $M\Sigma^{1/2}$ (say, by thresholding the KS-scores one by one); note that $M \Sigma^{1/2}  = [\mu_1, \mu_2, \ldots, \mu_K]'$.  However,    useful features are so sparse that  it is impossible for any methods  to have consistent clustering.
\item  {\it Clustering and feature selection are possible but non-trivial}.    $n^{1/3} \ll s \ll p / n^{2/3}$ and $u_0 = (r \log(p) /n)^{1/6}$, where $r$ is a constant. In this range, feature selection is indispensable and  there is a region where
IF-PCA may yield a consistent clustering but Classical PCA may not.  A similar conclusion can be drawn if the purpose is to recover the support of $M\Sigma^{1/2}$ by thresholding the KS-scores.
\item {\it Clustering is trivial but feature selection is impossible}.  $s \gg p /n^{2/3}$ and $\sqrt{p/(ns)} \leq  u_0  \ll n^{-1/6}$. In this range, the sparsity level is low and  Classical PCA is able to yield consistent clustering,  but the useful features are individually too weak that it is impossible to fully recover the support of  $M \Sigma^{1/2}$ by using all $p$ different KS-scores.
\end{itemize}
In \cite{JinWang3}, we investigate the phase transition with much more refined studies (in a slightly different setting).

\subsection{Tail probability of KS statistic}
\label{subsec:KStail} 
IF-PCA consists of a screening step  (IF-step) and a PCA step.  In the IF-step,  the key is to study the tail behavior of  the KS statistic $\psi_{n,j}$, defined in (\ref{DefineKS}).  Fix $1 \leq j \leq p$.  Recall that in our model, 
$X_i \sim N(\bar{\mu}  + \mu_k, \Sigma)$ if $i \in \mbox{Class $k$}$, $1 \leq i \leq n$, and that $j$ is a useless feature if and only if  $\mu_1(j) = \mu_2(j) = \ldots = \mu_K(j) = 0$.

Recall that  
$a_0  = \sqrt{(\pi -2)/(4\pi)}$.  
Theorem \ref{thm:KSnull} addresses the tail behavior of $\psi_{n,j}$ when feature $j$ is useless. 
\begin{thm} \label{thm:KSnull}
Fix $\theta \in (0,1)$ and let $n  = n_p = p^{\theta}$.
Fix $1 \leq j \leq p$. If the $j$-th feature is a useless feature, then as $p \goto \infty$,  for any sequence $t_p$ such that $t_p  \goto \infty$ and $t_p/\sqrt{n} \goto 0$,
\[
1  \lesssim \frac{P(\psi_{n,j} \geq t_p)}{(\sqrt{2}a_0)^{-1}\mathrm{exp}\bigl(- t_p^2/(2 a_0^2) \bigr)} \lesssim 2.
\]
\end{thm}
We conjecture that $P(\psi_{n,j} \geq t_p)  \sim 2 \cdot \frac{1}{\sqrt{2} a_0}\mathrm{exp}(- t_p^2/(2a_0^2))$, 
with possibly a more sophisticated proof than that in the paper.    

Recall that  $\tau$ is defined in (\ref{Definetau}).  Theorem \ref{thm:KSalt} addresses the tail behavior of $\psi_{n,j}$ when feature $j$ is useful.  
\begin{thm} \label{thm:KSalt}
Fix $\theta \in (0,1)$.  Let $n = n_p  = p^{\theta}$, and $\tau(j)$ be as in (\ref{Definetau}), where $j$ is a useful feature.
Suppose \eqref{Condition3} and \eqref{KSaltCondition} hold, and the threshold $t_p$ is such that $t_p \goto \infty$, that $t_p /\sqrt{n} \goto 0$, and that $\tau(j) \geq (1 + C) t_p$ for some constant $C > 0$. Then as $p\to\infty$,
\[
P(\psi_{n,j} \leq t_p)  \leq C  \biggl( K \mathrm{exp}\bigl(- \frac{1}{2 K a_0^2} (\tau(j)  - t_p)^2 \bigr) +  O(p^{-3}) \biggr).
\]
\end{thm}
Theorems~\ref{thm:KSnull}-\ref{thm:KSalt} are proved in the appendix. Combining two theorems,   roughly saying, we have  that 
\begin{itemize}
\item if $j$ is a useless feature, then the right tail of $\psi_{n,j}$ behaves like that of $N(0,a_0^2)$,
\item if $j$ is a useful feature, then the left tail of $\psi_{n,j}$ is bounded by that of $N(\tau(j), K a_0^2)$. 
\end{itemize}
These  suggest that the feature selection using the KS statistic in the current setting is very similar to feature selection with a Stein's normal means model; the latter is more or less well-understood (e.g., \cite{ABDJ}). 

As a result, the most interesting range for $\tau(j)$ is $\tau(j)  \geq  O(\sqrt{\log(p)})$.
If we threshold the KS-scores at $t_p(q)=\sqrt{2q\log(p)}$, by similar argument as in feature selection with a  Stein's normal means setting, we expect that 
\begin{itemize}
\item  All useful features are retained, except  for a fraction $\leq C p^{-[(\sqrt{r} - \sqrt{q})_+]^2/K}$,   
\item  No more than $(1 + o(1)) \cdot  p^{1 - q}$ useless features are (mistakenly) retained, 
\item  $\#\{\mbox{retained features}\}  = |\hat{S}_p(t_p(q))| \leq C  [p^{1 - \vartheta} +  p^{1-q} + \log(p)]$. 
\end{itemize}
These facts pave  the way for the PCA step; see Sections below.

{\bf Remark}.  Theorem \ref{thm:KSalt}  hinges on $\tau(j)$, which is a quantity proportional to the ``third moment"  $ \sum_{k = 1}^K \delta_k m_k^3(j)$ and can be viewed as the ``effective signal strength" of the KS statistic.  In the symmetric case (say, $K = 2$ and $\delta_1 = \delta_2 = 1/2$), the third moment (which equals to $0$) is no longer  the right quantity for  calibrating the effective signal strength of the KS statistic, and we must use the fourth moment.   In such cases,  for $1 \leq j \leq p$, let 
\[
\omega(j) =  \sqrt{n} \sup_{-\infty<y<\infty} \bigg[\frac{1}{8}y(1-3y^2) \phi(y) \cdot  \big( \sum_{k = 1}^K  \delta_k m_k^2(j)\big)^2 + \frac{1}{24}\phi^{(3)}(y) \cdot  \sum_{k = 1}^K  \delta_k m_k^4(j) \bigg],
\]
where $\phi^{(3)}(y)$ is the third derivative of the standard normal density $\phi(y)$.  Theorem \ref{thm:KSalt}   
  continues to hold provided that (a) $\tau(j)$ is replaced by $\omega(j)$,   (b)  the condition (\ref{Condition3}) of  $\tau_{min} \geq a_0 \sqrt{2 r \log(p)}$ is replaced by that of $\omega_{min} \geq  a_0 \sqrt{2 r \log(p)}$, where $\omega_{min} = \min_{j \in S_p(M)} \{ \omega(j)\}$, and (c)   the first part of condition (\ref{KSaltCondition}),  $\max \limits_{j\in S_p(M)}$   
$\big\{ \frac{\sqrt{n}}{\tau(j)}\sum\limits_{k=1}^K \delta_k m_k^4(j)\big\} \leq Cp^{-\delta}$, is replaced by that of $\max\limits_{j\in S_p(M)}\big\{ \frac{\sqrt{n}}{\omega(j)}\sum\limits_{k=1}^K \delta_k |m_k(j)|^5\big\} \leq Cp^{-\delta}$.   This is consistent with that in \cite{ACV}, which studies the  clustering problem in a similar setting (especially on the symmetric case) with great details.

In the literature, tight bounds of this kind are only available for the case where $X_i$ are iid samples from a known distribution (especially, parameters---if any---are known).  In this case, the bound is derived by \cite{Kolmogorov}; also see \cite{Wellner86}.
The setting considered here is more complicated, and how to derive tight bounds is an interesting but rather challenging problem. The main difficulty lies in that, any estimates of the unknown parameters $(\bar{\mu}(j), \mu_1(j), \ldots, \mu_k(j), \sigma(j))$ have stochastic fluctuations at the same order of that of the stochastic fluctuation of the empirical CDF, but
two types of fluctuations are correlated in a complicated way, so it is hard to derive the right constant $a_0$ in the exponent.
There are two existing approaches, one is due to   \cite{Durbin} which
approaches the problem by approximating the stochastic process  by a Brownian bridge, the other is due to \cite{Loader} (see also   \cite{Siegmund, Woodroofe}) on the local Poisson approximation of the boundary crossing probability. It is argued in   \cite{Loader} that the second approach is  more accurate.
Our proofs follow the idea in \cite{Siegmund, Loader}.

\subsection{Post-selection eigen-analysis}  \label{subsec:postselect}
For the PCA step, as in Section \ref{subsec:mainthm},   we let $W^{\hat{S}_p(t_p(q))}$ be the $n \times p$ matrix where the $j$-th column is the same as that of $W$ if $j \in \hat{S}_p(t_p(q))$ and is the zero vector otherwise. With such notations,
\begin{equation} \label{Whatadd} 
W^{\hat{S}_p(t_p(q))} = L M \Lambda + L (M - M^{\hat{S}_p(t_p(q))}) \Lambda   + (Z \Sigma^{-1/2} \Lambda + R)^{\hat{S}_p(t_p(q))}.
\end{equation} 
We analyze the there terms on the right hand side separately.

Consider the first term $L M \Lambda$. 
Recall that $L \in R^{n, K}$ with the $i$-th row being $e_k'$ if and only if $i  \in \mbox{Class $k$}$, $1 \leq i \leq n, 1 \leq k \leq K$,  and $M \in R^{K, p}$ with the $k$-th row being $m_k'=(\Sigma^{-1/2} \mu_k)'$, $1 \leq k \leq K$.
Also, recall that $A = \diag(\sqrt{\delta_1} \|m_1\|, \ldots, \sqrt{\delta_K} \|m_K\|)$ and $\Omega \in R^{K, K}$ with $\Omega(k, \ell) = m_k' \Lambda^2 m_{\ell}/(\|m_k\|\cdot\|m_{\ell}\|)$, $1 \leq k, \ell \leq K$.
Note that $rank(A \Omega A) = rank(LM) =  K-1$.
Assume all nonzero eigenvalues of $A \Omega A$ are simple, and denote them by
$\lambda_1 > \lambda_2 > \ldots > \lambda_{K-1}$.
Write
\begin{equation} \label{SVDDM}
A \Omega A   =  Q  \cdot   \diag(\lambda_1, \lambda_2, \ldots, \lambda_{K-1})  \cdot Q',  \qquad Q \in R^{K, K-1},  
\end{equation}
where the $k$-th column of $Q$ is the $k$-th eigenvector of $A \Omega A$,  and let
\begin{equation} \label{SVDLM}
LM\Lambda  = U D V'
\end{equation}
be an SVD of $LM\Lambda$.  Introduce
\begin{equation} \label{DefineG}
G = \diag(\sqrt{\delta_1}, \sqrt{\delta_2}, \ldots, \sqrt{\delta_K}) \in R^{K,K}.
\end{equation}
The following lemma is proved in Section \ref{app:lemma}.
\begin{lemma} \label{lemma:LM}
The matrix $LM\Lambda$ has $(K-1)$ nonzero singular values which are $\sqrt{n \lambda_1},  \ldots, \sqrt{n \lambda_{K-1}}$.  Also, there is a matrix  $H  \in {\cal H}_{K-1}$  (see (\ref{eqn:defineHk})) such that
\[
U  = n^{-1/2} L [G^{-1} Q  H] \in R^{n, K-1}.
\]
For the matrix $G^{-1} Q  H$, the $\ell^2$-norm of the $k$-th row is
$(\delta_k^{-1} - 1)^{1/2}$,
and the $\ell^2$-distance between the $k$-th row and the $\ell$-th row is $(\delta_k^{-1} + \delta_{\ell}^{-1})^{1/2}$, which is no less than $2$, $1 \leq k < \ell \leq K$.
\end{lemma}
By Lemma \ref{lemma:LM} and definitions, it follows that
\begin{itemize}
\item For any $1 \leq i \leq n$ and $1 \leq k \leq K-1$,  the $i$-th row of $U$ equals to the $k$-th row of $n^{-1/2} G^{-1} Q H$ if and only if Sample $i$ comes from Class $k$.
\item The matrix $U$ has $K$ distinct rows, according to which the rows of $U$ partition into $K$ different groups.   This partition coincides with the partition of the $n$ samples into $K$ different classes.  Also, the $\ell^2$-norm between each pair of the $K$ 
distinct rows is no less than $2/\sqrt{n}$.
\end{itemize}

Consider the second term on the right hand side of (\ref{Whatadd}).  This is the `bias' term caused by useful features which we may fail to select.
\begin{lemma} \label{lemma:Urem}
Suppose the conditions of Theorem \ref{thm:Frob} hold. As $p \goto \infty$, with probability at least $1 - o(p^{-2})$,
\[
\|L (M - M^{\hat{S}_p(t_p(q))}) \Lambda \|  \leq  C \|\kappa \|\sqrt{n} \cdot \Big[ p^{-(1-\vartheta)/2}\sqrt{\rho_1(L,M)}\cdot\sqrt{\log(p)} + p^{-[(\sqrt{r}-\sqrt{q})_+]^2/(2K)}\Big].
\]
\end{lemma}

Consider the last term on the right hand side of (\ref{Whatadd}).  This is the `variance' term consisting of two parts, the part from original measurement noise matrix $Z$ and the remainder term due to normalization.
\begin{lemma} \label{lemma:Zrem}
Suppose the conditions of Theorem \ref{thm:Frob} hold. As $p \goto \infty$, with probability at least $1 - o(p^{-2})$,
\[
\|(Z \Sigma^{-1/2} \Lambda  + R)^{\hat{S}_p(t_p(q))}\| \leq C \Big[ \sqrt{n}+  \Big( p^{(1-\vartheta\wedge q)/2} + \|\kappa\| p^{(\vartheta-q)_+/2} \sqrt{\rho_1(L,M)} \Big) \cdot (\sqrt{\log(p)})^3 \Big].
\]
\end{lemma}
Combining Lemmas \ref{lemma:Urem}-\ref{lemma:Zrem} and using the definition of $err_p$,
\begin{equation}\label{eqn:errbound}
\|W^{\hat{S}_p(t_p(q))} - L M \Lambda\|  \leq L_p err_p \cdot \frac{\sqrt{n}\|\kappa\|}{\rho_2(L,M)}.
\end{equation}

\subsection{Proofs of  the main results}
\label{subsec:mainpf}
We now show Theorem \ref{thm:Frob} and Corollary \ref{corollary1}. Proof of Theorem \ref{thm:Clu} is very similar to that of Theorem 2.2 in \cite{SCORE} and proof of Corollary \ref{corollary2} is elementary, so we omit them. 

Consider Theorem \ref{thm:Frob}. Let
\[
T = LM\Lambda^2M'L', \qquad \hat{T} = \sW(\sW)'.
\]
Recall that $U$ and $\hat{U}^{(t_p(q))}$ contain the $(K-1)$ leading eigenvectors of $T$ and $\hat{T}$, respectively.  
Using the sine-theta theorem \citep{sin-theta} (see also Proposition 1 in \cite{RankDetect}), 
\beq \label{Unorm-equ1}
\| \hat{U}^{(t_p(q))}(\hat{U}^{(t_p(q))})' - UU'\|\leq 2s^{-1}_{\min}(T)\|\hat{T}-T\|;
\eeq
in \eqref{Unorm-equ1}, we have used the fact that $T$ has a rank of $K-1$ so that the gap between the $(K-1)$-th and $K$-th largest eigenvalues is equal to the minimum nonzero singular value $s_{\min}(T)$.   
The following lemma is proved in Section \ref{app:lemma}. 
\begin{lemma} \label{lem:project}
For any integers $1\leq m\leq p$ and two $p\times m$ matrices $V_1,V_2$ satisfying $V_1'V_1=V_2'V_2=I$,  
there exists an orthogonal matrix $H \in R^{m,m}$ such that $\|V_1-V_2H\|_F\leq \| V_1V_1'-V_2V_2' \|_F$. 
\end{lemma}
\noindent
Combine \eqref{Unorm-equ1} with Lemma~\ref{lem:project} and note that $\hat{U}^{(t_p(q))}(\hat{U}^{(t_p(q))})' - UU'$ has a rank of $2K$ or smaller. It follows that there is an  $H\in \mathcal{H}_{K-1}$ such that
\beq \label{Unorm-equ2}
\|\hat{U}^{(t_p(q))}-UH\|_F\leq 2\sqrt{2K}s^{-1}_{\min}(T)\|\hat{T}-T\|. 
\eeq
First,  $\|\hat{T}-T\|  \leq   2 \|LM\Lambda\| \cdot \|\sW - LM\Lambda\| + \|\sW - LM\Lambda\|^2$.
From Lemmas \ref{lemma:Urem}-\ref{lemma:Zrem} and \eqref{Condition4}, $\|LM\Lambda\| \gg  \|\sW - LM\Lambda\|$. Therefore,
\[
\|\hat{T}-T\|  \lesssim   2 \|LM\Lambda\|  \|\sW - LM\Lambda\|\leq 2\sqrt{n} \|\kappa\| \cdot \|\sW - LM\Lambda\|.
\]
Second, by Lemma \ref{lemma:LM}, 
\[
s_{\min}(T)=n \cdot s_{\min}(A\Omega A') = n\| \kappa\|^2 /\rho_2(L,M).
\]
Plugging in these results into \eqref{Unorm-equ2}, we find that
\beq\label{Unorm-equ3}
\|\hat{U}^{(t_p(q))}-UH\|_F\leq 4\sqrt{2K}\frac{\rho_2(L,M)}{\sqrt{n}\|\kappa\|}\|\sW - LM\Lambda\|, 
\eeq
where by Lemmas \ref{lemma:Urem}-\ref{lemma:Zrem}, the right hand side equals to $L_p err_p$ . The claim then follows by combining  \eqref{Unorm-equ3} and \eqref{eqn:errbound}.

Consider Corollary \ref{corollary2}.   For each $j\in S_p(M)$, it can be deduced that $\kappa(j)\geq \eps(M)$, using especially (\ref{Condition2}). Therefore,   $\|\kappa\|\geq L_p p^{\frac{(1-\vartheta)}{2}}n^{-1/6}=L_p p^{\frac{1-\vartheta}{2} - \frac{\theta}{6}}$. The error bound in Corollary \ref{corollary1} reduces to
\begin{equation} \label{corotemp}
L_p p^{-[(\sqrt{r}-\sqrt{q})_+]^2/(2K)} + L_p  \left\{
\begin{array}{lr}
p^{-\theta/3+ (\vartheta-q)_+/2}, & \theta < 1-\vartheta,\\
p^{\theta/6 -(1-\vartheta)/2+ (1-\theta-q)_+/2}, & \theta \geq 1-\vartheta.
\end{array}\right.
\end{equation}
Note that \eqref{corotemp} is lower bounded by $L_pp^{\theta/6-(1-\vartheta)/2}$ for any $q\geq 0$; and it is upper bounded by $L_p p^{-\theta/3+\vartheta/2}$ when taking $q=0$. The first and third claims then follow immediately.
Below, we show the second claim.

First,  consider the case  $\theta<1-\vartheta$. If $r>\vartheta$, we can take any $q\in (\vartheta, r)$ and the error bound is $o(1)$. If $r\leq \vartheta$, noting that $(\vartheta-r)/2<\theta/3$, there exists $q<r$ such that $(\vartheta-q)/2<\theta/3$, and the corresponding error bound is $o(1)$.

In particular, if $r>(\sqrt{2K\theta/3} + \sqrt{\vartheta})^2$, we have $(\sqrt{r}-\sqrt{\vartheta})^2/(2K) > \theta/3$; then for $q\geq \vartheta$, the error bound is $L_pp^{-\theta/3}+L_pp^{-(\sqrt{r}-\sqrt{q})^2/(2K)}$; for $q<\vartheta$, the error bound is $L_p p^{-\theta/3+(\vartheta-q)/2}$; so the optimal $q^*=\vartheta$ and the corresponding error bound is $L_pp^{-\theta/3}=L_pn^{-1/3}$.

Next,  consider the case $1-\vartheta\leq \theta < 3(1-\vartheta)$. If $r>1-\theta$, for any $q\in (1-\theta, r)$, the error bound is $o(1)$;  note that $\theta/6<(1-\vartheta)/2$. If $r\leq 1-\theta$, noting that $(1-\theta-r)/2 < (1-\vartheta)/2-\theta/6$, there is a  $q<r$ such that $(1-\theta-q)/2 < (1-\vartheta)/2-\theta/6$, and the corresponding error bound is $o(1)$.

In particular, if $r>(\sqrt{K(1-\vartheta)-K\theta/3} + \sqrt{1-\theta})^2$, we have that $(\sqrt{r}-\sqrt{1-\theta})^2/(2K) > (1-\vartheta)/2 - \theta/6$; then for $q\geq 1-\theta$, the error bound is $L_pp^{\theta/6-(1-\vartheta)/2}+L_pp^{-(\sqrt{r}-\sqrt{q})^2/(2K)}$; for $q<1-\theta$, the error bound is $L_pp^{\theta/6-(1-\vartheta)/2 + (1-\theta-q)/2}$; so the optimal $q^*=1-\theta$ and the corresponding error bound is $L_pp^{\theta/6-(1-\vartheta)/2}=L_p n^{1/6}s_p^{-1/2}$.

\section{Simulations}
\label{sec:Simul}
\setcounter{equation}{0}
We conducted a small-scale simulation study to investigate the numerical performance of IF-PCA.   We consider two variants of IF-PCA, denoted by 
IF-PCA(1) and IF-PCA(2). In IF-PCA(1), the threshold is chosen using HCT (so the choice is data-driven), and in IF-PCA(2), the threshold $t$ is given. In both variants, we skip the normalization step on KS scores (that step is designed for microarray data only). The pseudocodes of IF-PCA(2) and IF-PCA(1) are given in Table \ref{tab:code2} (Section \ref{sec:main})  and Table \ref{tab:code3}, respectively.   
We compared IF-PCA(1) and IF-PCA(2) with $4$ other different methods: classical $k$-means (kmeans), $k$-means++ (kmeans$+$), classical hierarchical
clustering (Hier), and SpectralGem (SpecGem; same as classical PCA).   In hierarchical clustering, we only consider the linkage type of  ``complete"; other choices of linkage have very similar results.  
\begin{table}[ht!]
\caption{Pseudocode for IF-PCA(1) (for simulations; threshold set by Higher Criticism)} 
\scalebox{0.83}{ 
{\begin{tabular}{ll}\\ \hline
& \underline{Input}: data matrix $X$, number of classes $K$.      \underline{Output}: class label vector $\hat{y}^{IF}_{HC}$. \\
1. &Rank features: Let $\psi_{n,j}$ be the KS-scores as in (\ref{DefineKS}), and $F_0$ be the CDF of  $\psi_{n,j}$ under null, $1 \leq j \leq p$.      \\
2. &Threshold choice by HCT: Calculate $P$-values by  $\pi_j = 1 - F_0(\psi_{n,j})$, $1 \leq j \leq p$  and sort them by   \\ 
     & $\pi_{(1)} <  \pi_{(2)}  < \ldots < \pi_{(p)}$.   Define $HC_{p, j} = \sqrt{p}  ( j/p  -  \pi_{(j)} )  / \sqrt{ \max\{ \sqrt{n} ( j/p  -  \pi_{(j)}) , 0 \}  + j/p}$, and let  \\
     &  $\hat{j} = \margmax_{\{ j: \pi_{(j)} > \log(p)/p, j < p/2\} } \{HC_{p, j}\}$.
    HC threshold  $t_p^{HC}$  is the $\hat{j}$-largest KS-score.  \\ 
3. & Post-selection PCA: Define post-selection data matrix  $W^{(HC)}$ (i.e., sub-matrix of $W$ consists of  all   \\ 
& column $j$ of $W$ with $\psi_{n,j} > t_p^{HC}$). Let $U \in R^{n, K-1}$ be the matrix of the first $(K - 1)$ left singular \\
& vectors of  $W^{(HC)}$. Cluster by $\hat{y}^{IF}_{HC} = kmeans(U, K)$. \\
 \hline
\end{tabular}}
} 
\label{tab:code3} 
\end{table}

In each experiment, we fix parameters $(K, p, \theta, \vartheta, r, rep)$, two probability mass vectors $\delta=(\delta_1,\cdots,\delta_K)'$ and $\gamma=(\gamma_1,\gamma_2,\gamma_3)'$, and three probability densities $g_\sigma, g_{\mu}$ defined over $(0,\infty)$ and $g_{\bar{\mu}}$ defined over $(-\infty,\infty)$. With these parameters, we let $n = n_p = p^{\theta}$ and $\eps_p = p^{1 - \vartheta}$; $n$ is the sample size,  $\eps_p$ is roughly the fraction of useful features, and $rep$ is the number of repetitions.\footnote{For each parameter setting, we generate the $X$ matrix for $rep$ times,  and at each time, we apply all the six algorithms. The clustering errors are averaged over all the repetitions.} 
We generate the $n\times p$ data matrix $X$ as follows. 
\begin{itemize}
\item Generate the class labels $y_1, y_2, \ldots, y_{n}$ $iid$ from  $MN(K, \delta)$\footnote{We say $X \sim MN(K, \delta)$ if $P(X = k) = \delta_k$, $1 \leq k \leq K$; MN stands for multinomial.}, and let  $L$ be the $n  \times K$ matrix such that the $i$-th row of $L$ equals to $e_k'$ if and only if
$y_i = k$, $1 \leq k \leq K$.
\item Generate the overall mean vector $\bar{\mu}$ by $\bar\mu(j) \stackrel{iid}{\sim} g_{\bar\mu}$, $1 \leq j \leq p$.   
\item Generate the contrast mean vectors $\mu_1,\cdots,\mu_K$ as follows. First, generate $b_1,b_2,\ldots, b_p$ $iid$ from $\mathrm{Bernoulli}(\eps_p)$. Second, for each $j$ such that $b_j=1$, generate the $iid$ signs $\{\beta_k(j)\}_{k=1}^{K-1}$ such that $\beta_k(j)=-1,0,1$ with probability $\gamma_1,\gamma_2,\gamma_3$, respectively, and generate the feature magnitudes $\{h_k(j)\}_{k = 1}^{K-1}$ $iid$ from
$g_{\mu}$. Last, for $1 \leq k \leq K-1$, set $\mu_k$ by (the factor $72 \pi$ is chosen to be consistent with (\ref{Definetau})) 
\[
\mu_k(j) = \bigl[72 \pi  \cdot  (2 r \log(p))   \cdot  n^{-1}  \cdot  h_k(j) \bigr]^{1/6}   \cdot b_j  \cdot \beta_k(j),
\]
and let
$\mu_K = - \frac{1}{\delta_K} \sum_{k = 1}^{K-1} \delta_k \mu_k$.

\item Generate the noise matrix $Z$ as follows. First, generate a $p \times 1$ vector $\sigma$ by $\sigma(j) \stackrel{iid}{\sim} g_{\sigma}$. Second, generate
the $n$ rows of $Z$ $iid$ from $N(0, \Sigma)$, where $\Sigma = \diag(\sigma^2(1),$ $\sigma^2(2), \cdots, \sigma^2(p))$. 
\item Let $X = {\bf 1}\bar\mu' + L [\mu_1,\cdots,\mu_K] + Z$.
\end{itemize}

In the simulation settings, $r$ can be viewed as the parameter of (average) signal strength. The density $g_\sigma$ characterizes noise heteroscedasticity; when $g_\sigma$ is a point mass at $1$, the noise variance of all the features are equal. The density $g_{\mu}$ controls the strengths of useful features; when $g_{\mu}$ is a point mass at $1$, all the useful features have the same strength. The signs of useful features are captured in the probability vector $\gamma$; when $K=2$, we always set $\gamma_2=0$ so that $\mu_k(j)\neq 0$ for a useful feature $j$; when $K \geq 3$, for a useful feature $j$, we allow $\mu_k(j)=0$ for some $k$. 

For IF-PCA(2), the theoretical threshold choice as in \eqref{Definetp} is $t=\sqrt{2\tilde{q}\log(p)}$ for some $0<\tilde{q}<(\pi-2)/(4\pi)\approx .09$. We often set $\tilde{q} \in \{.03, .04, .05, .06\}$, depending on the signal strength parameter $r$. 
 
The simulation study contains $5$ experiments, which we now describe.

{\it Experiment 1.} In this experiment, we study the effect of signal strength over clustering performance, and compare two cases: the classes have unequal or equal number of samples. 
We set $(K, p, \theta, \vartheta, rep) = (2, 4\times 10^4, .6, .7, 100)$, and $\gamma = (.5, 0, .5)$ (so that the useful features have equal probability to have positive and negative signs). Denote by $U(a,b)$ the uniform distribution over $(a-b,a+b)$. 
We set $g_{\mu}$ as $U(.8, 1.2)$, $g_{\sigma}$ as $U(1, 1.2)$, and  $g_{\bar\mu}$ as $N(0,1)$. We investigate two choices of $\delta$: $(\delta_1, \delta_2) = (1/3,  2/3)$ and $(\delta_1, \delta_2) = (1/2,  1/2)$; we call them ``asymmetric" and ``symmetric" case, respectively. In the latter case, the two classes roughly have equal number of samples. The threshold in IF-PCA(2) is taken to be $t= \sqrt{2\cdot .06 \cdot \log(p)}$.

\begin{figure}[ht!]
\begin{center}
\includegraphics[height = 1.8 in]{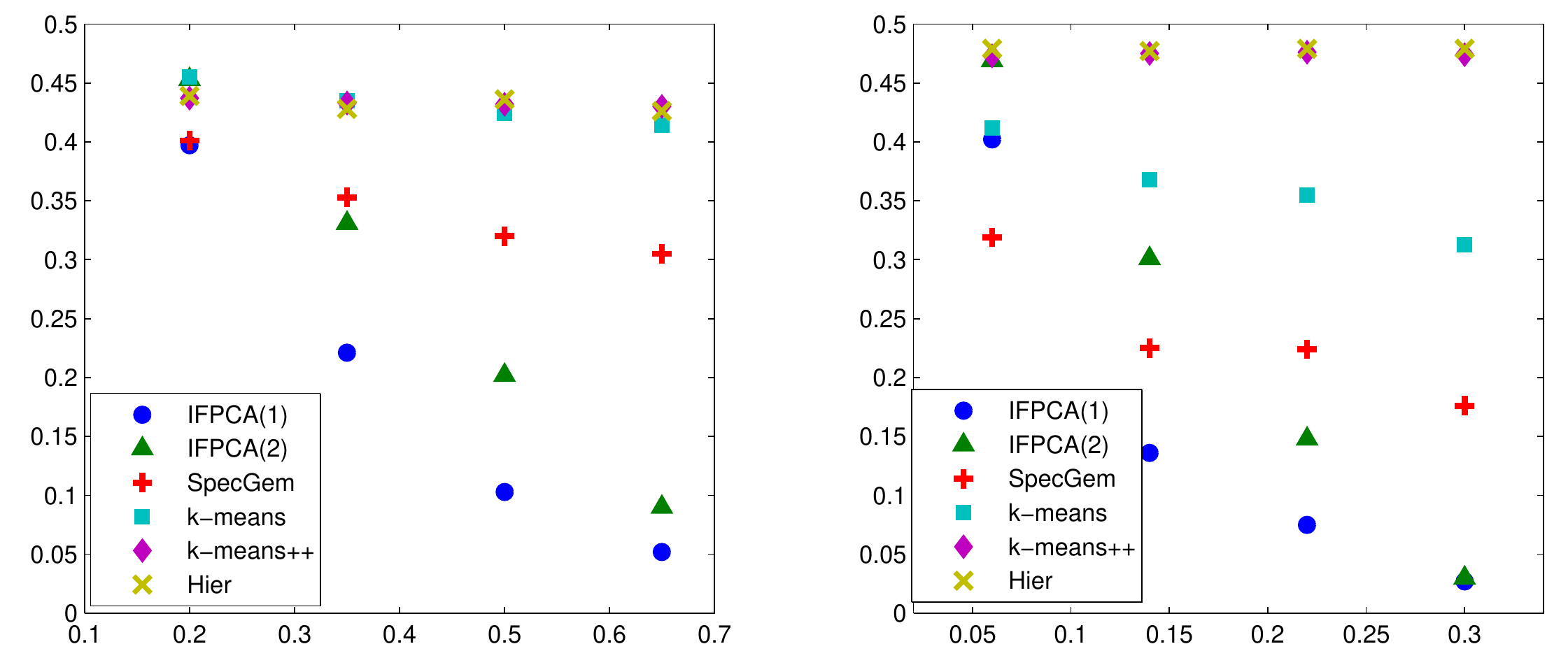}
\end{center}
\caption{Comparison of clustering error rates (Experiment 1a). x-axis:  signal strength parameter $r$. y-axis: error rates.  Left:   $\delta = (1/3, 2/3)$. Right:  $\delta = (1/2, 1/2)$.} 
\label{fig:exp1}
\end{figure}
In Experiment 1a, 
we let the signal strength parameter $r\in \{ .20, .35, .50, .65\}$ for the asymmetric case, and $r \in \{ .06, .14, .22, .30\}$ for the symmetric case.
The results are summarized in Figure \ref{fig:exp1}. 
We find that two versions of IF-PCA outperform the other methods in most settings, increasingly so when the signal strength increases.
Moreover, two versions of IF-PCA have similar performance, with those of IF-PCA(1) being slightly better. This suggests that our threshold choice by HCT is not only data-driven but also yields satisfactory clustering results.  On the other hand, it also suggests  that
IF-PCA is relatively insensitive to different choices of the threshold, as long as they are  in a certain range.

In Experiment 1b, we make a more careful comparison between the asymmetric and symmetric cases. Note that for the same parameter $r$, the actual signal strength in the symmetric case is stronger because of normalization. As a result, for $\delta=(1/3,2/3)$, we still let $r \in \{0.20, 0.35, 0.50, 0.65\}$, but for $\delta=(1/2,1/2)$, we take $r' = c_0 \times \{0.20, 0.35, 0.50, 0.65\}$, 
where $c_0$ is a constant chosen such that for any $r>0$, $r$ and $c_0r$ yield the same value of $\kappa(j)$ (see (\ref{Definekappa})) in the asymmetric and symmetric cases, respectively; we note that $\kappa(j)$ can be viewed as the effective signal-to-noise ratio of Kolmogorov-Smirnov statistic. The results are summarized in Table \ref{table:exp1b}. Both versions of IF-PCA  have better clustering results when $\delta = (1/3, 2/3)$,  suggesting that the clustering task is more difficult in the symmetric case. This is consistent with the theoretical results; see for example \cite{ACV, JinWang3}.
\begin{table}[ht!]
	\caption{Comparison of average clustering error rates (Experiment 1).  Number in the brackets are standard deviations of the error rates.}
	\begin{center}
		\begin{tabular}{|c|cc|cc|}
			\hline
			&\multicolumn{2}{c|}{$(\delta_1, \delta_2) = (1/2, 1/2)$}  & \multicolumn{2}{c|}{$(\delta_1, \delta_2) = (1/3, 2/3)$}\\
			\hline
			$r$ & IF-PCA(1) & IF-PCA(2) & IF-PCA(1) & IF-PCA(2)\\
			\hline
			$.20$  &.467(.04)  &.481(.01)  &.391(.11) 	&.443(.08)\\
			$.35$  &.429(.08)  &.480(.02)  &.253(.15) 	&.341(.16)\\
			$.50$  &.368(.13)  &.466(.05)  &.144(.14) 	&.225(.18)\\
			$.65$  &.347(.13)  &.459(.07)  &.099(.12) 	&.098(.11)\\
			\hline
		\end{tabular}
		\label{table:exp1b}
	\end{center}
\end{table}

{\it Experiment 2.} In this experiment, we allow feature sparsity to vary (Experiment 2a), and investigate the effect of unequal feature strength (Experiment 2b). We set $(K, p, \theta, r, rep) = (2, 4\times 10^4, .6, .3, 100)$ (so $n = 577$), $\gamma = (.5, 0, .5)$ and $(\delta_1, \delta_2) = (1/3,  2/3)$. The threshed for IF-PCA(2)  is $t = \sqrt{2 \cdot .05 \cdot \log(p)}$.

\begin{figure}[ht!]
\begin{center}
\includegraphics[height = 2 in]{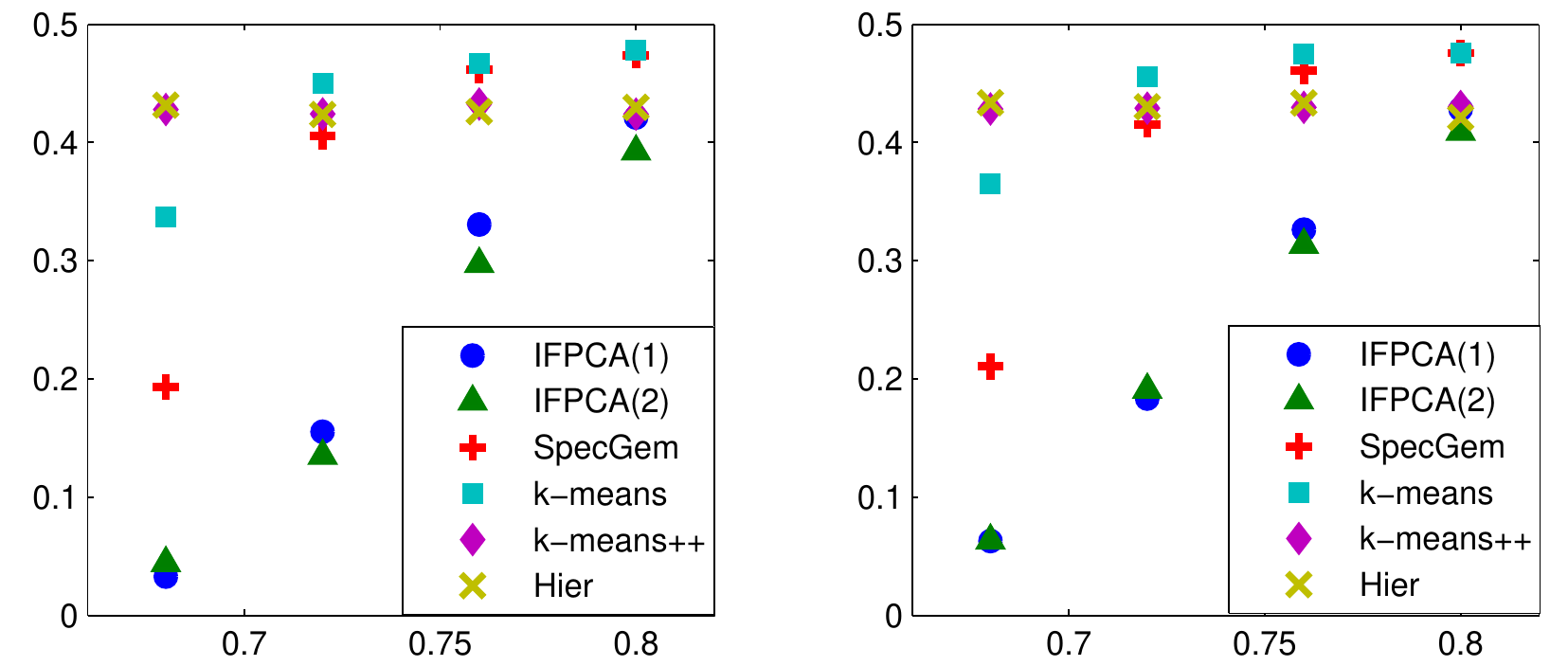}
\end{center}
\caption{Comparison of average clustering error rates (Experiment 2). x-axis: sparsity parameter  $\vartheta$.  y-axis: error rates.    Left:   $g_{\mu}$ is $\widetilde{TN}(1, .1^2, .2)$ and $g_\sigma$ is $\widetilde{TN}(1, .1^2, .1)$. Right:  $g_{\mu}$ is $\widetilde{TN}(1, .1, .7)$ and $g_\sigma$ is point mass at 1. } 
\label{fig:exp2}
\end{figure}

In Experiment 2a, we let $\vartheta$ range in $\{.68, .72, .76, .80\}$. Since the number of useful features is roughly $p^{1-\vartheta}$, a larger $\vartheta$ corresponds to a higher sparsity level. For any $\mu$ and $a,b>0$, let $\widetilde{TN}(u,b^2,a)$ be the conditional distribution of $(X | u - a \leq X \leq u + a)$ for $ X\sim N(u,b^2)$, where TN stands for ``Truncated Normal". 
We take $g_{\bar{\mu}}$ as $N(0,1)$, $g_{\mu}$ as $\widetilde{TN}(1, .1^2, .2)$, and $g_{\sigma}$ as $\widetilde{TN}(1, .1^2, .1)$.   The results are summarized in  the left panel of Figure \ref{fig:exp2}, where for all sparsity levels,
two versions of IF-PCA have similar performance, and each of them significantly outperforms the other methods.

In Experiment 2b, we use the same setting except that $g_{\mu}$ is $\widetilde{TN}(1, .1, .7)$ and $g_{\sigma}$ is the point mass at $1$.  Note that in Experiment 2a, the support of $g_\mu$ is $(.8, 1.2)$, and in the current setting, the support is $(.3, 1.7)$ which is wider.  As a result, the  strengths of useful features in the current setting  have more variability.  
At the same time, we force the noise variance of all features to be $1$, for a fair comparison. The results are summarized in  the right panel of Figure \ref{fig:exp2}. They are similar to those in Experiment 2a, suggesting that IF-PCA  continues to work well even when the feature strengths are unequal.

{\it Experiment 3.} In this experiment, we study how different threshold choices affect the performance of IF-PCA.  With the same as those in Experiment 2b, we investigate four threshold choices for IF-PCA(2): $t = \sqrt{2 \tilde{q} \log(p)}$ for  $\tilde{q}\in  
\{.03, .04, .05, .06\}$, where we recall that the theoretical choice of threshold \eqref{Definetp} suggests $0<\tilde{q}<.09$. The results are summarized in Table \ref{table:exp2b}, which suggest that IF-PCA(1) and IF-PCA(2) have comparable performances, and that IF-PCA(2) is relatively insensitive to
different threshold choices,  as long as they fall in a certain range.  However,  the best threshold choice does  depend on $\vartheta$. From a practical view point,  since $\vartheta$ is unknown, it is preferable to set the threshold in a data-driven fashion; this is what we use in IF-PCA(1).
\begin{table}[ht!]
\centering
\caption{Comparison of average clustering error rates (Experiment 3).  Numbers in the brackets are the standard deviations of the error rates. }

\begin{tabular}{|c|c|cccc|}
\hline
 & Threshold ($\tilde{q}$) & $\vartheta = .68$ & $\vartheta = .72$ & $\vartheta = .76$ & $\vartheta = .80$\\
\hline
IF-PCA(1) &HCT (stochastic)& .053(.08)  & .157(.16)  &.337(.14)  &.433(.10)\\
\hline
\multirow{4}{*}{IF-PCA(2)}
& $.03$	&.038(.05)  	& .152(.12)	& .345(.13)	&.449(.06)\\
& $.04$ 	&.045(.08)  	& .122(.12)	& .312(.15)	&.427(.09)\\
& $.05$    & .068(.12)   		&.154(.15)  &.303(.16) 	&.413(.12)\\
& $.06$     & .118(.15)  		&.237(.17)  &.339(.16) 	&.423(.10)\\
\hline
\end{tabular}
\label{table:exp2b}
\end{table}
 
\begin{figure}[ht!]
\begin{center}
\includegraphics[height = 2 in]{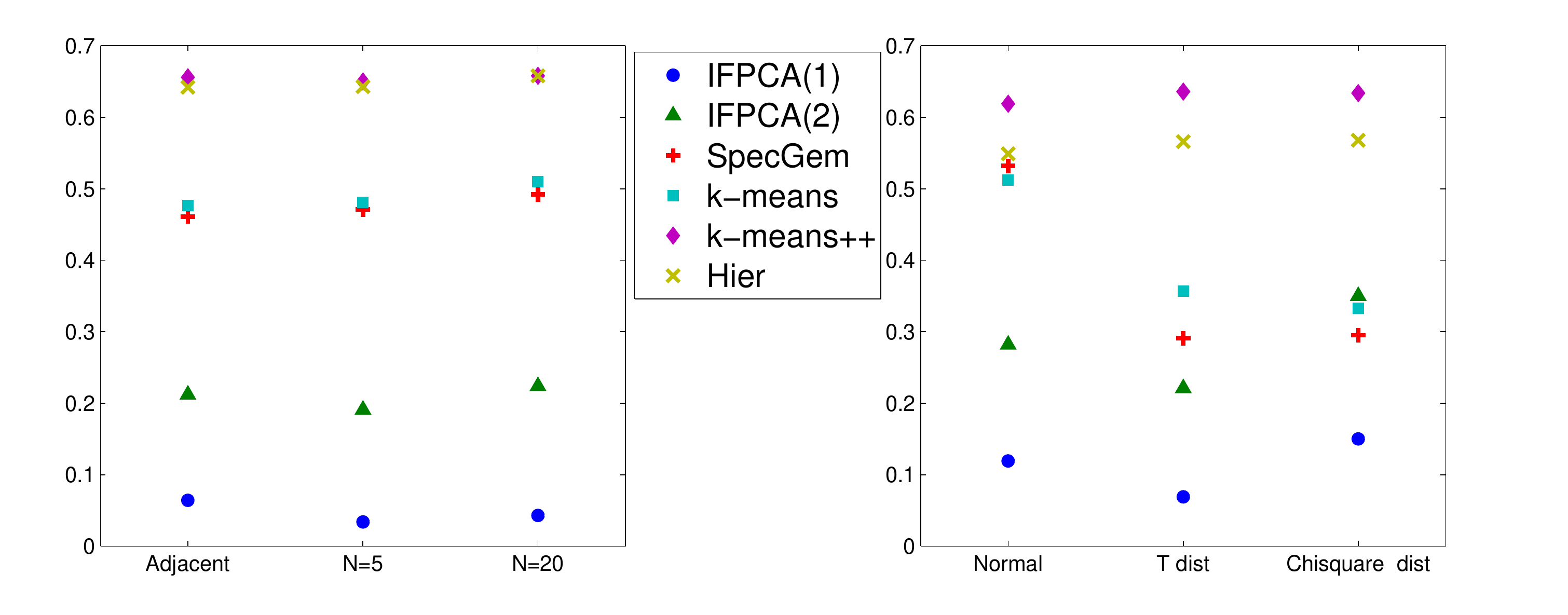}
\end{center}
\caption{Comparison of average clustering error rates for Experiment 4 (left panel) and Experiment 5 (right panel). y-axis: error rates}
\label{fig:exp45}
\end{figure}

{\it Experiment 4.}
In this experiment, we investigate the effects of correlations among the noise over the clustering results. 
We generate the data matrix $X$ the same as before, except for that the noise matrix $Z$ is replaced by $Z A$, for a matrix  $A \in R^{p,p}$. Fixing a number $d \in (-1,1)$,  we consider three choices of $A$, (a)-(c). In (a),   $A(i,j) = 1 \{i = j\} + d \cdot 1\{j = i + 1\}$, $1 \leq i, j \leq p$.  
In (b)-(c), fixing an integer $N > 1$,  for each $j = 1, 2, \ldots, p$, we randomly generate a size $N$ subset of $\{1,2, \ldots,p\} \setminus \{j\}$, denoted by $I_N(j)$. We then let $A(i,j) = 1 \{i = j\} + d  \cdot 1\{i \in I_N(j)\}$. 
For (b), we take $N = 5$ and for (c), we take $N = 20$. 
We set $d= .1$ in (a)-(c). 
We set $(K, p, \theta, \vartheta, r, rep) = (4, 2\times 10^4, .5, .6, .7, 100)$ (so $n = 141$), and $(\delta_1, \delta_2, \delta_3, \delta_4) = (1/4,  1/4, 1/4, 1/4)$, $\gamma = (.3, .05, .65)$. For an exponential random variable $X \sim Exp(\lambda)$, denote the density of $\big[ b + X | a_1 \leq b + X \leq a_2 \big]$
by $\widetilde{TSE}(\lambda, b, a_1, a_2)$, where $TSE$ stands for  `Truncated Shifted Exponential'.
We take $g_{\bar{\mu}}$ as $N(0,1)$, $g_{\mu}$ as $\widetilde{TSE}(.1, .9, -\infty, \infty)$ (so it has a mean $1$), and $g_{\sigma}$ as $\widetilde{TSE}(.1, .9, .9, 1.2)$. The threshold for IF-PCA(2) is $t=\sqrt{2\cdot .03\cdot\log(p)}$. 
The results are summarized in the left panel of Figure \ref{fig:exp45},
which suggest that IF-PCA continues to work in the presence of correlations among the noise:  IF-PCA significantly outperforms the other $4$ methods, especially for the randomly selected correlations.

{\it Experiment 5.}
 In this experiment, we study how different noise distributions  affect the clustering results.  We generate the data matrix $X$ the same  as before, except for  the distribution of the noise matrix $Z$ is different.
We consider three different settings for the noise matrix $Z$:  (a) for a vector $a = (a_1, a_2, \ldots, a_K)$,  generate row $i$ of $Z$ by  $Z_i \stackrel{iid}{\sim} N(0, a_k I_p)$ if   Sample $i$ comes from Class $k$, $1 \leq k \leq K$, $1 \leq i \leq n$,   (b)   $Z = \sqrt{2/3} \tilde{Z}$, where all entries of $\tilde{Z}$ are $iid$ samples from $t_6(0)$, where $t_6(0)$ denotes the central  $t$-distribution with $df = 6$,  (c)   $Z = [\tilde{Z} - 6]/\sqrt{12}$,  where the entries of $\tilde{Z}$ are $iid$ samples from  the chi-squared distribution with $df = 6$ (in (b)-(c),  the constants of $\sqrt{2/3}$ and $\sqrt{12}$ are chosen so that each entry of $Z$ has zero mean and unit variance). We set $(K, p, \theta, \vartheta, r, rep) = (4, 2\times 10^4, .5, .55, 1, 100)$, $(\delta_1, \delta_2, \delta_3, \delta_4) = (1/4,  1/4, 1/3, 1/6)$, and $\gamma = (.4, .1, .5)$. We take $g_{\bar{\mu}}$ to be $N(0,1)$. In case (a), we take $(a_1, a_2, a_3, a_4) = (0.8, 1, 1.2, 1.4)$. The threshold for IF-PCA(2) is set as  $t= \sqrt{2\cdot .03\cdot \log(p)}$.
The results are summarized in the right panel of Figure \ref{fig:exp45}, which suggest that IF-PCA continues  to  outperform  the other $4$ clustering methods.
 
\section{Connections and extensions}
\label{sec:Discu}
\setcounter{equation}{0} 
We propose IF-PCA as a new spectral clustering method, and we have
successfully applied the method to  clustering using gene microarray data.
IF-PCA is a two-stage method which consists of a marginal screening step and 
a post-selection clustering step. The methodology contains three important ingredients: using the KS statistic for marginal screening, post-selection PCA, and threshold choice by HC.
 
The KS statistic can be viewed as  an omnibus test  or a  goodness-of-fit measure.  
The methods and theory we developed on the KS statistic can be useful in many other settings, where it is of interest to find a powerful yet robust test. For example, they can be used for nonGaussian detection of the Cosmic Microwave Background (CMB)  or can be used for detecting rare and weak signals or small cliques in  large graphs (e.g.,  \cite{DJ13}).

The KS statistic can also be viewed as a marginal screening procedure. 
Screening is a well-known approach in high dimensional analysis.  For example, in variable selection, we use marginal screening for dimension reduction \citep{FanLv}, and in cancer classification, we use screening to adapt Fisher's LDA and QDA  to modern settings \citep{DJ08,  EfronClassification,   QUADRO}.    However, the setting here is very different.  

Of course, another important reason that we choose to use the KS-based marginal screening   in IF-PCA is for  simplicity and practical feasibility: with such a screening method,   we are able to  (a) use Efron's proposal of empirical null  to correct the null distribution,  and (b) set the threshold by Higher Criticism; (a)-(b) are especially important  as we wish to have a tuning-free and yet effective procedure for subject clustering with gene microarray data. 
In more complicated situations,  it is possible that marginal screening is sub-optimal, and it is desirable to use a more sophisticated screening method. We mention two possibilities below.

In the first possibility, we might use the recent approaches by \cite{BJNP, PJ}, where the primary interest is signal recovery or feature estimation. The point here is that, while the two problems---subject clustering and feature estimation---are very different, we  still hope that a better feature estimation method may improve the results of subject clustering. In these papers, the authors proposed {\it Augmented sparse PCA (ASPCA)} as a new approach to feature estimation and showed that under certain sparse settings,  ASPCA may have advantages over marginal screening methods, and that ASPCA  is asymptotically minimax. This suggests an alternative to IF-PCA, where in the IF step, we replace the marginal KS screening by some augmented feature screening approaches.  However, the open question is, how to develop such an approach that is tuning-free and practically feasible. We leave this to the future work. 

Another possibility is to combine the KS statistic with the recent innovation of Graphlet Screening (\cite{gs, KJF}) in variable selection.  
This is particularly appropriate if the columns of the noise matrix $Z$ are correlated, 
where it is desirable to exploit the graphic structures of the correlations to  improve the
screening efficiency. 
Graphic Screening
is a graph-guided multivariate screening procedure and has advantages
over the better known method of marginal screening and the lasso. At the heart of Graphlet Screening is 
a graph, which in our setting is defined as follow: 
each feature $j$, $1 \leq j \leq p$, is a node,  and there is an edge between nodes $i$ and $j$ if and only if 
row $i$ and row $j$ of the normalized data matrix $W$ are strongly correlated 
(note that for a useful feature, the means of the corresponding row of  $W$ are nonzero; in our range of interest,  these nonzero means are at the order of $n^{-1/6}$, and so have negligible effects over the correlations). In this sense, adapting Graphlet Screening in the screening step helps to solve highly correlated data. We leave this to the future work.

The post-selection PCA is a flexible idea that can be adapted to address many other problems.   Take model (\ref{model1}) for example. The method can be adapted to address the problem of testing whether $LM = 0$ or $LM \neq 0$ (that is, whether the data matrix consists of a low-rank structure or not),   the problem of estimating $M$, or the problem of  estimating $LM$. The latter is connected to recent interest on sparse PCA and low-rank matrix recovery. Intellectually, the PCA approach is connected to SCORE for community detection on social networks \citep{SCORE}, but is very different.

Threshold choice by HC is a recent innovation, and was first proposed in \citep{DJ08}
(see also  \citep{FJY}) in the context of classification.  However, our focus here is on clustering, and
  the method and theory we need are very different from those in \citep{DJ08, FJY}.
In particular, this paper requires sophisticated post-selection Random Matrix Theory (RMT), which we do not need in \citep{DJ08,  FJY}.  Our study on RMT is connected to \citep{Johnstone, Paul, BS2006, Zeitouni2000, Lee2010} but is very different.

In a high level, IF-PCA is connected to the approaches by \citep{ASW, CH} in that all three approaches  are two-stage methods that consist of a screening step and a post-selection clustering step. However,  the screening step   and the post-selection step in all three approaches are significantly different from each other.  Also,  IF-PCA is connected to the spectral graph partitioning  algorithm by \citep{NJ}, but it is very different, especially in feature selection and threshold choice by HC.  
 
In this paper,   we have assumed
that the first $(K-1)$ contrast mean vectors $\mu_1, \mu_2, \ldots, \mu_{K-1}$ are linearly independent (consequently,   the rank   
of the matrix $M$ (see (\ref{modelmatrix1})) is $(K-1)$), and that $K$ is known (recall that $K$ is the number of classes).  
In the gene microarray examples we discuss in this paper, 
a class is a patient group (normal, cancer, cancer sub-type) 
so $K$ is usually known to us as a priori. 
Moreover, it is believed that different cancer sub-types can be distinguished from each other  by one or more genes (though we do not know which) so $\mu_1, \mu_2, \ldots, \mu_{K-1}$ are 
linearly independent.  Therefore, both assumptions are reasonable.  

On the other hand, in a broader context,  either of these two assumptions could be violated. 
Fortunately, at least to some extent, the main ideas in this paper can be extended. 
We consider two cases. In the first one,  we assume $K$ is known  but  $r = \mathrm{rank}(M) < (K-1)$. In this case, the main results in this paper continue 
to hold, provided that some mild regularity conditions hold. In detail, let $U \in R^{n,r}$ be the matrix consisting the first $r$ left singular vectors of $L M \Lambda$ as before; it can be shown that,  as before, $U$ has $K$ distinct rows.  The additional regularity condition we need here is that,  the $\ell^2$-norm between   any pair  of the $K$ distinct rows has a reasonable lower bound. 
In the second case, we assume $K$ is unknown and has to be estimated. 
In the literature, this is a well-known hard problem. To tackle this problem,  
one might utilize the recent developments on rank detection \citep{KN} (see also \citep{RankDetect, BJNP}), where in a similar 
setting, the authors constructed a confident lower bound for the number of classes $K$.  A problem of interest is then to investigate how to combine the methods in these papers with IF-PCA to deal with the more challenging case of unknown $K$; we leave this for future study.


\section*{Acknowledgements}
The authors would like to thank David Donoho, Shiqiong Huang, Tracy Zheng Ke, Pei Wang,   and anonymous referees for valuable  pointers and discussion.  


\appendix

\setcounter{equation}{0}
We now prove Theorems \ref{thm:KSnull}--\ref{thm:KSalt}  and Lemmas \ref{lemma:LM}--\ref{lem:project}.
Note that Theorems \ref{thm:Frob}-\ref{thm:Clu} and Corollaries
\ref{corollary1}-\ref{corollary2} are proved in Section \ref{subsec:mainpf}.

\section{Proof of Theorem 2.3} \label{app:KSnull}
We use the techniques developed by   \cite{Loader}.
For short, write $u = \bar{\mu}(j)$, $\sigma = \sigma(j)$,  $n = n_p$,   $\bar{X} = \bar{X}(j)$, $X_i=X_i(j)$, $\psi_n=\psi_{n_p,j}$ and
$s = \hat{\sigma}(j)$.  Under these notations,
\[
\psi_n = \sqrt{n} \sup_{ - \infty < v < \infty} \{| \Phi( (v - \bar{X})/s)  - F_n(v)] | \}.
\]
Define
\begin{equation}\label{eqn:KSplus}
\psi_n^{\pm} = \sqrt{n} \sup_{ - \infty < v < \infty} \{ \mp [\Phi( (v - \bar{X})/s)  - F_n(v)]\}.
\end{equation}
Writing  $P(\psi_n \geq t_p) \leq  P(\psi_n^{-} \geq t_p) + P(\psi_n^+ \geq t_p)$,  and noting  that by symmetry and time reversal, the two terms on the right hand side equal to each other, it follows that
\begin{equation} \label{KSpf1}
 P(\psi_n^{-} \geq t_p) \leq P(\psi_n \geq t_p) \leq 2 P(\psi_n^{-}  \geq t_p).
\end{equation}
At the same time, note that $\psi_n^-$ is an ancillary statistic to the parameters $(u, \sigma)$,
so it is independent of the sufficient statistics $(\bar{X}, s^2)$. Therefore,
\begin{equation} \label{KSpf2}
P(\psi_n^-  \geq t_p)  =  P\bigl(\psi_n^{-}  \geq t_p  | \bar{X} = 0, s^2 = 1 \bigr).
\end{equation}
Combining (\ref{KSpf1})-(\ref{KSpf2}) and comparing the result with the theorem,  all we need to show is
\begin{equation}\label{eqn:ks-}
P( \psi_n^-  \geq  t_p |  \bar{X} = 0, s^2 = 1)   \sim  \sqrt{\frac{2\pi}{ \pi - 2 }} \mathrm{exp}(- \frac{2\pi}{\pi - 2} t_p^2).
\end{equation}

We now show (\ref{eqn:ks-}).   Denote for short $q_v = \Phi(v) - t_p/\sqrt{n}$. It follows from (\ref{eqn:KSplus}) and basic algebra that
\begin{align}
P\bigl(\psi_n^-  \geq t_p  |  \bar{X} = 0, s^2 = 1  \bigr)
 \equiv P \bigl( \inf_{-\infty < v < \infty} \{ F_n(v) - q_v \}   \leq 0   \bigl|   \bar{X} = 0, s^2 = 1 \bigr).
\label{KSeq1}
\end{align}
Introduce the {\it first boundary crossing time} by
\[
\tau = \inf \{ v:  F_n(v) <  q_v  \}.
\]
Let $v_j$ be the solution of
\[
q_v = j/n,  \qquad  j  = 0, 1, \ldots, n-1.
\]
Since $F_n(v)$ is a monotone staircase function taking values from $\{0, 1/n, 2/n, \ldots, 1\}$ and $q_v$ is strictly increasing in $v$,  it is seen that $\{ \tau < \infty\} = \{ v_0, v_1,  \ldots, v_{n-1}\}$ and that  $F_n(v_j) = j/n$ given $\tau = v_j$.
As a result,
\begin{align}
& P \bigl( \inf_{-\infty < v < \infty} \{  F_n(v) -   q_v \}   \leq 0   \bigl|  \bar{X} = 0, s^2 = 1 \bigr)  =
\sum_{j = 0}^{n-1} P(\tau = v_j, F_n(v_j) =\frac{j}{n} | \bar{X} = 0, s^2 = 1 )   \nonumber  \\
& \qquad = \sum_{j = 0}^{n - 1} P \bigl( \tau = v_j  \bigl|  F_n(v_j) = \frac{j}{n},  \bar{X} = 0, s^2 = 1 \bigr)  \cdot P\bigl( F_n(v_j) = \frac{j}{n}     \bigl|  \bar{X} = 0, s^2 = 1 \bigr).  \label{firstpassage}
\end{align}
Introduce
\[
g_0(v) = \Phi(v)\Phi(-v)  -  \phi^2(v) (1 + v^2/2),   \qquad g_1(v) = \Phi(-v) + v \phi(v) (1 + v^2/2 ).
\]
The following lemma is proved in the appendix, using results from  (\cite{Loader}) and (\cite{BR}).
\begin{lemma} \label{lemma:firstpassage}
With $t_p$ in Theorem \ref{thm:KSnull},   for each $0 \leq j \leq n-1$,
\[
P \bigl( \tau = v_j  \bigl|  F_n(v_j) = \frac{j}{n},  \bar{X} = 0, s^2 = 1 \bigr) \sim  \frac{g_1(v_j)}{ g_0(v_j)}  \frac{t_p}{\sqrt{n}},
\]
and that
\[
P\bigl( F_n(v_j) = \frac{j}{n}     \bigl|  \bar{X} = 0, s^2 = 1 \bigr) \sim   \frac{1}{\sqrt{2\pi n  g_0(v_j)} } \exp(- \frac{t_p^2}{2g_0(v_j)}).
\]
\end{lemma}
Combining (\ref{firstpassage}) with Lemma \ref{lemma:firstpassage}, we have
\begin{equation}\label{sum}
P(\psi_n^{-} \geq t_p)  \sim  t_p  \biggl[  \frac{1}{n}  \sum_{j = 0}^{n - 1}  \frac{g_1(v_j)}{\sqrt{2\pi g_0(v_j)^3}} \exp(-\frac{t_p^2}{2g_0(v_j)})\biggr].
\end{equation}
Moreover, recall that $v_j$ is the solution of
$\Phi(v) =  j/n + t_p/\sqrt{n}$,  so
$(v_{j+1} - v_{j}) \phi(v_j) \sim  1/n$.  Inserting this into (\ref{sum}) gives
\[
P(\psi_n^{-} \geq t_p)  \sim t_p  \sum_{j = 0}^{n - 1}  \frac{g_1(v_j) }{\sqrt{2\pi g_0(v_j)^3}} \exp(-\frac{t_p^2}{2g_0(v_j)}) \phi(v_j)  (v_{j+1} - v_j).
\]
Note that the right hand side can be approximated by a Riemman integral and
\begin{equation} \label{ksin}
\sim  t_p  \int_{ v_0 }^{ v_{n-1}}    \frac{g_1(v)}{\sqrt{2\pi} g_0^{3/2}(v)}  \exp(-\frac{t_p^2}{2 g_0(v)})  \phi(v)  dv.
\end{equation}
The following lemma is proved in the appendix.
\begin{lemma} \label{lemma:RI}
With $t_p$ in Theorem \ref{thm:KSnull},
\begin{equation} \label{KSpf5}
t_p \int_{ v_0 }^{ v_{n-1}}   \frac{g_1(v)}{\sqrt{2\pi}  g_0^{3/2}(v)}  \exp(-\frac{t_p^2}{2 g_0(v)})  \phi(v)  dv
 \sim \sqrt{\frac{2\pi}{\pi - 2}} \exp(- \frac{2\pi}{\pi-2}  t_p^2) .
\end{equation}
\end{lemma}
Inserting (\ref{KSpf5}) into (\ref{sum})-(\ref{ksin}) gives (\ref{eqn:ks-}).  \qed

\subsection{Proof of Lemma \ref{lemma:firstpassage}}
The following lemma is proved  in  \cite{BR},     \cite{Woodroofe},   or   \cite{Loader}.
Given $n$ samples $X_1, X_2, \ldots, X_n$ from an exponential family
\[
f(x; \theta)  =  h_0(x) \exp(\theta'x - \eta(\theta)), \qquad  x \in R^d, \; \theta \in R^d.
\]
Let $\hat{\theta}$ be the Maximum Likelihood Estimator (MLE) for $\theta$.  Note that a sufficient statistic for $\theta$ is $\frac{1}{n}\sum_{i = 1}^n  X_i$, and then $\hat{\theta}$ is a  function of $\frac{1}{n}\sum_{i = 1}^n X_i$.  Denote the density function
of $\frac{1}{n} \sum_{i = 1}^n X_i$ by $f_0^{(n)}$.
\begin{lemma} \label{lemma:BR}
$f_0^{(n)}(x) =  (1 + o(1)) \cdot   (2\pi n)^{-d/2}   \cdot |det(\eta''(\hat\theta))|^{-1/2}   \exp(- n \ell(\hat{\theta}, x))$,
where
$\ell(\hat{\theta}, x)  =  (\hat\theta - \theta)' x - (\eta(\hat\theta) - \eta(\theta))$.
\end{lemma}
For preparations, we need some calculations related to density associated with $n$ samples from a normal distribution.
Let $f(x)$ be the density of $N(u, \sigma^2)$. If we let
\begin{equation}\label{eqn:parameter}
y = x^2,  \qquad  \alpha = - \frac{1}{2\sigma^2}, \qquad \beta = \frac{u}{\sigma^2},
\end{equation}
then we have
\[
f(x)  =  \exp \bigl( \alpha y +\beta x + \frac{\beta^2}{4\alpha} - \frac{1}{2}\log(-\frac{\pi}{\alpha}) \bigr).
\]
Let $\hat{\theta}^{(1)}$ be the MLE, which is a function of $(\frac{1}{n} \sum_{i =1}^n X_i^2, \frac{1}{n} \sum_{i = 1}^n X_i)$.  Applying Lemma \ref{lemma:BR} with $d = 2$,  $\hat{\theta} = \hat{\theta}^{(1)}$, $\eta =  \eta^{(1)} = -\frac{\beta^2}{4 \alpha} + \frac{1}{2} \log(-\frac{\pi}{\alpha})$,  and
$x = x_0^{(1)} =  (1, 0)$ (corresponding to  $\frac{1}{n} \sum_{i = 1}^n X_i^2 =1$ and $\frac{1}{n} \sum_{i = 1}^n X_i = 0$),  we have
\[
\ell(\hat{\theta}^{(1)}, x_0^{(1)})  = -1/2 - \alpha - \beta^2/4\alpha + \log(-2\alpha)/2,
\]
and so
\begin{equation} \label{ApplyBR1}
P\bigl( \bar{X} = 0, s^2 = 1 \bigr)  = (1 + o(1)) \cdot   \frac{1}{2\pi n \sqrt{2}} \exp([\frac{1}{2} + \alpha + \frac{\beta^2}{4\alpha}  - \frac{\log(-2\alpha)}{2}] n).
\end{equation}

Alternatively,  writing $v = v_j$ for short,  we can embed the above normal density into a {\it three-parameter exponential} family
\begin{equation}\label{eqn:embed}
f_v(x;  \alpha, \beta, \delta) = \exp(\alpha x^2 + \beta x + \delta 1\{ x> v\} - \eta^{(2)} (\alpha, \beta, \delta)),
\end{equation}
where
\begin{equation}\label{eqn:psi}
\eta^{(2)}(\alpha, \beta, \delta) = - \frac{\beta^2}{4\alpha} + \frac{1}{2}\log \bigl(-\frac{\pi}{\alpha}) + \log(\Phi(  { \sqrt{ - 2\alpha} }v  -  \frac{\beta}{\sqrt{ - 2\alpha}} ) + e^\delta \Phi(  - { \sqrt{ - 2\alpha} }v  +  \frac{\beta}{\sqrt{ - 2\alpha}}  ) \bigr).
\end{equation}
Note that when $\delta = 0$, $f_{v}(x; \alpha, \beta, \delta) \equiv f(x)$.
 We let
\begin{equation}\label{eqn:boundary}
h_v= h_v(\alpha, \beta) =  - { \sqrt{ - 2\alpha} }v  +  \frac{\beta}{\sqrt{ - 2\alpha}}.
\end{equation}
Denote the MLE for $\theta \equiv (\alpha, \beta, \delta)$  by $\hat{\theta}^{(2)}$. We note that $\hat{\theta}^{(2)}$  is a function of $(\frac{1}{n} \sum_{i = 1}^n X_i^2, \frac{1}{n} \sum_{i = 1}^n X_i,  \frac{1}{n} \sum_{i = 1}^n 1\{X_i > v\})$.  Let $x_0^{(2)} = (1, 0,  1 - q_v)$, corresponding to $\frac{1}{n} \sum_{i = 1}^n X_i^2 = 1$,  $ \frac{1}{n} \sum_{i = 1}^n X_i  = 0$, and
$\frac{1}{n} \sum_{i = 1}^n 1\{X_i > v\} = 1 - q_v$,  and denote
\[
(\alpha^*_n(v), \beta_n^*(v), \delta_n^*(v)) = \hat{\theta}^{(2)} \biggr|_{x = x_0^{(2)}}.
\]
Then $(\alpha^*_n(v), \beta_n^*(v), \delta_n^*(v))$ is the solution of the following equation system:
\begin{eqnarray}\label{eqn:mle}
\left\{\begin{array}{l}
\beta \Phi(h_v) \Phi(- h_v) - \sqrt{(1 - \beta v)} [ q_v - \Phi(-h_v) ]  \phi(h_v) = 0,\\
\alpha = (\beta v - 1)/2,\\
e^\delta = \frac{1 - q_v}{q_v}  \Phi(-h_v)/\Phi(h_v).\\
\end{array}
\right.
\end{eqnarray}
Applying Lemma \ref{lemma:BR} with $d = 3$, $\hat{\theta} = \hat{\theta}^{(2)}$, $\eta = \eta^{(2)}(\alpha^*_n(v), \beta^*_n(v), \delta^*_n(v))$,  and $x = x_0^{(2)}$ gives
\begin{equation}\label{eqn:ell}
\ell(\hat{\theta}^{(2)}, x_0^{(2)}) = (\alpha^*_n(v) - \alpha) + \delta^*_n(v) (1 - q_v) - \eta^{(2)}(\alpha^*_n(v), \beta^*_n(v), \delta^*_n(v)) + \eta(\alpha, \beta),
\end{equation}
and  so
\begin{equation} \label{ApplyBR2}
P\bigl( F_n(v_j) = \frac{j}{n},      \bar{X} = 0, s^2 = 1 \bigr) = \frac{(1 + o(1)) e^{ - n \ell(\hat{\theta}^{(2)}, x_0^{(2)})}}{(2\pi n)^{3/2} (\det(H_{{\eta}^{(2)}}(\alpha_n^*(v), \beta_n^*(v), \delta_n^*(v)))^{1/2}}   ,
\end{equation}
 where  $H_{{\eta}^{(2)}}( \alpha^*_n(v), \beta^*_n(v), \delta^*_n(v))$ is the $3 \times 3$ Hessian matrix of $\eta^{(2)}(\alpha, \beta, \delta)$, evaluated at the point $( \alpha^*_n(v), \beta^*_n(v), \delta^*_n(v))$.

Introduce
\begin{equation}\label{eqn:mu}
\mu(v, q_v) = \frac{\exp( \alpha^*_n(v) v^2 + \beta^*_n(v) v - \eta^{(1)}(\alpha^*_n(v), \beta^*_n(v)))}{ \Phi( - h_v(\alpha^*_n(v), \beta^*_n(v))) + e^{\delta^*_n(v)} \Phi(h_v(\alpha^*_n(v), \beta^*_n(v)))},
\end{equation}
and
\begin{equation}\label{eqn:dell}
\tilde{\ell}(v)  = \alpha^*_n(v) + 1/2 + \delta^*_n(v) (1 - q_v) - \eta^{(2)}(\alpha^*_n(v), \beta^*_n(v), \delta^*_n(v)) + \log(\sqrt{2\pi}).
\end{equation}
The following lemma is proved in the appendix.
\begin{lemma}\label{lemma:taylor}
When $t_p/\sqrt{n} \goto 0$, we have the following approximations for the functions of MLE estimators,
\[
\tilde{\ell}(v) = \frac{ 1 }{ 2g_0(v) } t_p^2/n + O(t_p^3/n^{3/2}),
\]
\[
\phi(v) - \mu(v, q_v) =  \phi(v) \frac{ g_1(v)}{ g_0(v)} t_p/\sqrt{n} + O(t_p^2/n),
\]
and
\[
\det(H_{{\eta}^{(2)}}(\alpha^*_n(v), \beta^*_n(v), \delta^*_n(v)) = 2 g_0(v) + O(t_p/\sqrt{n}).
\]
\end{lemma}

We now proceed to prove Lemma \ref{lemma:firstpassage}.
 Consider the first claim.  Now we show the approximation for $P \bigl( \tau = v_j  \bigl|  F_n(v_j) = \frac{j}{n},  \bar{X} = 0, s^2 = 1 \bigr)$.
By Loader \cite[(20), (21) and Lemma B.2]{Loader},
\begin{equation} \label{loader}
P \bigl( \tau = v_j  \bigl|  F_n(v_j) = \frac{j}{n},  \bar{X} = 0, s^2 = 1 \bigr)  =  (\frac{\partial q_v}{\partial v})^{-1} (\frac{\partial q_v}{\partial v} - \mu(v, q_v))   \biggr|_{v = v_j}(1 + o(1)) ,
\end{equation}
where $\mu(v, q_v)$ is defined in (\ref{eqn:mu}).  Using Lemma \ref{lemma:taylor}  and recalling that $q_v = \Phi(v) - t_p/\sqrt{n}$, it follows from definitions and direct calculations that
\[
P \bigl( \tau = v_j  \bigl|  F_n(v_j) = \frac{j}{n},  \bar{X} = 0, s^2 = 1 \bigr)  = \frac{ g_1(v_j)}{ g_0(v_j) } t_p/\sqrt{n} (1 + o(1)),
\]
and the claim follows.

Consider the second claim.
Write
\begin{equation} \label{ApplyBR0}
P\bigl( F_n(v_j) = \frac{j}{n}     \bigl|  \bar{X} = 0, s^2 = 1 \bigr)  = \frac{P\bigl( F_n(v_j) = \frac{j}{n},      \bar{X} = 0, s^2 = 1 \bigr)}{P\bigl( \bar{X} = 0, s^2 = 1 \bigr)},
\end{equation}
where both the denominator and numerator are thought of as the density function at that point; this is a slight misuse of notations.
Inserting (\ref{ApplyBR1}) and (\ref{ApplyBR2}) into (\ref{ApplyBR0}) and note that $g_0(v) > 0$ for all $v$, then
\begin{equation} \label{ApplyBR4}
P(F_n(v_j) = \frac{j}{n}|\bar{X} = 0, s^2 = 1) =  (1 + o(1)) \frac{e^{ - n \tilde{\ell}(v)}}{\sqrt{\pi n}  [\det(H_{{\eta}^{(2)}}(\alpha_n^*(v_j), \beta_n^*(v_j), \delta_n^*(v_j) )) ]^{1/2}}  .
\end{equation}
Using Lemma \ref{lemma:taylor}, the right hand side reduces to
\[
(1 + o(1)) \cdot  \frac{\exp(- t_p^2/2 g_0(v_j))}{ \sqrt{2\pi n g_0(v_j) } },
\]
and the claim follows.   \qed


\subsection{Proof of Lemma \ref{lemma:RI}}
Recall that $g_0(v) = \Phi(v)\Phi(-v)  -  \phi^2(v) (1 + v^2/2)$ and $g_1(v) = \Phi(-v) + v\phi(v) (1 + v^2/2 )$. Denote for short
\[
h(v) = \frac{g_1(v) \phi(v)}{\sqrt{2 \pi} g_0^{3/2}(v)}.
\]
What we need to show is
\[
t_p  \int_{v_0}^{v_{n-1}}   h(v) \exp(-\frac{t_p^2}{2g_0(v)}) dv  \sim \sqrt{\frac{2\pi}{\pi - 2}} \exp(-\frac{2\pi}{\pi - 2} t_p^2).
\]
The following results follow from elementary calculus.
\begin{description}
\item[(a).]   Since $t_p \goto \infty$ and $t_p/\sqrt{n} \goto 0$, it is seen that $v_{n-1} = (\sqrt{n}/t_p)^{1/2}  (1 + o(1))$ and $v_0 = -(\log(\sqrt{n}/t_p))^{1/2}  (1+o(1))$.
\item[(b).]  Since that for all $v > 0$,  $g_0'(v) = \phi(v)[(v + v^3) \phi(v) - (2 \Phi(v) - 1)] < 0$, and that $\lim_{v \goto \infty} g_0(v) /\Phi(-v) \sim 1$,  $g_0(v)$ is a symmetric and positive function over $(-\infty, \infty)$.  Moreover,
\[
\frac{1}{g_0(0)}   =  \frac{ 4\pi }{ \pi - 2 },   \qquad   \frac{d}{dv}(\frac{1}{g_0(v)}) \biggr|_{v= 0} =0,  \qquad  \frac{d^2}{dv^2} (\frac{1}{g_0(v)}) \biggr|_{v = 0}   =  \frac{ 8\pi }{ (\pi - 2)^2 },
\]
\item[(c).] $h(v)$ is a positive function with $h(0)  =  \sqrt{4\pi} /  (\pi - 2)^{3/2}$, $h'(0) = 0$, and $|h''(v)| \leq C $ for some constant $C$ when $|v| \leq 1/2$.
\end{description}
Denote $b_n = t_p^{-5/6}$. Note that $b_n \goto 0$, $t_p b_n^{3/2} \goto 0$, but $t_p b_n \goto \infty$.
We write
\begin{equation} \label{RIpf}
t_p  \int_{v_0}^{v_{n-1}}   h(v) \exp(-\frac{t_p^2}{2 g_0(v)}) dv  = I + II + III,
\end{equation}
where
\[
I = \int_{|v| \leq b_n} t_p  h(v)  \exp(- \frac{t_p^2}{2 g_0(v)}) dv,   \qquad
II = \int_{b_n \leq |v| \leq 1} t_p  h(v)  \exp(- \frac{t_p^2}{2 g_0(v)}) dv,
\]
and
\[
III = \int_{|v| >  1, v_0 \leq v \leq v_{n-1}} t_p  h(v)  \exp(- \frac{t_p^2}{2 g_0(v)}) dv,
\]
where in $I$ and $II$, we have used $|v_0| > 1$ and $|v_{n-1}| > 1$.

Consider $I$.  By elementary calculus,
It follows that
\[
I =  \int_{-b_n}^{b_n} t_p  h(v) e^{- \frac{t_p^2}{2g_0(v)}} dv =  t_p e^{- \frac{t_p^2}{2 g_0(0)}}  \int_{-b_n}^{b_n} (h(0) + O( b_n^2))e^{- \frac{2\pi}{(\pi - 2)^2} t_p^2 v^2 + O(t_p^2 b_n^3)} dv.
\]
Recall that $b_n = o(1)$ and $t_p^2 b_n^3  = o(1)$, it follows from elementary calculus that
\begin{equation} \label{RIpf1}
I \sim \sqrt{\frac{2\pi}{\pi - 2}} \exp(- \frac{2\pi}{\pi - 2} t_p^2).
\end{equation}

Consider $II$. It is seen that $h(v) \leq C$ for $b_n \leq |v| \leq 1$. Recall that $g_0(v)$ is symmetric and monotone on $[0, \infty]$,
\[
II \leq  C t_p  \exp(-\frac{t_p^2}{2 g_0(b_n)}).
\]
Moreover, by the first and second derivative of $1/g_0(v)$ in (b),  there is a constant $c_0 > 0$ such that
\[
\frac{1}{g_0(b_n)} \geq \frac{1}{g_0(0)} + c_0 b_n^2.
\]
Inserting this into
\begin{equation} \label{RIpf2}
II \leq \exp(-\frac{2\pi}{\pi-2}t_p^2) \cdot [C t_p \exp(- \frac{c_0}{2} t_p^2 b_n^2]  = o(1)  \exp(-\frac{2\pi}{\pi-2}t_p^2).
\end{equation}
Consider $III$. By symmetry and elementary calculus,   there is a constant $C_1$, such that
\[
h(v) \leq  C_1 |v| / \sqrt{g_0(v)},  \qquad |v| \geq 1
\]
so
\[
h(v) \exp(-\frac{t_p^2}{2 g_0(v)}) \leq   \frac{C_1 |v|}{\sqrt{g_0(v)}} \exp(-\frac{t_p^2}{2g_0(v)}), \qquad |v| \geq 1.
\]
Note that,   first,  there is a constant $C >0$ such that
\[
\frac{1}{g_0(v)}  \geq \frac{1}{g_0(1)} + C (|v| - 1),
\]
and second,   for sufficiently large $t_p$, the function $\sqrt{x} e^{- (t_p^2/2) x }$ is monotonely decreasing
in $[1/g_0(1), \infty)$,
so
\[
\frac{|v|}{\sqrt{g_0(v)}} \exp(-\frac{t_p^2}{2g_0(v)}) \leq  |v| \sqrt{\frac{1}{g_0(1)} + C(|v|-1)} \exp(-\frac{t_p^2}{2} (\frac{1}{g_0(1)}   + C (|v|-1) )).
\]
Combining these,
\begin{multline*}
III \leq C_1 t_p  \int_{|v| \geq 1}  |v| \sqrt{\frac{1}{g_0(1)} + C(|v|-1)} \exp(-\frac{t_p^2}{2} (\frac{1}{g_0(1)} \\  
+ C (|v|-1) )) dv  \leq C t_p \exp(-\frac{t_p^2}{2 g_0(1)}).
\end{multline*}
Since $1/g_0(1) > 4\pi/(\pi-2)$, it follows that
\begin{equation} \label{RIpf3}
III = o(1) \exp(-\frac{2\pi}{\pi-2}t_p^2).
\end{equation}
Inserting (\ref{RIpf1}), (\ref{RIpf2}), and (\ref{RIpf3}) into (\ref{RIpf}) gives the claim.  \qed

\subsection{Proof of Lemma \ref{lemma:taylor}}
We need some preparations.
Throughout this subsection, $\eps_n = t_p/\sqrt{n}$ for short.  First, we study
$(\alpha^*_n(v), \beta_n^*(v), \delta_n^*(v))$, which satisfies the equations
\begin{eqnarray}\label{eqn:mle2}
\left\{\begin{array}{l}
\beta \Phi(h_v) \Phi(- h_v) - \sqrt{(1 - \beta v)} [ q_v - \Phi(- h_v) ]  \phi(h_v) = 0,\\
\alpha = (\beta v - 1)/2,\\
e^\delta = \frac{1 - q_v}{q_v}  \Phi(-h_v)/\Phi(h_v).\\
\end{array}
\right.
\end{eqnarray}
We solve for $\beta_n^*(v)$ first. Recall that
\[
q_v = \Phi(v) - \eps_n, \qquad   h_v(\alpha, \beta) =  - { \sqrt{ - 2\alpha} }v  +  \beta/\sqrt{ - 2\alpha}.
\]
Inserting this and the second equation in (\ref{eqn:mle2}) into the first equation of (\ref{eqn:mle2}),
\begin{equation}\label{eqn:betaapp}
\beta \Phi(h_v(\beta)) \Phi( - h_v(\beta) ) - \sqrt{1 - \beta v}[ \Phi(v) - \eps_n -  \Phi(- h_v(\beta))] \phi(h_v(\beta)) = 0,
\end{equation}
and
\[
h_v(\beta)  =\frac{\beta(v^2 + 1) - v}{\sqrt{1 - \beta v}}.
\]
It is seen that $\beta_n^*(v) = o(1)$, we expand this in the neighborhood of $\beta = 0$. Denote $h_v = h_v(\beta)|_{\beta = \beta_n^*(v)}$ for short.
\begin{equation}\label{eqn:happ}
h_v  = - v + \beta_n^*(v)(1 + \frac{v^2}{2}) + O((\beta_n^*(v))^2),\,\,
\Phi(h_v) = \Phi(-v) +  \phi(v) \beta_n^*(v)(\frac{v^2}{2} + 1) + O((\beta_n^*(v))^2).
\end{equation}
Reorganizing this gives
\[
\beta_n^*(v) \Phi(-v)\Phi(v) - \phi(v) ( \beta_n^*(v)(v^2/2 + 1)\phi(v)  -  \eps_n)  + O((\beta_n^*(v))^2)  = 0,
\]
and so
\begin{equation}\label{eqn:beta}
\beta^*_n(v)   \sim  - \frac{ \phi(v) }{ \Phi(v)\Phi(-v)  -  (v^2/2 + 1)\phi^2(v) } \eps_n.
\end{equation}
Inserting this back into (\ref{eqn:mle2}) gives
\begin{equation} \label{alphabeta}
\left\{\begin{array}{l}
 \alpha^*_n(v) = (\beta^*_n(v) v - 1)/2, \\
e^{\delta^*_n(v)} = \frac{1 - q_v}{q_v} \frac{\Phi( - h_v(\beta^*_n(v))) }{\Phi( h_v(\beta^*_n(v))) }.
\end{array}
\right.
\end{equation}

We now show the results. Consider the first claim. By definitions,
\[
\tilde{\ell}(v) = \alpha^*_n(v) + 1/2 + \delta^*_n(v) (1 - q_v) - \eta^{(2)}(\alpha^*_n(v), \beta^*_n(v), \delta^*_n(v)) + \log(\sqrt{2\pi}).
\]
Combining this with (\ref{alphabeta}),
\begin{multline*}
\tilde{\ell}(v)  = 
\frac{ \beta^*_n(v) v}{2} + \frac{1}{2}\log(1 -  \beta^*_n(v) v) - \frac{ \beta^*_n(v)^2}{2(1 - \beta^*_n(v) v)}\\
 + (1-q_v)\log \frac{1 - q_v}{\Phi( h_v) }+ q_v \log \frac{q_v}{\Phi( - h_v ) }.
\end{multline*}
Recall that $\beta_n^*(v) = O(\eps_n)$. Using Taylor expansion and (\ref{eqn:betaapp}),
\begin{eqnarray*}
2\tilde{\ell}(v) & = &\frac{((v^2/2 + 1)\phi(v)\beta^*_n(v) - \epsilon_n)^2}{\Phi(-v)\Phi(v)} - (v^2/2+1)\beta^*_n(v)^2 + O(\epsilon_n^3)\\
& = &\frac{ 1 }{ \Phi(v)\Phi(-v) - \phi^2(v) (1 + v^2/2)} \epsilon_n^2 + O(\epsilon_n^3),
\end{eqnarray*}
and the claim follows  by recalling $\eps_n = t_p/\sqrt{n}$.

Consider the second claim. Recall  that in  $\phi(v)  -  \mu(v, q_v)$,
\begin{equation} \label{proofreadadd1}
\mu(v, q_v) = \frac{\exp( \alpha^*_n(v) v^2 + \beta^*_n(v) v - \eta^{(1)}(\alpha^*_n(v), \beta^*_n(v)))}{ \Phi( - h_v) + e^{\delta^*_n(v)} \Phi(h_v)},
\end{equation}
where $\eta^{(1)}(\alpha^*_n(v), \beta^*_n(v))) = - \frac{(\beta^*_n(v))^2}{4 \alpha^*_n(v)} +\frac{1}{2} \log(-\pi/\alpha^*_n(v))$.
Inserting the second equation of (\ref{alphabeta}) into (\ref{proofreadadd1}) and noting that the numerator is the density function for normal distribution with parameter $(\alpha = \alpha^*_n(v), \beta = \beta^*_n(v))$ at point $v$, it gives  that
\[
\mu(v, q_v)|_{ \alpha^*_n(v), \beta^*_n(v), \delta^*_n(v)} =
\frac{q_v}{\Phi( - h_v)} \phi( h_v ).
\]
Combining with (\ref{eqn:happ}) and (\ref{eqn:betaapp}), we have
\begin{eqnarray*}
\mu(v, q_v)  & = & \phi(v) - \phi(v) \frac{ \Phi(-v) + v \phi(v) (1 + v^2/2)}{ \Phi(v)\Phi(-v) - \phi^2(v) (1 + v^2/2)} \epsilon_n + O(\epsilon_n^2),
\end{eqnarray*}
it follows that
\[
\phi(v) - \mu(v, q_v)
= \phi(v) - \mu(v, q_v) \phi(v) \frac{ \Phi(-v) + v \phi(v) (1 + v^2/2)}{ \Phi(v)\Phi(-v) - \phi^2(v) (1 + v^2/2)} \epsilon_n + O(\epsilon_n^2),
\]
and the claim follows  by recalling $\eps_n = t_p/\sqrt{n}$.

Consider the last claim.  Recall that $\det(H_{{\eta}^{(2)}}(\alpha^*_n(v), \beta^*_n(v), \delta^*_n(v)))$  is the determinant of the
$3 \times 3$ Hessian matrix of $\eta^{(2)}(\alpha, \beta, \delta)$ evaluated at the point  $(\alpha, \beta, \delta) = (\alpha^*_n(v)$, $\beta^*_n(v), \delta^*_n(v))$.
By definition and direct calculations,  the top left entry of the matrix $H_{\eta^{(2)}}(\alpha, \beta, \delta)$ is
\begin{align*}
\frac{\partial^2 \eta^{(2)}(\alpha, \beta,   \delta)}{\partial\alpha^2}      & =  \frac{ \alpha - \beta^2 }{ 2 \alpha^3 }     
\frac{ (-\beta/2\alpha + v)^2 (e^\delta - 1)^2 \phi^2( h_v(\alpha, \beta) )}{ 2\alpha ( \Phi( - h_v(\alpha, \beta) )
     + e^\delta \Phi(  h_v(\alpha, \beta) )   )^2}     \\
& + \frac{ \phi( h_v(\alpha, \beta)) (e^\delta - 1) ( (v - \frac{\beta}{2\alpha})^2(v + \frac{\beta}{2\alpha}) +  \frac{ 3\beta }{ 4 \alpha^2 }  -  \frac{v}{2\alpha}  )}{ \sqrt{ - 2 \alpha }  ( \Phi( - h_v(\alpha, \beta) ) + e^\delta \Phi(  h_v(\alpha, \beta) )   )},
\end{align*}
which is $2 + O(\eps_n)$ when evaluated at the point $(\alpha, \beta, \delta) = (\alpha^*_n(v)$, $\beta^*_n(v), \delta^*_n(v))$.  By similar calculations,
\[
H_{\eta^{(2)}}(\alpha_n^*(v), \beta_n^*(v), \delta_n^*(v)) = \left(\begin{array}{lll}
2  	&0 		& \phi(v)v \\
0	& 1  & \phi(v)   \\
v \phi(v)	& \phi(v) 	& \Phi(v)\Phi(-v)\\
\end{array}
\right) + \eps_n \cdot Rem,
\]
where $Rem$ is a $3 \times 3$ matrix each entry of which is $O(1)$. As a result,
\[
\det(H_{\eta^{(2)}}(\alpha^*_n(v), \beta^*_n(v), \delta^*_n(v)))  = 2(\Phi(v)\Phi(-v) - (1 + v^2/2)\phi^2(v)) + O(\epsilon_n),
\]
and the claim follows by recalling $\eps_n = t_p/\sqrt{n}$.     \qed

\section{Proof of Theorem 2.4}\label{app:KSalt}
For notational simplicity,
we fix $j$ and suppress the dependence of $j$ in all notations.
In this section,  $m_k = m_k(j)$, $\tau=\tau(j)$, $\bar{X} = \bar{X}(j)$,    $X_i = X_i(j)$, and $\hat{\sigma} = \hat{\sigma}(j)$,   all of them  represent a number instead of a vector. Similarly,
$X$ denotes the $j$-th column of the original data matrix, so it is now an $n \times 1$ vector instead of an $n \times p$ matrix.
 This is a slight misuse of the notation.
When the $j$-th feature is useful,
all samples of the $j$-th feature partition into $K$ different groups, and sample $i$ belongs to group $k$ if and only if $y_i =k$. Let $\bar{X}^{(k)}$ and $\hat{\sigma}_X^{(k)}$ be the sample mean and sample standard deviation of group $k$. Decompose
\[
F_n(t) - \Phi(\frac{t - \bar{X}}{\hat{\sigma}_X}) = (I) + (II),
\]
where
\[
(I) = F_n(t) - \sum_{k = 1}^K \delta_k \Phi(\frac{t - \bar{X}^{(k)}}{\hat{\sigma}_X^{(k)}}),
\qquad (II) =  \sum_{k = 1}^K \delta_k \Phi(\frac{t - \bar{X}^{(k)}}{\hat{\sigma}_X^{(k)}})  -  \Phi(\frac{t - \bar{X}}{\hat{\sigma}_X}).
\]
Consider $(I)$. Introduce
\[
M_{n}^{(k)}(t) \equiv \sqrt{n\delta_k}  \bigl[  \frac{1}{n\delta_k}\sum_{i: y_i = k} 1\{X_i < t\} - \Phi(\frac{t - \bar{X}^{(k)}}{\hat{\sigma}^{(k)}})  \bigr].
\]
We can rewrite
\[
(I) = \frac{1}{\sqrt{n}}\sum_{k=1}^K \sqrt{\delta_k} \cdot M_n^{(k)}(t),
\]
where we note that $\frac{1}{n \delta_k} \sum_{i: y_i = k} 1\{X_i < t\}$ is the empirical CDF for the observations in group $k$. As a result,   $\sup_{-\infty<t<\infty}|M_n^{(k)}(t)|$ is the KS statistic for group $k$.
It follows from Theorem \ref{thm:KSnull} that $P(\sup_{-\infty<t<\infty}|M_n^{(k)}(t)|>\tilde{\eta}_n)\leq 2\sqrt{2\pi/(\pi - 2)} \exp(-\frac{2\pi}{\pi - 2} \tilde{\eta}_n^2)\cdot [1+o(1)]$, for any sequence $\tilde{\eta}_n\to\infty$ and $\tilde{\eta}_n/\sqrt{n}\to 0$. We then have
\begin{align} \label{(I)result}
P(\sqrt{n} \sup_{-\infty < t < \infty }|(I)| > \tilde{\eta}_n) & \leq   P(\sum_{k = 1}^K \sqrt{\delta_k}\sup_{-\infty<t<\infty}|M_n^{(k)}(t)| > \tilde{\eta}_n) \cr
& \leq \sum_{k=1}^K P( \sup_{-\infty<t<\infty} |M_n^{(k)}(t)|> \frac{\tilde{\eta}_n}{\sum_{k=1}^K \sqrt{\delta_k}} )\cr
&\leq \sum_{k=1}^K P( \sup_{-\infty<t<\infty} |M_n^{(k)}(t)|> \frac{\tilde{\eta}_n}{\sqrt{K}} )\cr
& \leq  2 K \sqrt{\frac{2\pi}{\pi - 2}} \exp(-\frac{2\pi}{(\pi - 2)K} \tilde{\eta}_n^2)\cdot [1+o(1)],
\end{align}
where the third inequality follows from  $\sum_{k=1}^K\sqrt{\delta_k}\leq \sqrt{K}(\sum_{k=1}^K\delta_k)^{1/2}=\sqrt{K}$, by Cauchy-Schwarz inequality.

Consider $(II)$. The following lemma is proved below:
\begin{lemma}  \label{lem:(II)}
Under conditions of Theorem \ref{thm:KSalt}, with probability at least $1-O(p^{-3})$,
\[
\big| \sup_{-\infty < t<\infty }|(II)| - \frac{1}{\sqrt{n}}\tau \big|\leq C \sum_{k=1}^K \delta_km_k^4.
\]
\end{lemma}
By Lemma \ref{lem:(II)} and \eqref{KSaltCondition}, with probability at least $1-O(p^{-3})$,
\begin{equation}  \label{(II)result}
\sqrt{n}\sup_{-\infty<t<\infty} |(II)| = \tau [1+O(p^{-\delta})].
\end{equation}

Combining \eqref{(I)result}-\eqref{(II)result}, when $\tau \geq (1+C)t_p$,
\begin{eqnarray*}
P(\psi_n \leq t_p)& \leq & P\big(\sqrt{n} \sup_{-\infty<t<\infty}|(I)| \geq  \tau [1+O(p^{-\delta})]-t_p \big)  + O(p^{-3})\cr
& \leq& 2K\sqrt{\frac{2\pi}{\pi - 2}}\exp\Big\{-\frac{2\pi}{(\pi - 2)K} \big( \tau[1+O(p^{-\delta})] - t_p\big)^2 \Big\} + O(p^{-3})\cr
&\leq &  2K\sqrt{\frac{2\pi}{\pi - 2}}\exp\Big\{-\frac{2\pi}{(\pi - 2)K} \big( \tau - t_p\big)^2 \Big\}[1+o(1)] + O(p^{-3}),
\end{eqnarray*}
where the last inequality follows from that $\tau \leq L_p$ (recall that $L_p$ is a generic multi-$\log(p)$ term).  This gives the claim. \qed

\subsection{Proof of Lemma \ref{lem:(II)}}
Let $L_n^{(k)}(t) = \frac{t - \bar{X}^{(k)}}{\hat{\sigma}_X^{(k)}}   -  \frac{t - \bar{X}}{\hat{\sigma}_X}$.
We apply Taylor expansion to $(II)$ and obtain
\begin{align} \label{taylor}
(II) &= \phi(\frac{t - \bar{X}}{\hat{\sigma}_X})\sum_{k=1}^K\delta_k L_n^{(k)}(t) + \frac{1}{2} \phi^{(1)}(\frac{t - \bar{X}}{\hat{\sigma}_X})\sum_{k=1}^K \delta_k [L_n^{(k)}(t)]^2 \cr
& + \frac{1}{6} \phi^{(2)}(\frac{t - \bar{X}}{\hat{\sigma}_X}) \sum_{k=1}^K \delta_k [L_n^{(k)}(t)]^3 + \frac{1}{24} \sum_{k=1}^K \phi^{(3)}(\xi_k)\cdot \delta_k [L_n^{(k)}(t)]^4,
\end{align}
where $\phi^{(m)}$ denotes the $m$-th derivative of the standard normal density function $\phi$ and $\xi_k$ falls between $\frac{t-\bar{X}}{\hat{\sigma}_X}$ and $\frac{t-\bar{X}}{\hat{\sigma}_X} + L_n^{(k)}(t)$.

To simplify \eqref{taylor}, we rewrite $L_n^{(k)}(t)$ as
\[
L_n^{(k)}(t) =  \frac{\bar{X} - \bar{X}^{(k)}}{\hat{\sigma}_X} +  \frac{t - \bar{X}}{\hat{\sigma}_X} \Big(\frac{\hat{\sigma}_X}{\hat{\sigma}_X^{(k)}} - 1 \Big)  +  \frac{\bar{X} - \bar{X}^{(k)}}{\hat{\sigma}_X}\Big(\frac{\hat{\sigma}_X}{\hat{\sigma}_X^{(k)}} - 1\Big).
\]
Furthermore,
\begin{align*}
\frac{\hat{\sigma}_X}{\hat{\sigma}_X^{(k)}} - 1 &= \frac{\hat{\sigma}_X^2 - (\hat{\sigma}_X^{(k)})^2}{2 \hat{\sigma}_X^2} + \frac{2\hat{\sigma}_X^2 (2\hat{\sigma}_X+\hat{\sigma}_X^{(k)})}{(\hat{\sigma}_X^{(k)} + \hat{\sigma}_X )^2  \hat{\sigma}_X^{(k)}} \Big(\frac{\hat{\sigma}^2_X - (\hat{\sigma}_X^{(k)})^2}{2\hat{\sigma}^2_X} \Big)^2\cr
&=  \frac{\hat{\sigma}_X^2 - (\hat{\sigma}_X^{(k)})^2}{2 \hat{\sigma}_X^2} + \frac{3}{2} \Big(\frac{\hat{\sigma}^2_X - (\hat{\sigma}_X^{(k)})^2}{2\hat{\sigma}^2_X} \Big)^2[1+o(1)],
\end{align*}
where in the last inequality, we have used the fact that $|\hat{\sigma}_X-\sigma|=o(1)$ and $|\hat{\sigma}^{(k)}_X-\sigma|=o(1)$, which is easily seen from \eqref{term1}-\eqref{term2} below. Together, we have
\begin{align}  \label{Lapprox}
L_n^{(k)}(t) &=  \frac{\bar{X} - \bar{X}^{(k)}}{\hat{\sigma}_X}
+ \frac{t - \bar{X}}{\hat{\sigma}_X} \frac{\hat{\sigma}_X^2 - (\hat{\sigma}_X^{(k)})^2}{2 \hat{\sigma}_X^2}
+ \frac{\bar{X} - \bar{X}^{(k)}}{\hat{\sigma}_X}\frac{\hat{\sigma}_X^2 - (\hat{\sigma}_X^{(k)})^2}{2 \hat{\sigma}_X^2} +  \eps_n^{(k)}(t),
\end{align}
where
\begin{equation}  \label{higherterm}
\sum_{k=1}^K \delta_k |\eps_n^{(k)}(t)| \leq \sum_{k=1}^K \delta_k \frac{|t-\bar{X}|+|\bar{X}-\bar{X}^{(k)}|}{\hat{\sigma}_X}\cdot \frac{3}{2}\Big(\frac{\hat{\sigma}^2_X - (\hat{\sigma}_X^{(k)})^2}{2\hat{\sigma}^2_X} \Big)^2 \cdot [1+o(1)].
\end{equation}
Plugging \eqref{Lapprox} into \eqref{taylor} gives
\begin{equation} \label{proofreadadd3}
(II) = (II_1) + (II_2) +(II_3) + err,
\end{equation}
where
\begin{align*}
(II_1) & = \phi(\frac{t - \bar{X}}{\hat{\sigma}_X})\sum_{k=1}^K\delta_k \frac{\bar{X}-\bar{X}^{(k)}}{\hat{\sigma}_X},\cr
(II_2) &= \phi(\frac{t - \bar{X}}{\hat{\sigma}_X}) \frac{t - \bar{X}}{\hat{\sigma}_X}\sum_{k=1}^K \delta_k \frac{\hat{\sigma}_X^2 - (\hat{\sigma}_X^{(k)})^2}{2 \hat{\sigma}_X^2} + \frac{1}{2}\phi^{(1)}(\frac{t - \bar{X}}{\hat{\sigma}_X}) \sum_{k=1}^K \delta_k \frac{(\bar{X}-\bar{X}^{(k)})^2}{\hat{\sigma}^2_X},\cr
(II_3)
&= \phi(\frac{t - \bar{X}}{\hat{\sigma}_X}) \sum_{k=1}^K \delta_k \frac{\bar{X} - \bar{X}^{(k)}}{\hat{\sigma}_X}\frac{\hat{\sigma}_X^2 - (\hat{\sigma}_X^{(k)})^2}{2 \hat{\sigma}_X^2}
+ \frac{1}{6} \phi^{(2)}(\frac{t - \bar{X}}{\hat{\sigma}_X})\sum_{k=1}^K \delta_k \Big( \frac{\bar{X}-\bar{X}^{(k)}}{\hat{\sigma}_X} \Big)^3 \cr
& + \frac{1}{2}\phi^{(1)}(\frac{t - \bar{X}}{\hat{\sigma}_X})\sum_{k=1}^K 2\delta_k \frac{\bar{X} - \bar{X}^{(k)}}{\hat{\sigma}_X}\cdot\frac{t - \bar{X}}{\hat{\sigma}_X}\frac{\hat{\sigma}_X^2 - (\hat{\sigma}_X^{(k)})^2}{2 \hat{\sigma}_X^2}.
\end{align*}

Denote $y=(t-\bar{X})/\hat{\sigma}_X$.
We show that $(II_1) = (II_2) = 0$, and that
\begin{equation} \label{(II3)}
(II_3)  = [\phi(y) + y\phi^{(1)}(y)] \sum_{k=1}^K \delta_k \frac{\bar{X} - \bar{X}^{(k)}}{\hat{\sigma}_X}\frac{\hat{\sigma}_X^2 - (\hat{\sigma}_X^{(k)})^2}{2 \hat{\sigma}_X^2}  +\frac{1}{6}\phi^{(2)}(y) \sum_{k=1}^K \delta_k \Big( \frac{\bar{X}-\bar{X}^{(k)}}{\hat{\sigma}_X} \Big)^3.
\end{equation}
The last claim follows by basic algebra, so we only show the first two claims.
Consider the first claim. By definition and elementary calculation,
\[
\bar{X} =\sum_{k=1}^K \delta_k \bar{X}^{(k)}, \qquad \hat{\sigma}_X^2 =  \sum_{k = 1}^K \delta_k (\hat{\sigma}_X^{(k)})^2 + \sum_{k=1}^K  \delta_k (\bar{X}^{(k)} - \bar{X})^2.
\]
In particular, this implies that
\begin{equation} \label{orderequal}
\sum_{k=1}^K \delta_k \big((\hat{\sigma}^{(k)}_X)^2 - \hat{\sigma}_X^2\big) = \sum_{k=1}^K \delta_k (\bar{X}^{(k)}-\bar{X})^2.
\end{equation}
 It follows that   $(II_1)=\hat{\sigma}_X^{-1}\phi(y)[\bar{X}-\sum_{k=1}^K\delta_k\bar{X}^{(k)}]=\hat{\sigma}_X^{-1}\phi(y)\cdot 0=0$, and the first claim follows.
At the same time, using \eqref{orderequal},
\begin{align*}
(II_2) &= y \phi(y) \sum_{k=1}^K \delta_k \frac{\hat{\sigma}_X^2 - (\hat{\sigma}_X^{(k)})^2}{2 \hat{\sigma}_X^2} + \frac{1}{2}\phi^{(1)}(y) \sum_{k=1}^K \delta_k \frac{(\bar{X}-\bar{X}^{(k)})^2}{\hat{\sigma}^2_X}\cr
&= [y\phi(y) + \phi^{(1)}(y)] \sum_{k=1}^K \delta_k  \frac{(\bar{X}-\bar{X}^{(k)})^2}{2\hat{\sigma}^2_X},
\end{align*}
where the right hand side is $0$ as $y\phi(y)+\phi^{(1)}(y)=0$ for any $y$.   This shows the second claim.

Combining the above with (\ref{proofreadadd3}) gives
\begin{equation} \label{proofreadadd2}
(II) = (II_3) + err,
\end{equation}
where $(II_3)$ is given by (\ref{(II3)}).

We now study $(II_3)$.  By basics of the normal distribution, with probability at least $1-O(p^{-3})$, $|\bar{X}^{(k)}-\mu_k-\bar{\mu}|\leq C\sigma\sqrt{\log(p)/n}$ and $|(\hat{\sigma}^{(k)}_X)^2 - \sigma^2|\leq C\sigma^2\sqrt{\log(p)/n}$, for all $1\leq k\leq K$. It follows that $|\bar{X}-\bar{\mu}|\leq C\sigma\sqrt{\log(p)/n}$ and
\begin{equation}  \label{term1}
\bar{X}^{(k)}-\bar{X} =  \mu_k + \sigma\cdot O\big(\sqrt{\log(p)/n}\big).
\end{equation}
In addition,
\[
(\hat{\sigma}^{(k)}_X)^2 - \hat{\sigma}_X^2 = \sum_{\ell \neq k} \delta_{\ell} \big( (\hat{\sigma}^{(k)}_X)^2 - (\hat{\sigma}^{(\ell)}_X)^2\big) - \sum_{\ell=1}^K \delta_{\ell} (\bar{X}-\bar{X}^{(\ell)})^2
 = - \sum_{\ell=1}^K \delta_{\ell}\mu_{\ell}^2 + err^{(k)},
\]
where $|err^{(k)}|\leq C\sigma^2 \sqrt{\log(p)/n} + \sum_{\ell=1}^K \delta_{\ell}|\mu_{\ell}||\bar{X}^{(\ell)}-\bar{X}-\mu_{\ell}|\leq C\sigma^2(1+\sum_{\ell=1}^K\delta_{\ell}|m_{\ell}|) \sqrt{\log(p)/n}$.
Noting that $\max_{\ell}|m_{\ell}|\to 0$, we have $\sum_{\ell=1}^K\delta_{\ell}|m_{\ell}|=o(1)$. As a result,
\begin{equation} \label{term2}
\hat{\sigma}_X^2 - (\hat{\sigma}^{(k)}_X)^2 = \sum_{\ell=1}^K \delta_{\ell}\mu_{\ell}^2 + \sigma^2\cdot O\big(\sqrt{\log(p)/n}\big).
\end{equation}
From \eqref{term1}-\eqref{term2},
\[
\sum_{k=1}^K \delta_k \frac{\bar{X} - \bar{X}^{(k)}}{\hat{\sigma}_X}\frac{\hat{\sigma}_X^2 - (\hat{\sigma}_X^{(k)})^2}{2 \hat{\sigma}_X^2} = -\sum_{k=1}^K \delta_k\mu_k\frac{\sum_{\ell=1}^K \delta_{\ell}\mu^2_{\ell}}{2\hat{\sigma}^3_X} + err_1.
\]
Since $\sum_{k=1}^K\delta_k\mu_k=0$, the first term is equal to $0$. In addition, $|err_1|\leq C\sigma \sqrt{\log(p)/n}\frac{\sum_{\ell=1}^K \delta_{\ell}\mu_{\ell}^2}{2\hat{\sigma}^3_X} + C\sum_{k=1}^K\delta_k |\mu_k|\frac{\sigma^2 \sqrt{\log(p)/n}}{2\hat{\sigma}_X^3}$. Since $\sigma^{-1} \max_{\ell}|\mu_{\ell}|=\max_{\ell}|m_{\ell}|\to 0$, the second term in $|err_1|$ dominates. Therefore, we have
\begin{equation} \label{(II3)term1}
\sum_{k=1}^K \delta_k \frac{\bar{X} - \bar{X}^{(k)}}{\hat{\sigma}_X}\frac{\hat{\sigma}_X^2 - (\hat{\sigma}_X^{(k)})^2}{2 \hat{\sigma}_X^2} = O\big( \sqrt{\log(p)/n}\big)\cdot \frac{\sigma^3\sum_{k=1}^K \delta_k|m_k|}{\hat{\sigma}_X^3}.
\end{equation}
Similarly, from \eqref{term1},
\[
\sum_{k=1}^K \delta_k \Big( \frac{\bar{X} - \bar{X}^{(k)}}{\hat{\sigma}_X}\Big)^3 = \sum_{k=1}^K \delta_k \frac{\mu_k^3}{\hat{\sigma}_X^3} + err_2,
\]
where $|err_2|\leq C\sum_{k=1}^K \delta_k|\mu_k|\cdot \sigma^2 \log(p)/n+C\sum_{k=1}^K \delta_k\mu_k^2\cdot \sigma\sqrt{\log(p)/n}$, equivalent with $C \sigma^3 [\sum_{k=1}^K \delta_k|m_k|\cdot  \log(p)/n+ \sum_{k=1}^K \delta_k m_k^2\cdot \sqrt{\log(p)/n}]$. 
By \eqref{KSaltCondition}, $|m_k|$ is either 0 or lower bounded by $C\sqrt{\log(p)/n}$, so there is $C\sqrt{\log(p)/n} |m_k| \leq   m_k^2$ for $1 \leq k \leq K$, and consequently 
$C\sqrt{\log(p)/n} \sum_{k=1}^K \delta_k|m_k| \leq \sum_{k=1}^K \delta_k m_k^2$. Combining it with $|err_2|$, the second term in $|err_2|$ dominates.
So we have
\begin{equation}\label{(II3)term2}
\sum_{k=1}^K \delta_k \Big( \frac{\bar{X} - \bar{X}^{(k)}}{\hat{\sigma}_X}\Big)^3 = \frac{\sum_{k=1}^K \delta_k \mu_k^3}{\hat{\sigma}_X^3} + O\big( \sqrt{\log(p)/n}\big)\cdot \frac{\sigma^3\sum_{k=1}^K \delta_km_k^2}{\hat{\sigma}_X^3}
\end{equation}
Plugging \eqref{(II3)term1}-\eqref{(II3)term2} into \eqref{(II3)}, we obtain
\begin{equation} \label{thirdterm}
(II_3) = \frac{\sigma^3}{\hat{\sigma}_X^3}\cdot \Big[ \frac{1}{6}\phi^{(2)}(y) \sum_{k=1}^K \delta_k m_k^3 + O\big( \sqrt{\log(p)/n}\big) \sum_{k=1}^K \delta_k|m_k|\Big].
\end{equation}
By our assumptions, $\sqrt{\log(p)/n} \leq C |\sum_{k = 1}^K \delta_k m_k^3| \leq C \sum_{k = 1}^K \delta_k |m_k|^3$. Combining this with Cauchy-Schwarz inequality,  $\sqrt{\log(p)/n} \cdot  \sum_{k = 1}^K  \delta_k  |m_k|   \leq  C  (\sum_{k = 1}^K \delta_k |m_k|) \cdot (\sum_{k = 1}^K \delta_k |m_k|^3) \leq C   (\sum_{k = 1}^K \delta_k m_k^2)^{1/2} \sqrt{(\sum_{k = 1}^K \delta_k m_k^2) (\sum_{k = 1}^K \delta_k m_k^4)}$, which is  $C (\sum_{k = 1}^K \delta_k m_k^2) \sqrt{\sum_{k = 1}^K \delta_k m_k^4} \leq C \sum_{k = 1}^K \delta_k m_k^4$, again by Cauchy-Schwarz inequality. Combining this with (\ref{thirdterm}),
\begin{equation}  \label{thirdtermA}
(II_3) = \frac{\sigma^3}{\hat{\sigma}_X^3}\cdot \Big[\frac{1}{6}\phi^{(2)}(y) \sum_{k=1}^K \delta_k m_k^3 + O\big(\sum_{k=1}^K \delta_k m_k^4 \big)\Big].
\end{equation}
Note that $|\sigma^3/\hat{\sigma}_X^3-1|\leq C|\hat{\sigma}-\sigma|/\sigma$. From \eqref{term2}, $|\hat{\sigma}-\sigma|\leq C\sigma\big(\sum_{k=1}^K\delta_km_k^2 + \sqrt{\log(p)/n}\big)$, where $\sum_{k=1}^K\delta_km_k^2\cdot \sum_{k=1}^K\delta_k|m_k|^3\ll (\sum_{k=1}^K\delta_km_k^2)^2\leq \sum_{k=1}^K\delta_km_k^4$ and $\sqrt{\log(p)/n}\sum_{k=1}^K\delta_k|m_k|^3\leq C\sum_{k=1}^K\delta_k m_k^4$ as $C\sqrt{\log(p)/n} |m_k| \leq   m_k^2$, $1 \leq k \leq K$. Therefore,
\begin{equation}  \label{thirdtermB}
|\sigma^3/\hat{\sigma}_X^3-1|\cdot \sum_{k=1}^K \delta_k|m_k|^3\leq C \sum_{k=1}^K\delta_km_k^4.
\end{equation}
Combining \eqref{thirdtermA}-\eqref{thirdtermB}, with probability at least $1-O(p^{-3})$,
\begin{equation} \label{thirdtermnew}
(II_3) = \frac{1}{6}\phi^{(2)}(y) \sum_{k=1}^K \delta_k m_k^3 + O\big(\sum_{k=1}^K \delta_k m_k^4 \big).
\end{equation}

Now, we bound $err$. It has three parts: (i) the last term in the Taylor expansion \eqref{taylor}, (ii) those terms related to $\eps_n^{(k)}(t)$ in \eqref{Lapprox}, and (iii) those terms included in the first three terms of the Taylor expansion but excluded from $(II_1)$-$(II_3)$.
First, consider (i). From \eqref{Lapprox} and \eqref{term1}-\eqref{term2},
$|L_n^{(k)}(t)|\leq C\sigma(|m_k|+ |y|\sqrt{\log(p)/n}+|y|\sum_{k=1}^K\delta_km_k^2)$. Noting that $\sum_{k = 1}^K \delta_k m_k^2 \geq \sum_{k = 1}^K \delta_k |m_k|^3 \geq C\sqrt{\log(p)/n}$ and $(1+|y|)^4\phi^{(3)}(y)$ are uniformly bounded, we have
\[
\frac{1}{24}\sum_{k=1}^K \phi^{(3)}(\xi_k)\cdot \delta_k [L_n^{(k)}(t)]^4 \leq C \Big[\sum_{k=1}^K \delta_km_k^4 + \big(\sum_{k=1}^K \delta_k m_k^2\big)^4\Big] \leq C\sum_{k=1}^K \delta_k m_k^4.
\]
Second, consider (ii). Those terms related to $\eps_n^{(k)}(t)$ will not exceed $\phi(y)\sum_{k=1}^K\delta_k|\eps_n^{(k)}(t)|$. Combining \eqref{higherterm} and \eqref{term1}-\eqref{term2},
\begin{multline*}
\sum_{k=1}^K\delta_k|\eps_n^{(k)}(t)|\leq C\big[\sum_{k=1}^K\delta_k m_k^2 + \sqrt{\log(p)/n}\big]^2\leq C\Big[ \big(\sum_{k=1}^K \delta_km_k^2\big)^2 + \log(p)/n\Big]\\
\leq C\sum_{k=1}^K \delta_k m_k^4,
\end{multline*}
where the last inequality comes from $\log(p)/n\leq (\sum_{k=1}^K\delta_k|m_k|)^2\leq \sum_{k=1}^K\delta_km_k^2$ and $(\sum_{k=1}^K\delta_km_k^2)^2\leq \sum_{k=1}^K\delta_km_k^4$.
Last, consider (iii). We can easily figure out that the dominating terms are the following:
\begin{align*}
&\sum_{k=1}^K \delta_k \Big(\frac{\hat{\sigma}^2_X - (\hat{\sigma}_X^{(k)})^2}{2\hat{\sigma}^2_X} \Big)^2
 = \frac{\sigma^4}{\hat{\sigma}_X^4}\Big[\frac{1}{4} \big(\sum_{k=1}^K \delta_k m_k^2 \big)^2 + O\big(\sqrt{\log(p)/n}\big) \cdot \sum_{k=1}^K \delta_km_k^2 \Big], \cr
&\sum_{k=1}^K \delta_k \Big(\frac{\bar{X}-\bar{X}^{(k)}}{\hat{\sigma}_X} \Big)^4
 = \frac{\sigma^4}{\hat{\sigma}_X^4}\Big[ \sum_{k=1}^K \delta_k m_k^4 + O\big(\sqrt{\log(p)/n}\big) \cdot \sum_{k=1}^K \delta_k|m_k|^3\Big], \cr
&\sum_{k=1}^K \delta_k \frac{\hat{\sigma}_X^2 - (\hat{\sigma}_X^{(k)})^2}{2 \hat{\sigma}_X^2}\Big( \frac{\bar{X} - \bar{X}^{(k)}}{\hat{\sigma}_X} \Big)^2
= \frac{\sigma^4}{\hat{\sigma}_X^4}\Big[ \frac{1}{2}\big(\sum_{k=1}^K \delta_k m_k^2\big)^2 + O\big(\sqrt{\frac{\log(p)}{n}}\big) \cdot \sum_{k=1}^K \delta_k m_k^2\Big].
\end{align*}
Since $C\sqrt{\log(p)/n} |m_k| \leq   m_k^2$, $\sqrt{\log(p)/n}\sum_{k=1}^K\delta_k |m_k|^3\leq \sum_{k=1}^K\delta_km_k^4$. By Cauchy-Schwarz inequality, $(\sum_{k=1}^K \delta_km_k^2)^2\leq \sum_{k=1}^K\delta_km_k^4$. Moreover, $\sqrt{\log(p)/n}\sum_{k=1}^K\delta_km_k^2\ll \sqrt{\log(p)/n}\sum_{k=1}^K\delta_k|m_k|\leq C\sum_{k=1}^K\delta_km_k^4$, as we have seen in deriving \eqref{thirdtermnew}. So these terms are bounded by
$C\frac{\sigma^4}{\hat{\sigma}_X^4} \sum_{k=1}^K\delta_km_k^4$,
where $\hat{\sigma}=\sigma[1+o(1)]$.
Combining the results for (i)-(iii), with probability $1-O(p^{-3})$,
\begin{equation} \label{taylorerr}
|err|\leq C \sum_{k=1}^K \delta_k m_k^4.
\end{equation}

Now, inserting \eqref{thirdtermnew} and \eqref{taylorerr} into (\ref{proofreadadd2}) gives that with probability $1-O(p^{-3})$,
\begin{equation} \label{proofreadadd5}
(II)   = \frac{1}{6}\phi^{(2)}(y) \sum_{k=1}^K \delta_k m_k^3 + O\big(\sum_{k=1}^K \delta_k m_k^4 \big), \qquad y = (t-\bar{X})/\hat{\sigma}_X.
\end{equation}
Note that $\sum_{k=1}^K \delta_k m_k^3$ does not depend on $t$,
and that  $O(\sum_{k=1}^K \delta_k m_k^4)$  represents a term that $\leq C\sum_{k=1}^K \delta_k m_k^4$ in magnitude,   where $C$ does not depend on $(t, \bar{X}, \hat{\sigma}_X)$; this is because $y^a\phi^{(b)}(y)$ are always uniformly bounded for all $y$ and any fixed integers $a,b\geq 0$. It follows that with probability $1-O(p^{-3})$,
\[
\sup_{-\infty<t<\infty}|(II)| = \sup_{-\infty<t<\infty} \bigl\{  \frac{1}{6}|\phi^{(2)}(y)| \bigr\}  \cdot  |\sum_{k=1}^K \delta_k m_k^3|  + O\big(\sum_{k=1}^K \delta_k m_k^4 \big).
\]
By elementary calculus,  $\sup_{-\infty<t<\infty} \{  \frac{1}{6} |\phi^{(2)}(y)| \} = \frac{1}{6\sqrt{2\pi}}$.
Recalling
$\tau = \frac{1}{6\sqrt{2\pi}} \sqrt{n} |\sum_{k = 1}^K \delta_k m_k^3|$ gives the claim.
\qed
\section{Proof of Lemmas 2.1--2.4} \label{app:lemma}

\subsection{Proof of Lemma \ref{lemma:LM}}
For simplicity, we drop the subscripts of the matrices as long as there is no confusion.
Consider the first two claims.  By basic algebra,  for any two SVDs of $LM\Lambda$,
\[
L M \Lambda  =  U D  V'  = \widetilde{U}  \widetilde{D}  \widetilde{V}',
\]
if we require the diagonal entries of $D$ and $\widetilde{D}$ to be arranged in the descending order, then  there is a matrix $H \in {\cal H}_{K-1}$ such that
\begin{equation} \label{LMpf1}
D  = \widetilde{D}, \qquad U  = \widetilde{U}  H.
\end{equation}
At the same time, we write $A \Omega A = (GM\Lambda) (GM\Lambda)'$, where
$G = G_{K,K}$ is as in (\ref{DefineG}).  For any SVD of $GM \Lambda$, say,
$G M \Lambda = Q^* D^* (V^*)'$, by the way $Q$ is defined, there is a matrix $H \in {\cal H}_{K-1}$, $Q = Q^* H$.  This says \begin{equation} \label{SVDDMnew}
GM\Lambda = Q H D^* (V^*)'
\end{equation}

Writing  $LM\Lambda  = (n^{-1/2} L G^{-1})  (\sqrt{n} G M \Lambda )$ and using (\ref{SVDDMnew}),
\begin{equation} \label{LMpf2}
LM\Lambda = (n^{-1/2} L G^{-1})  QH  \diag(\sqrt{n \lambda_1}, \ldots, \sqrt{n \lambda_{K-1}}) (V^*)'.
\end{equation}
Direct calculations show that $(n^{-1/2} L G^ {-1} Q H)' (n^{-1/2} L G^{-1} Q H)  = I_{K-1}$. As a result,  the right hand side of (\ref{LMpf1}) is a SVD of  $LM$, and $\sqrt{n\lambda_1}, \ldots, \sqrt{n \lambda_{K-1}}$ are all the nonzero singular values of $LM\Lambda$. Moreover, by (\ref{LMpf2}),   there is an  $H \in {\cal H}_{K-1}$  such that
\[
U_{n, K-1} = n^{-1/2} L G^{-1} Q  H, \qquad Q = Q_{K, K-1}.
\]
These prove the first two claims.

Consider the last two claims.  Denote $\xi$ by the unit-norm $K \times 1$  vector $(\sqrt{\delta_1}, \ldots, \sqrt{\delta_K})'$ and denote $\widetilde{Q}$ by the $K \times K$ matrix
\[
\widetilde{Q} = [Q, \xi].
\]
On one hand, by basics on SVD, for any $1 \leq k \leq K-1$, the $k$-th  column of $Q$ is an eigenvector of the matrix $GM\Lambda^2 M'G'$, with $\lambda_k$ being the associated eigenvalue. On the other hand,
recall that $G = \diag(\sqrt{\delta_1}, \ldots, \sqrt{\delta_K})$ and that the $k$-th row of $M \Lambda $ is $\mu_k' \widetilde{\Sigma}^{-1/2}$,
\[
\xi' GM \Lambda =  (\sum_{k = 1}^K \delta_k \mu_k')  \widetilde{\Sigma}^{-1/2} = 0,
\]
where in the last equality, we have used $\delta_k \mu_k = 0$; see (\ref{sumofmu}).  It follows that $\xi$ is an eigenvector of $GM \Lambda^2 M'G'$, with $0$ being the associated eigenvalue. Combining these and noting that $\lambda_1 > \lambda_2 > \ldots \lambda_{K-1} > 0$,    it follows from basic algebra that $\widetilde{Q}$ is an orthogonal matrix.

Now, for any $1 \leq k \leq K$, denote the $k$-th row of $Q$ by $q_k'$.  Since $\widetilde{Q}$ is orthogonal,
$\|q_k\|^2 = (1 - \delta_k)$, and
 the $\ell^2$-norm of the $k$-th row of $GQH$ is
\[
\|\delta_k^{-1/2} q_k\| = \delta_k^{-1/2} \|q_k\| = (\delta_k^{-1} - 1)^{1/2}.
\]
Moreover, for any $1 \leq  \ell \leq K$ and $k \neq \ell$,  again by the orthogonality of $\widetilde{Q}$,
$q_k' q_{\ell} = - \sqrt{\delta_k \delta_{\ell}}$.
The  $\ell^2$-distance between the $k$-th row and the $\ell$-th row  of $GQH$ is then
\[
\|\delta_k^{-1/2}  q_k - \delta^{-1/2} q_{\ell} \|   =  \bigl[ \frac{1}{\delta_k} (1 - \delta_k) + \frac{1}{\delta_{\ell}}(1 - \delta_{\ell})  -
\frac{2}{\sqrt{\delta_k \delta_{\ell}}}  (- \sqrt{\delta_k \delta_{\ell}}) \bigr]^{1/2}  = \bigl[\frac{1}{\delta_k} + \frac{1}{\delta_{\ell}} \bigr]^{1/2},
\]
where the right hand side  $\geq 2/\sqrt{(\delta_k + \delta_{\ell})}  \geq 2$. Combining these give the  claims.  \qed
\subsection{Proof of Lemma \ref{lemma:Urem}}
Write $\hat{S}  =  \hat{S}_p(t_p(q))$ and $t_p=t_p(q)$ for short. Note that
\[
\Vert L(M-M^{\hat{S}})\Lambda\Vert \leq \Vert L(M-M^{\hat{S}})\Vert \cdot\Vert\Lambda\Vert \leq \Vert L(M-M^{\hat{S}})\Vert_F\cdot \Vert\Lambda\Vert.
\]
The $j$-th coordinate of the diagonal matrix $\Lambda$ is $\sigma(j)/\sqrt{E[\hat{\sigma}^2(j)]}$. By direct calculation (see \eqref{hatsigma} for details), $E[\hat{\sigma}^2(j)]=\sigma^2(j)[1+\kappa^2(j)]\geq \sigma^2(j)$. So
\begin{equation} \label{equ-Lambda}
\Vert \Lambda\Vert\leq 1.
\end{equation}
Therefore, it suffices to show that with probability $\geq 1-o(p^{-2})$,
\begin{equation} \label{lemLMmiss-equ0}
\| L(M-M^{\hat{S}}) \|_F \leq C \|\kappa \|\sqrt{n}\cdot \big[ p^{-(1-\vartheta)/2}\sqrt{\rho_1(L,M)\log(p)} + p^{-[(\sqrt{r}-\sqrt{q})_+]^2/(2K)}\big].
\end{equation}

Now, we show \eqref{lemLMmiss-equ0}. By simple algebra,
\begin{equation}  \label{lemLMmiss-equ3}
\Vert L(M - M^{\hat{S}}) \Vert^2_F = n \sum_{j\in S_p(M)} \kappa^2(j) \cdot 1\{ \psi_{n_p,j}\leq t_p\} \equiv n \sum_{j\in S_p(M)} R_j.
\end{equation}
Here, $R_j$'s are independent and either $R_j=\kappa^2(j)$ or $R_j=0$. By Theorem \ref{thm:KSalt} and the fact that $\tau(j)\geq \tau_{min}\geq a_0\cdot \sqrt{2r\log(p)}$,
\[
P(R_j\neq 0) \leq C[ p^{-[(\sqrt{r}-\sqrt{q})_+]^2/K} + p^{-3}].
\]
It follows that
\begin{equation}  \label{lemLMmiss-equ1}
\sum_{j\in S_p(M)}E[R_j] \leq C[ p^{-[(\sqrt{r}-\sqrt{q})_+]^2/K} + p^{-3}] \sum_{j\in S_p(M)} \kappa^2(j) = C[ p^{-[(\sqrt{r}-\sqrt{q})_+]^2/K} + p^{-3}]\cdot \| \kappa \|^2.
\end{equation}

To control $\sum_{j\in S_p(M)}(R_j - E[R_j])$, we use Bennet's lemma (see \cite[Page 851]{Wellner86}): If $R_j\leq b$ and $\sum_{j\in S_p(M)}\mathrm{Var}(R_j)\leq v$, then for any $x\geq 0$
\[
P\Big( \sum_{j\in S_p(M)} ( R_j - E[R_j]) \geq x \Big) \leq \left\{
\begin{array}{lcr}
\exp\big( -\frac{c x^2}{2v}\big), &&  xb \leq v,\\
\exp\big( -\frac{c x}{2b} \big) && xb > v,
\end{array} \right.
\]
where $c=2\log(2)-1\approx .773$. Taking $x=\sqrt{6v\log(p)/c}$ when $6 b^2\log(p)\leq cv$, and $x=6b\log(p)/c$ when $6b^2\log(p)>cv$, we find that with probability $\geq 1-O(p^{-3})$
\begin{equation}  \label{Bennett}
\sum_{j\in S_p(M)} (R_j - E[R_j]) \leq C\big[\sqrt{v\log(p)}+b \log(p)\big].
\end{equation}
Note that $R_j\leq \Vert \kappa \Vert^2_\infty$ and $\mathrm{Var}(R_j)\leq \kappa^4(j)\cdot 2P(R_j\neq 0)\leq C\| \kappa \|^2_\infty[ p^{-[(\sqrt{r}-\sqrt{q})_+]^2/K} + p^{-3}]\cdot \kappa^2(j)$. We take
\[
b=\Vert \kappa \Vert^2_\infty, \qquad v=C\| \kappa \|^2_\infty[ p^{-[(\sqrt{r}-\sqrt{q})_+]^2/K} + p^{-3}] \cdot \Vert\kappa \Vert^2.
\]
It follows that with probability $\geq 1-O(p^{-3})$,
\begin{equation}  \label{lemLMmiss-equ2}
\sum_{j\in S_p(M)} ( R_j - E[R_j]) \leq C \Big[ \Vert\kappa \Vert^2_\infty \log(p) + \Vert \kappa\Vert_\infty\Vert \kappa\Vert \sqrt{(p^{-[(\sqrt{r}-\sqrt{q})_+]^2/K} + p^{-3})\log(p)} \Big].
\end{equation}

Combining \eqref{lemLMmiss-equ1} and \eqref{lemLMmiss-equ2} gives
\begin{align*}
\sum_{j\in S_p(M)} R_j & \leq C \Big( \|\kappa \|_\infty \sqrt{\log(p)} + \|\kappa \|\sqrt{p^{-[(\sqrt{r}-\sqrt{q})_+]^2/K} + p^{-3}}  \Big)^2\cr
&= C \|\kappa \|^2 \Big( p^{-(1-\vartheta)/2}\sqrt{\rho_1(L,M)\log(p)} + \sqrt{p^{-[(\sqrt{r}-\sqrt{q})_+]^2/K} + p^{-3}}  \Big)^2\cr
&\leq C \|\kappa \|^2 \big[ p^{-(1-\vartheta)/2}\sqrt{\rho_1(L,M)\log(p)} + p^{-[(\sqrt{r}-\sqrt{q})_+]^2/(2K)} \big]^2,
\end{align*}
where the last inequality is due to that $\rho_1(L,M)\geq 1$ and $1-\vartheta<3$. Inserting this into \eqref{lemLMmiss-equ3} gives \eqref{lemLMmiss-equ0}, and the claim follows.
\qed

\subsection{Proof of Lemma \ref{lemma:Zrem}}
Before we show Lemma \ref{lemma:Zrem}, we show  that with probability at least $1-O(p^{-3})$,
\begin{equation} \label{lemsize-equ0}
|\hat{S}(t_p(q))|\leq C\big[p^{1-\vartheta} +p^{1-q} + \log(p)\big] .
\end{equation}
Write for short $\hat{S} = \hat{S}(t_p(q))$ and $t_p=t_p(q)$. Noting that $|\hat{S}\cap S_p(M)|\leq |S_p(M)| = p^{1-\vartheta}$, we only need to bound $|\hat{S}\setminus S_p(M)|$.
Write
\[
|\hat{S}\setminus S_p(M)| = \sum_{j\notin S_p(M)} 1\{\psi_{n_p,j}\geq t_p \}.
\]
Note that $\psi_{n_p, j}$'s are independent. In addition, by Theorem \ref{thm:KSnull}, for $j\notin S_p(M)$,
\[
P (\psi_{n_p,j}\geq t_p) \leq C p^{-q}.
\]
So $E[|\hat{S}\setminus S_p(M)|]\leq Cp^{1-q}$. We now apply Bennett's lemma as in \eqref{Bennett} with $b=1$ and $v=p\cdot 2Cp^{-q}\cdot (1-Cp^{-q})$, it follows that with probability at least $1-O(p^{-3})$,
\[
|\hat{S}\setminus S_p(M)| - E[|\hat{S}\setminus S_p(M)|]\leq C\big[\log(p) + p^{(1-q)/2} \sqrt{\log(p)}\big].
\]
Combining the above gives $|\hat{S}\setminus S_p(M)|\leq C[p^{1-q}+\log(p)]$, and \eqref{lemsize-equ0} follows immediately.

We now proceed to show Lemma \ref{lemma:Zrem}. Write $m_p=m_p(q)=p^{1-\vartheta}+p^{1-q}+\log(p)$ and let ${\cal B}={\cal B}(t_p)$ be the collection of subsets of $\{1,\cdots,p\}$ with size $\leq Cm_p$. Given \eqref{lemsize-equ0}, to show the claim, it suffices to show that with probability at least $1-O(p^{-3})$,
\begin{equation}  \label{lemnormbound-equ0}
\max_{B\in {\cal B}} \Vert (Z\Sigma^{-1/2}\Lambda + R)^B \Vert  \leq
C\Big[\sqrt{n}+\sqrt{m_p\log(p)}+\|\kappa\|\cdot p^{-\frac{1-\vartheta}{2}}  \sqrt{m_p\rho_1(L,M)\log(p)}\Big].
\end{equation}
But by triangle inequality,  to show (\ref{lemnormbound-equ0}),  it suffices to show that with probability at least $1-O(p^{-3})$,
\begin{equation}  \label{lemnormbound-equ0A}
\max_{B\in {\cal B}} \Vert (Z\Sigma^{-1/2}\Lambda)^B    \Vert   \leq  C\Big[\sqrt{n}+\sqrt{m_p\log(p)}\Big],
\end{equation}
and
\begin{equation}   \label{lemnormbound-equ0B}
\max_{B\in {\cal B}} \Vert R^B \Vert  \leq   C \Big[\sqrt{n}+\sqrt{m_p\log(p)} + \|\kappa\|\cdot p^{-(1-\vartheta)/2}  \sqrt{m_p\rho_1(L,M)\log(p)} \Big].
\end{equation}

First, we consider (\ref{lemnormbound-equ0A}). The random matrix $Z\Sigma^{-1/2}$ has {\it iid} entries with the standard normal distribution. The following lemma is proved in \cite[Corollary 5.35]{Vershynin}.
\begin{lemma} \label{lemma:vershynin}
Let $A$ be an $N \times n$ matrix whose entries are independent standard normal random variables. Then for every $x \geq 0$, with probability at least $1 - 2\exp(- x^2/2)$, one has
\[
\sqrt{N} - \sqrt{n} - x \leq s_{\min}(A) \leq s_{\max}(A) \leq \sqrt{N} + \sqrt{n} + x,
\]
where $s_{\min}(A)$ and $s_{\max}(A)$ are correspondingly minimum and maximum eigenvalue of random matrix $A$.
\end{lemma}
We apply Lemma \ref{lemma:vershynin} to $A=(Z\Sigma^{-1/2})^B$ (as an $n\times |B|$ matrix by removing zero columns) and $x=\sqrt{2(3+Cm_p)\log(p)}$. Then for each fixed $B$, with probability $\geq 1-2p^{-(3+Cm_p)}$,
\[
\Vert (Z\Sigma^{-1/2})^B \Vert \leq \sqrt{n}+\sqrt{|B|}+\sqrt{2(3+Cm_p)\log(p)}.
\]
Noting that $|{\cal B}|\leq p^{Cm_p}$ and $m_p\geq 1$, we obtain
$\max_{B\in {\cal B}} \| (Z\Sigma^{-1/2})^B\| \leq \sqrt{n} + C\sqrt{m_p\log(p)}$.
In addition, $\| (Z\Sigma^{-1/2}\Lambda)^B\|=\| (Z\Sigma^{-1/2})^B\|\cdot \|\Lambda\|\leq \| (Z\Sigma^{-1/2})^B\|$, where we have used \eqref{equ-Lambda}. It follows that
\[
\max_{B\in {\cal B}} \| (Z\Sigma^{-1/2}\Lambda)^B\| \leq \sqrt{n} + C\sqrt{m_p\log(p)},
\]
and (\ref{lemnormbound-equ0A}) follows.

Next, consider (\ref{lemnormbound-equ0B}). Let $G=\Sigma^{1/2}(\hat{\Sigma}^{-1/2} - \tilde{\Sigma}^{-1/2})$, and we can write
\begin{align*}
R & = (LM + Z\Sigma^{-1/2})\Sigma^{1/2}(\hat{\Sigma}^{-1/2} - \tilde{\Sigma}^{-1/2}) - {\bf 1}_n(\bar{X}-\bar{\mu})'\hat{\Sigma}^{-1/2}\cr
&= (LM + Z\Sigma^{-1/2})G - {\bf 1}_n(\bar{X}-\bar{\mu})'\Sigma^{-1/2} (\Lambda + G).
\end{align*}
Therefore, $\max_{B\in{\cal B}}\| R^B\|$ does not exceed
\begin{equation} \label{lemnormbound-equ1}
 \|G\|\big(\max_{B\in{\cal B}}\|LM^B\| + \max_{B\in{\cal B}}\| (Z\Sigma^{-1/2})^B\|\big)  + (1 + \| G\|) \max_{B\in{\cal B}} \|[{\bf 1}_n(\bar{X}-\bar{\mu})'\Sigma^{-1/2}]^B\|.
\end{equation}
We now bound $\|G\|$, $\max_{B\in{\cal B}} \|[{\bf 1}_n(\bar{X}-\bar{\mu})'\Sigma^{-1/2}]^B\|$, and
$\max_{B\in {\cal B}}\|LM^{B}\|$ separately; a bound for $\max_{B\in{\cal B}}\| (Z\Sigma^{-1/2})^B\|$ is already given in (\ref{lemnormbound-equ0A}).

We first bound $\|G\|$. For sample $i$ that belongs to Class $k$, $X_i(j)=\bar{\mu}+\mu_k(j)+Z_i(j)$; also, $\bar{X}(j)=\bar{\mu}+\bar{Z}(j)$. Therefore,
\begin{align} \label{hatsigma}
\hat{\sigma}^2(j) &= \frac{1}{n}\sum_{i=1}^n\big[ X_i(j) - \bar{X}(j) \big] = \frac{1}{n}\sum_{k=1}^K \sum_{y_i=k} [\mu_k(j)+Z_i(j)-\bar{Z}(j)]^2\cr
&= \sum_{k=1}^K \delta_k \mu^2_k(j) + \frac{1}{n}\sum_{i=1}^n [Z_i(j)-\bar{Z}(j)]^2 + \frac{2}{n}\sum_{k=1}^K \sum_{y_i=k} \mu_k(j)[Z_i(j)-\bar{Z}(j)].
\end{align}
The first term is equal to $\sigma^2(j)\kappa^2(j)$. From basic properties of normal distributions, the second term is $\sigma^2(j) + \sigma^2(j)\cdot O(\sqrt{\log(p)/n})$ with probability $1-O(p^{-4})$. The third term is equal to
\[
2\sum_{k=1}^K \delta_k\mu_k(j)[\bar{Z}^{(k)}(j) - \bar{Z}(j)], \qquad \mbox{where } \bar{Z}^{(k)}(j) = \frac{1}{n\delta_k}\sum_{y_i=k} Z_i(j).
\]
With probability $\geq 1-O(p^{-4})$, $|\bar{Z}^{(k)}(j) - \bar{Z}(j)|\leq C\sigma(j)\sqrt{\log(p)/(n\delta_k)}$. So the absolute value of this term is bounded by $C\sum_{k=1}^K\sqrt{\delta_k}|\mu_k(j)|\cdot \sigma(j)\sqrt{\log(p)/n}$. By Cauchy-Schwarz inequality, $\sum_{k=1}^K \sqrt{\delta_k}\leq \sqrt{K}$, and this term is further bounded by $C\sqrt{K}\cdot \sigma^2(j)\max_{1\leq k\leq K}|m_k(j)|\cdot \sqrt{\log(p)/n}\ll \sigma^2(j)\sqrt{\log(p)/n}$, recalling \eqref{Condition2}. As a result, with probability $\geq 1-O(p^{-3})$,
\[
\max_{1\leq j\leq p}\Big|\frac{\hat{\sigma}^2(j) -  \sigma^2(j)[1+\kappa^2(j)]}{\sigma^2(j)}\Big|\leq Cn^{-1/2}\sqrt{\log(p)}.
\]
In addition, $\tilde{\Sigma}(j,j) = E[\hat{\sigma}^2(j)]=\sigma^2(j)[1+\kappa^2(j)]$, hence, the $j$-th diagonal of $\Lambda$ is $1/\sqrt{1+\kappa^2(j)}$. It follows that
\begin{equation} \label{lemnormbound-equ6}
\| G\|  = \max_{1\leq j\leq p} \frac{|\hat{\sigma}(j) - \sigma(j)\sqrt{1+\kappa^2(j)}|}{\hat{\sigma}(j)\sqrt{1+\kappa^2(j)}}
 \leq C\Big|\frac{\hat{\sigma}^2(j) -  \sigma^2(j)[1+\kappa^2(j)]}{\sigma^2(j)}\Big| \leq C \sqrt{\frac{\log(p)}{n}}.
\end{equation}

Second, we bound $\max_{B\in{\cal B}} \|[{\bf 1}_n(\bar{X}-\bar{\mu})'\Sigma^{-1/2}]^B\|$. Let $\zeta(j)=\sqrt{n}[\bar{X}(j)-\bar{\mu}(j)]/\sigma(j)$. Noting that for a rank-$1$ matrix,
its maximum singular value is equal to its Frobenius norm, we obtain
\[
\Vert [{\bf 1}_n(\bar{X}-\bar{\mu})'\Sigma^{-1/2}]^B\Vert = \Vert ({\bf 1}_n(\bar{X}-\bar{\mu})'\Sigma^{-1/2})^B\Vert_F =  \Vert\zeta^B \Vert.
\]
Since $\bar{\mu}=\frac{1}{n}\sum_{i=1}^n E[X_i]$, we write
\[
\zeta(j) = \frac{1}{\sqrt{n}}\sum_{i=1}^n \frac{X_i(j)-E[X_i(j)]}{\sigma(j)}.
\]
Here, $\zeta(j)$'s are {\it iid} $N(0, 1)$ random variables, so $\Vert\zeta^B\Vert^2$ has a $\chi^2_{|B|}(0)$ distribution; hence, $E[\Vert \zeta^B \Vert]\leq \sqrt{E[\Vert \zeta^B \Vert^2]}=\sqrt{|B|}$. Furthermore, by Gaussian concentration theory (see \cite[Proposition 2.5]{Dudley}), for any $x\geq 0$,
\[
P\big( \Vert \zeta^B\Vert - E[\Vert \zeta^B \Vert] > x\big)\leq e^{-x^2/2}.
\]
Taking $t=\sqrt{2(3+Cm_p)\log(p)}$, for each fixed $B$, with probability at least $1-2p^{-(3+Cm_p)}$,
\[
\Vert \zeta^B\Vert \leq \sqrt{|B|} + \sqrt{2(3+Cm_p(q))\log(p)}.
\]
Combining the above results and noting that $|B|\leq Cm_p$ for $B\in{\cal B}$, with probability $\geq 1-O(p^{-3})$,
\begin{equation} \label{lemnormbound-equ4}
\max_{B\in{\cal B}} \Vert [{\bf 1}_n(\bar{X}-\bar{\mu})'\Sigma^{-1/2}]^B\Vert \leq C \sqrt{m_p \log(p)}.
\end{equation}

Last, we bound $\max_{B\in {\cal B}}\|LM^{B}\|$. Note that $\Vert LM^B\Vert \leq \Vert LM^B\Vert_F$. Therefore,
\[
\Vert LM^B\Vert^2 \leq n\sum_{k=1}^n \delta_k \sum_{j\in B}m_k^2(j) = n\sum_{j\in B}\kappa^2(j)\leq n|B|\|\kappa\|^2_\infty.
\]
By definition, $\|\kappa\|^2_\infty \leq \rho_1(L,M)\cdot p^{-(1-\vartheta)}\|\kappa\|^2$. For $B\in {\cal B}$, $|B|\leq Cm_p$. Together, we have
\begin{equation} \label{lemnormbound-equ7}
\max_{B\in{\cal B}} \Vert LM^B \Vert\leq C\|\kappa\|\sqrt{n}\cdot p^{-(1-\vartheta)/2}\sqrt{m_p\rho_1(L,M)}.
\end{equation}

Finally, with all these bounds, inserting
\eqref{lemnormbound-equ6} and \eqref{lemnormbound-equ4}-\eqref{lemnormbound-equ7} into \eqref{lemnormbound-equ1} gives
\begin{equation} \label{lemnormbound-equ3}
\max_{B\in {\cal B}}\| R^B\|  \leq C\Big[\sqrt{\log(p)} + \sqrt{m_p\log(p)} + \|\kappa\|\cdot p^{-(1-\vartheta)/2}\sqrt{m_p\rho_1(L,M)\log(p)} \Big],
\end{equation}
and \eqref{lemnormbound-equ0B} follows.
\qed

\subsection{Proof of Lemma \ref{lem:project}}
Let $M=V_1'V_2$. For any orthogonal matrix $H$, $\|V_1-V_2H\|_F^2=2\mathrm{tr}(I_K - MH)$. Moreover, $\| V_1V_1'-V_2V_2' \|_F^2=\mathrm{tr}(V_1V_1'+V_2V_2'-V_1(V_1'V_2)V_2'- V_2(V_2'V_1)V_1')=2\mathrm{tr}(I_K - MM')$.  Therefore, to show the claim, it suffices to show that there exists an orthogonal matrix $H$ such that
\beq \label{tmp-1}
\mathrm{tr}(MB-MM')\geq 0. 
\eeq
Let $M=PDQ'$ denote the singular value decomposition of $M$, where $P'P=Q'Q=I_K$ and $D=\mathrm{diag}(d_1,\cdots,d_K)$ is such that $d_1\geq d_2\geq \cdots\geq d_k\geq 0$. Introduce $H=QP'$. Then
\beq \label{tmp-2}
\mathrm{tr}(MH-MM') = \mathrm{tr}(PDP'-PD^2P')=\mathrm{tr}(D-D^2)=\sum_{k=1}^K (d_k-d_k^2). 
\eeq
Since $\|M\|\leq \|V_1\|\|V_2\|\leq 1$, we have $0\leq d_k\leq 1$ for $1\leq k\leq K$. Combining it with \eqref{tmp-2} gives \eqref{tmp-1}. 


\bibliographystyle{imsart-nameyear}

\bibliography{ifpca}

\begin{thebibliography}{52}

\bibitem[\protect\citeauthoryear{Abramovich et~al.}{2006}]{ABDJ}
\begin{barticle}[author]
\bauthor{\bsnm{Abramovich},~\bfnm{Felix}\binits{F.}},
  \bauthor{\bsnm{Benjamini},~\bfnm{Yoav}\binits{Y.}},
  \bauthor{\bsnm{Donoho},~\bfnm{David}\binits{D.}} \AND
  \bauthor{\bsnm{Johnstone},~\bfnm{Iain}\binits{I.}}
(\byear{2006}).
\btitle{Adapting to unknown sparsity by controlling the false discovery rate}.
\bjournal{Ann. Statist.}
\bvolume{34}
\bpages{584--653}.
\end{barticle}
\endbibitem

\bibitem[\protect\citeauthoryear{Amini and Wainwright}{2008}]{Wainwright}
\begin{barticle}[author]
\bauthor{\bsnm{Amini},~\bfnm{Arash}\binits{A.}} \AND
  \bauthor{\bsnm{Wainwright},~\bfnm{Martin}\binits{M.}}
(\byear{2008}).
\btitle{High-dimensional analysis of semidefinite relaxations for sparse
  principal components}.
\bjournal{Ann. Statist.}
\bvolume{37}
\bpages{2877-2921}.
\end{barticle}
\endbibitem

\bibitem[\protect\citeauthoryear{Arias-Castro, Lerman and Zhang}{2013}]{Ery1}
\begin{barticle}[author]
\bauthor{\bsnm{Arias-Castro},~\bfnm{Ery}\binits{E.}},
  \bauthor{\bsnm{Lerman},~\bfnm{Gilad}\binits{G.}} \AND
  \bauthor{\bsnm{Zhang},~\bfnm{Teng}\binits{T.}}
(\byear{2013}).
\btitle{Spectral clustering based on local PCA}.
\bjournal{arXiv:1301.2007}.
\end{barticle}
\endbibitem

\bibitem[\protect\citeauthoryear{Arias-Castro and Verzelen}{2014}]{ACV}
\begin{barticle}[author]
\bauthor{\bsnm{Arias-Castro},~\bfnm{Ery}\binits{E.}} \AND
  \bauthor{\bsnm{Verzelen},~\bfnm{Nicolas}\binits{N.}}
(\byear{2014}).
\btitle{Detection and feature selection in sparse mixture models}.
\bjournal{arXiv:1405.1478}.
\end{barticle}
\endbibitem

\bibitem[\protect\citeauthoryear{Arthur and Vassilvitskii}{2007}]{kmeans+}
\begin{binproceedings}[author]
\bauthor{\bsnm{Arthur},~\bfnm{David}\binits{D.}} \AND
  \bauthor{\bsnm{Vassilvitskii},~\bfnm{Sergei}\binits{S.}}
(\byear{2007}).
\btitle{k-means++: The advantages of careful seeding}.
In \bbooktitle{Proceedings of the eighteenth annual ACM-SIAM symposium on
  Discrete algorithms}
\bpages{1027--1035}.
\end{binproceedings}
\endbibitem

\bibitem[\protect\citeauthoryear{Azizyan, Singh and Wasserman}{2013}]{ASW}
\begin{binproceedings}[author]
\bauthor{\bsnm{Azizyan},~\bfnm{Martin}\binits{M.}},
  \bauthor{\bsnm{Singh},~\bfnm{Aarti}\binits{A.}} \AND
  \bauthor{\bsnm{Wasserman},~\bfnm{Larry}\binits{L.}}
(\byear{2013}).
\btitle{Minimax theory for high-dimensional Gaussian mixtures with sparse mean
  separation}.
In \bbooktitle{Advances in Neural Information Processing Systems}
\bpages{2139--2147}.
\end{binproceedings}
\endbibitem

\bibitem[\protect\citeauthoryear{Baik and Silverstein}{2006}]{BS2006}
\begin{barticle}[author]
\bauthor{\bsnm{Baik},~\bfnm{Jinho}\binits{J.}} \AND
  \bauthor{\bsnm{Silverstein},~\bfnm{Jack~W}\binits{J.~W.}}
(\byear{2006}).
\btitle{Eigenvalues of large sample covariance matrices of spiked population
  models}.
\bjournal{J. Multivariate Anal.}
\bvolume{97}
\bpages{1382--1408}.
\end{barticle}
\endbibitem

\bibitem[\protect\citeauthoryear{Birnbaum et~al.}{2013}]{BJNP}
\begin{barticle}[author]
\bauthor{\bsnm{Birnbaum},~\bfnm{Aharon}\binits{A.}},
  \bauthor{\bsnm{Johnstone},~\bfnm{Iain~M}\binits{I.~M.}},
  \bauthor{\bsnm{Nadler},~\bfnm{Boaz}\binits{B.}} \AND
  \bauthor{\bsnm{Paul},~\bfnm{Debashis}\binits{D.}}
(\byear{2013}).
\btitle{Minimax bounds for sparse PCA with noisy high-dimensional data}.
\bjournal{Ann. Statist.}
\bvolume{41}
\bpages{1055--1084}.
\end{barticle}
\endbibitem

\bibitem[\protect\citeauthoryear{Borovkov and Rogozin}{1965}]{BR}
\begin{barticle}[author]
\bauthor{\bsnm{Borovkov},~\bfnm{AA}\binits{A.}} \AND
  \bauthor{\bsnm{Rogozin},~\bfnm{BA}\binits{B.}}
(\byear{1965}).
\btitle{On the multi-dimensional central limit theorem}.
\bjournal{Theory Probab. Appl.}
\bvolume{10}
\bpages{55--62}.
\end{barticle}
\endbibitem

\bibitem[\protect\citeauthoryear{Cai, Ma and Wu}{2013}]{RankDetect}
\begin{barticle}[author]
\bauthor{\bsnm{Cai},~\bfnm{Tony}\binits{T.}},
  \bauthor{\bsnm{Ma},~\bfnm{Zongming}\binits{Z.}} \AND
  \bauthor{\bsnm{Wu},~\bfnm{Yihong}\binits{Y.}}
(\byear{2013}).
\btitle{Optimal estimation and rank detection for sparse spiked covariance
  matrices}.
\bjournal{Probab. Theory Related Fields}
\bvolume{161}
\bpages{781--815}.
\end{barticle}
\endbibitem

\bibitem[\protect\citeauthoryear{Chan and Hall}{2010}]{CH}
\begin{barticle}[author]
\bauthor{\bsnm{Chan},~\bfnm{Yao-ban}\binits{Y.-b.}} \AND
  \bauthor{\bsnm{Hall},~\bfnm{Peter}\binits{P.}}
(\byear{2010}).
\btitle{Using evidence of mixed populations to select variables for clustering
  very high-dimensional data}.
\bjournal{J. Amer. Statist. Soc.}
\bvolume{105}.
\end{barticle}
\endbibitem

\bibitem[\protect\citeauthoryear{Chen and Li}{2009}]{ChenJH}
\begin{barticle}[author]
\bauthor{\bsnm{Chen},~\bfnm{Jiahua}\binits{J.}} \AND
  \bauthor{\bsnm{Li},~\bfnm{Pengfei}\binits{P.}}
(\byear{2009}).
\btitle{Hypothesis test for normal mixture models: the EM approach}.
\bjournal{Ann. Statist.}
\bvolume{37}
\bpages{2523--2542}.
\end{barticle}
\endbibitem

\bibitem[\protect\citeauthoryear{Davis and Kahan}{1970}]{sin-theta}
\begin{barticle}[author]
\bauthor{\bsnm{Davis},~\bfnm{Chandler}\binits{C.}} \AND
  \bauthor{\bsnm{Kahan},~\bfnm{William~Morton}\binits{W.~M.}}
(\byear{1970}).
\btitle{The rotation of eigenvectors by a perturbation. III}.
\bjournal{SIAM Journal on Numerical Analysis}
\bvolume{7}
\bpages{1--46}.
\end{barticle}
\endbibitem

\bibitem[\protect\citeauthoryear{Dettling}{2004}]{Dettling}
\begin{barticle}[author]
\bauthor{\bsnm{Dettling},~\bfnm{Marcel}\binits{M.}}
(\byear{2004}).
\btitle{BagBoosting for tumor classification with gene expression data}.
\bjournal{Bioinformatics}
\bvolume{20}
\bpages{3583--3593}.
\end{barticle}
\endbibitem

\bibitem[\protect\citeauthoryear{Donoho}{2015}]{50years}
\begin{barticle}[author]
\bauthor{\bsnm{Donoho},~\bfnm{David}\binits{D.}}
(\byear{2015}).
\btitle{50 years of data science}.
\bjournal{Manuscript}.
\end{barticle}
\endbibitem

\bibitem[\protect\citeauthoryear{Donoho and Jin}{2004}]{DJ04}
\begin{barticle}[author]
\bauthor{\bsnm{Donoho},~\bfnm{David}\binits{D.}} \AND
  \bauthor{\bsnm{Jin},~\bfnm{Jiashun}\binits{J.}}
(\byear{2004}).
\btitle{Higher criticism for detecting sparse heterogeneous mixtures}.
\bjournal{Ann. Statist.}
\bpages{962--994}.
\end{barticle}
\endbibitem

\bibitem[\protect\citeauthoryear{Donoho and Jin}{2008}]{DJ08}
\begin{barticle}[author]
\bauthor{\bsnm{Donoho},~\bfnm{David}\binits{D.}} \AND
  \bauthor{\bsnm{Jin},~\bfnm{Jiashun}\binits{J.}}
(\byear{2008}).
\btitle{Higher criticism thresholding: Optimal feature selection when useful
  features are rare and weak}.
\bjournal{Proc. Natl. Acad. Sci.}
\bvolume{105}
\bpages{14790--14795}.
\end{barticle}
\endbibitem

\bibitem[\protect\citeauthoryear{Donoho and Jin}{2015}]{DJ13}
\begin{barticle}[author]
\bauthor{\bsnm{Donoho},~\bfnm{David}\binits{D.}} \AND
  \bauthor{\bsnm{Jin},~\bfnm{Jiashun}\binits{J.}}
(\byear{2015}).
\btitle{Higher criticism for large-scale inference, especially for rare and
  weak effects}.
\bjournal{Stat. Sci.}
\bvolume{30}
\bpages{1--25}.
\end{barticle}
\endbibitem

\bibitem[\protect\citeauthoryear{Dudley}{1999}]{Dudley}
\begin{bbook}[author]
\bauthor{\bsnm{Dudley},~\bfnm{Richard~M}\binits{R.~M.}}
(\byear{1999}).
\btitle{Uniform central limit theorems}
\bvolume{23}.
\bpublisher{Cambridge Univ Press}.
\end{bbook}
\endbibitem

\bibitem[\protect\citeauthoryear{Durbin}{1985}]{Durbin}
\begin{barticle}[author]
\bauthor{\bsnm{Durbin},~\bfnm{James}\binits{J.}}
(\byear{1985}).
\btitle{The first-passage density of a continuous Gaussian process to a general
  boundary}.
\bjournal{J. Appl. Probab.}
\bpages{99--122}.
\end{barticle}
\endbibitem

\bibitem[\protect\citeauthoryear{Efron}{2004}]{Efron}
\begin{barticle}[author]
\bauthor{\bsnm{Efron},~\bfnm{Bradley}\binits{B.}}
(\byear{2004}).
\btitle{Large-scale simultaneous hypothesis testing}.
\bjournal{J. Amer. Statist. Soc.}
\bvolume{99}
\bpages{96-104}.
\end{barticle}
\endbibitem

\bibitem[\protect\citeauthoryear{Efron}{2009}]{EfronClassification}
\begin{barticle}[author]
\bauthor{\bsnm{Efron},~\bfnm{Bradley}\binits{B.}}
(\byear{2009}).
\btitle{Empirical Bayes estimates for large-scale prediction problems}.
\bjournal{J. Amer. Statist. Soc.}
\bvolume{104}
\bpages{1015--1028}.
\end{barticle}
\endbibitem

\bibitem[\protect\citeauthoryear{Fan, Jin and Yao}{2013}]{FJY}
\begin{barticle}[author]
\bauthor{\bsnm{Fan},~\bfnm{Yingying}\binits{Y.}},
  \bauthor{\bsnm{Jin},~\bfnm{Jiashun}\binits{J.}} \AND
  \bauthor{\bsnm{Yao},~\bfnm{Zhigang}\binits{Z.}}
(\byear{2013}).
\btitle{Optimal classification in sparse Gaussian graphic model}.
\bjournal{Ann. Statist.}
\bvolume{41}
\bpages{2537--2571}.
\end{barticle}
\endbibitem

\bibitem[\protect\citeauthoryear{Fan and Lv}{2008}]{FanLv}
\begin{barticle}[author]
\bauthor{\bsnm{Fan},~\bfnm{Jianqing}\binits{J.}} \AND
  \bauthor{\bsnm{Lv},~\bfnm{Jinchi}\binits{J.}}
(\byear{2008}).
\btitle{Sure independence screening for ultrahigh dimensional feature space}.
\bjournal{J. Roy. Statist. Soc. B}
\bvolume{70}
\bpages{849--911}.
\end{barticle}
\endbibitem

\bibitem[\protect\citeauthoryear{Fan et~al.}{2015}]{QUADRO}
\begin{barticle}[author]
\bauthor{\bsnm{Fan},~\bfnm{Jianqing}\binits{J.}},
  \bauthor{\bsnm{Ke},~\bfnm{Zheng~Tracy}\binits{Z.~T.}},
  \bauthor{\bsnm{Liu},~\bfnm{Han}\binits{H.}} \AND
  \bauthor{\bsnm{Xia},~\bfnm{Lucy}\binits{L.}}
(\byear{2015}).
\btitle{QUADRO: A supervised dimension reduction method via Rayleigh quotient
  optimization}.
\bjournal{Ann. Statist.}
\bvolume{43}
\bpages{1498-1534}.
\end{barticle}
\endbibitem

\bibitem[\protect\citeauthoryear{Gordon et~al.}{2002}]{lungcancer}
\begin{barticle}[author]
\bauthor{\bsnm{Gordon},~\bfnm{Gavin~J}\binits{G.~J.}},
  \bauthor{\bsnm{Jensen},~\bfnm{Roderick~V}\binits{R.~V.}},
  \bauthor{\bsnm{Hsiao},~\bfnm{Li-Li}\binits{L.-L.}},
  \bauthor{\bsnm{Gullans},~\bfnm{Steven~R}\binits{S.~R.}},
  \bauthor{\bsnm{Blumenstock},~\bfnm{Joshua~E}\binits{J.~E.}},
  \bauthor{\bsnm{Ramaswamy},~\bfnm{Sridhar}\binits{S.}},
  \bauthor{\bsnm{Richards},~\bfnm{William~G}\binits{W.~G.}},
  \bauthor{\bsnm{Sugarbaker},~\bfnm{David~J}\binits{D.~J.}} \AND
  \bauthor{\bsnm{Bueno},~\bfnm{Raphael}\binits{R.}}
(\byear{2002}).
\btitle{Translation of microarray data into clinically relevant cancer
  diagnostic tests using gene expression ratios in lung cancer and
  mesothelioma}.
\bjournal{Cancer research}
\bvolume{62}
\bpages{4963--4967}.
\end{barticle}
\endbibitem

\bibitem[\protect\citeauthoryear{Guionnet and Zeitouni}{2000}]{Zeitouni2000}
\begin{barticle}[author]
\bauthor{\bsnm{Guionnet},~\bfnm{Alice}\binits{A.}} \AND
  \bauthor{\bsnm{Zeitouni},~\bfnm{Ofer}\binits{O.}}
(\byear{2000}).
\btitle{Concentration of the spectral measure for large matrices}.
\bjournal{Electron. Comm. Probab.}
\bvolume{5}
\bpages{119--136}.
\end{barticle}
\endbibitem

\bibitem[\protect\citeauthoryear{Hastie, Tibshirani and Friedman}{2009}]{HTF}
\begin{bbook}[author]
\bauthor{\bsnm{Hastie},~\bfnm{Trevor}\binits{T.}},
  \bauthor{\bsnm{Tibshirani},~\bfnm{Robert}\binits{R.}} \AND
  \bauthor{\bsnm{Friedman},~\bfnm{Jerome}\binits{J.}}
(\byear{2009}).
\btitle{The elements of statistical learning},
\bedition{2nd} ed.
\bpublisher{Springer}.
\end{bbook}
\endbibitem

\bibitem[\protect\citeauthoryear{Jin}{2015}]{SCORE}
\begin{barticle}[author]
\bauthor{\bsnm{Jin},~\bfnm{Jiashun}\binits{J.}}
(\byear{2015}).
\btitle{Fast community detection by SCORE}.
\bjournal{Ann. Statist.}
\bvolume{43}
\bpages{57--89}.
\end{barticle}
\endbibitem

\bibitem[\protect\citeauthoryear{Jin, Ke and Wang}{2015a}]{JinWang2}
\begin{barticle}[author]
\bauthor{\bsnm{Jin},~\bfnm{Jiashun}\binits{J.}},
  \bauthor{\bsnm{Ke},~\bfnm{Zheng~Tracy}\binits{Z.~T.}} \AND
  \bauthor{\bsnm{Wang},~\bfnm{Wanjie}\binits{W.}}
(\byear{2015}a).
\btitle{Optimal spectral clustering by Higher Criticism Thresholding}.
\bjournal{Manuscript}.
\end{barticle}
\endbibitem

\bibitem[\protect\citeauthoryear{Jin, Ke and Wang}{2015b}]{JinWang3}
\begin{barticle}[author]
\bauthor{\bsnm{Jin},~\bfnm{Jiashun}\binits{J.}},
  \bauthor{\bsnm{Ke},~\bfnm{Zheng~Tracy}\binits{Z.~T.}} \AND
  \bauthor{\bsnm{Wang},~\bfnm{Wanjie}\binits{W.}}
(\byear{2015}b).
\btitle{Phase transitions for high dimensional clustering and related
  problems}.
\bjournal{arXiv:1502.06952}.
\end{barticle}
\endbibitem

\bibitem[\protect\citeauthoryear{Jin and Ke}{2016}]{JK}
\begin{barticle}[author]
\bauthor{\bsnm{Jin},~\bfnm{Jiashun}\binits{J.}} \AND
  \bauthor{\bsnm{Ke},~\bfnm{Zheng~Tracy}\binits{Z.~T.}}
(\byear{2016}).
\btitle{Rare and weak effects in large-scale inference: methods and phase
  diagrams}.
\bjournal{Statistica Sinica}
\bvolume{26}
\bpages{1--34}.
\end{barticle}
\endbibitem

\bibitem[\protect\citeauthoryear{Jin, Zhang and Zhang}{2014}]{gs}
\begin{barticle}[author]
\bauthor{\bsnm{Jin},~\bfnm{Jiashun}\binits{J.}},
  \bauthor{\bsnm{Zhang},~\bfnm{Cun-Hui}\binits{C.-H.}} \AND
  \bauthor{\bsnm{Zhang},~\bfnm{Qi}\binits{Q.}}
(\byear{2014}).
\btitle{Optimality of Graphlet Screening in high dimensional variable
  selection}.
\bjournal{J. Mach. Learn. Res.}
\bvolume{15}
\bpages{2723--2772}.
\end{barticle}
\endbibitem

\bibitem[\protect\citeauthoryear{Johnstone}{2001}]{Johnstone}
\begin{barticle}[author]
\bauthor{\bsnm{Johnstone},~\bfnm{Iain}\binits{I.}}
(\byear{2001}).
\btitle{On the distribution of the largest eigenvalue in principal components
  analysis}.
\bjournal{Ann. Statist.}
\bpages{295--327}.
\end{barticle}
\endbibitem

\bibitem[\protect\citeauthoryear{Jung and Marron}{2009}]{Jung}
\begin{barticle}[author]
\bauthor{\bsnm{Jung},~\bfnm{Sungkyu}\binits{S.}} \AND
  \bauthor{\bsnm{Marron},~\bfnm{James}\binits{J.}}
(\byear{2009}).
\btitle{PCA consistency in high dimension, low sample size context}.
\bjournal{Ann. Statist.}
\bvolume{37}
\bpages{4104--4130}.
\end{barticle}
\endbibitem

\bibitem[\protect\citeauthoryear{Ke, Jin and Fan}{2014}]{KJF}
\begin{barticle}[author]
\bauthor{\bsnm{Ke},~\bfnm{Zheng}\binits{Z.}},
  \bauthor{\bsnm{Jin},~\bfnm{Jiashun}\binits{J.}} \AND
  \bauthor{\bsnm{Fan},~\bfnm{Jianqing}\binits{J.}}
(\byear{2014}).
\btitle{Covariance assisted screening and estimation}.
\bjournal{Ann. Statist.}
\bvolume{42}
\bpages{2202--2242}.
\end{barticle}
\endbibitem

\bibitem[\protect\citeauthoryear{Kolmogorov}{1933}]{Kolmogorov}
\begin{barticle}[author]
\bauthor{\bsnm{Kolmogorov},~\bfnm{Andrey~N}\binits{A.~N.}}
(\byear{1933}).
\btitle{Sulla determinazione empirica di una legge di distribuzione}.
\bjournal{Giornale dell'Istituto Italiano degli Attuari}
\bvolume{4}
\bpages{83--91}.
\end{barticle}
\endbibitem

\bibitem[\protect\citeauthoryear{Kritchman and Nadler}{2008}]{KN}
\begin{barticle}[author]
\bauthor{\bsnm{Kritchman},~\bfnm{Shira}\binits{S.}} \AND
  \bauthor{\bsnm{Nadler},~\bfnm{Boaz}\binits{B.}}
(\byear{2008}).
\btitle{Determining the number of components in a factor model from limited
  noisy data}.
\bjournal{Chemometr. Intell. Lab}
\bvolume{94}
\bpages{19--32}.
\end{barticle}
\endbibitem

\bibitem[\protect\citeauthoryear{Lee, Luca and Roeder}{2010}]{Lee}
\begin{barticle}[author]
\bauthor{\bsnm{Lee},~\bfnm{Ann~B}\binits{A.~B.}},
  \bauthor{\bsnm{Luca},~\bfnm{Diana}\binits{D.}} \AND
  \bauthor{\bsnm{Roeder},~\bfnm{Kathryn}\binits{K.}}
(\byear{2010}).
\btitle{A spectral graph approach to discovering genetic ancestry}.
\bjournal{Ann. Appl. Statist.}
\bvolume{4}
\bpages{179--202}.
\end{barticle}
\endbibitem

\bibitem[\protect\citeauthoryear{Lee, Zou and Wright}{2010}]{Lee2010}
\begin{barticle}[author]
\bauthor{\bsnm{Lee},~\bfnm{Seunggeun}\binits{S.}},
  \bauthor{\bsnm{Zou},~\bfnm{Fei}\binits{F.}} \AND
  \bauthor{\bsnm{Wright},~\bfnm{Fred~A}\binits{F.~A.}}
(\byear{2010}).
\btitle{Convergence and prediction of principal component scores in
  high-dimensional settings}.
\bjournal{Ann. Statist.}
\bvolume{38}
\bpages{3605}.
\end{barticle}
\endbibitem

\bibitem[\protect\citeauthoryear{Lei and Vu}{2015}]{Lei}
\begin{barticle}[author]
\bauthor{\bsnm{Lei},~\bfnm{Jing}\binits{J.}} \AND
  \bauthor{\bsnm{Vu},~\bfnm{Vincent~Q}\binits{V.~Q.}}
(\byear{2015}).
\btitle{Sparsistency and agnostic inference in sparse {PCA}}.
\bjournal{Ann. Stat.}
\bvolume{43}
\bpages{299--322}.
\end{barticle}
\endbibitem

\bibitem[\protect\citeauthoryear{Loader et~al.}{1992}]{Loader}
\begin{barticle}[author]
\bauthor{\bsnm{Loader},~\bfnm{Clive~R}\binits{C.~R.}} \betal{et~al.}
(\byear{1992}).
\btitle{Boundary crossing probabilities for locally Poisson processes}.
\bjournal{Ann. Appl. Probab.}
\bvolume{2}
\bpages{199--228}.
\end{barticle}
\endbibitem

\bibitem[\protect\citeauthoryear{Ma}{2013}]{Ma}
\begin{barticle}[author]
\bauthor{\bsnm{Ma},~\bfnm{Zongming}\binits{Z.}}
(\byear{2013}).
\btitle{Sparse principal component analysis and iterative thresholding}.
\bjournal{Ann. Statist.}
\bvolume{41}
\bpages{772--801}.
\end{barticle}
\endbibitem

\bibitem[\protect\citeauthoryear{Ng, Jordan and Weiss}{2002}]{NJ}
\begin{barticle}[author]
\bauthor{\bsnm{Ng},~\bfnm{Andrew~Y}\binits{A.~Y.}},
  \bauthor{\bsnm{Jordan},~\bfnm{Michael~I}\binits{M.~I.}} \AND
  \bauthor{\bsnm{Weiss},~\bfnm{Yair}\binits{Y.}}
(\byear{2002}).
\btitle{On spectral clustering: Analysis and an algorithm}.
\bjournal{Advances in neural information processing systems}
\bvolume{2}
\bpages{849--856}.
\end{barticle}
\endbibitem

\bibitem[\protect\citeauthoryear{Paul}{2007}]{Paul}
\begin{barticle}[author]
\bauthor{\bsnm{Paul},~\bfnm{Debashis}\binits{D.}}
(\byear{2007}).
\btitle{Asymptotics of sample eigenstructure for a large dimensional spiked
  covariance model}.
\bjournal{Statistica Sinica}
\bvolume{17}
\bpages{1617}.
\end{barticle}
\endbibitem

\bibitem[\protect\citeauthoryear{Paul and Johnstone}{2012}]{PJ}
\begin{barticle}[author]
\bauthor{\bsnm{Paul},~\bfnm{Debashis}\binits{D.}} \AND
  \bauthor{\bsnm{Johnstone},~\bfnm{Iain~M}\binits{I.~M.}}
(\byear{2012}).
\btitle{Augmented sparse principal component analysis for high dimensional
  data}.
\bjournal{arXiv:1202.1242}.
\end{barticle}
\endbibitem

\bibitem[\protect\citeauthoryear{Shorack and Wellner}{1986}]{Wellner86}
\begin{bbook}[author]
\bauthor{\bsnm{Shorack},~\bfnm{Galen}\binits{G.}} \AND
  \bauthor{\bsnm{Wellner},~\bfnm{Jon}\binits{J.}}
(\byear{1986}).
\btitle{Empirical processes with applications to statistics}.
\bpublisher{John Wiley \& Sons}.
\end{bbook}
\endbibitem

\bibitem[\protect\citeauthoryear{Siegmund}{1982}]{Siegmund}
\begin{barticle}[author]
\bauthor{\bsnm{Siegmund},~\bfnm{D}\binits{D.}}
(\byear{1982}).
\btitle{Large deviations for boundary crossing probabilities}.
\bjournal{Ann. Prob.}
\bvolume{10}
\bpages{581--588}.
\end{barticle}
\endbibitem

\bibitem[\protect\citeauthoryear{Vershynin}{2012}]{Vershynin}
\begin{barticle}[author]
\bauthor{\bsnm{Vershynin},~\bfnm{Roman}\binits{R.}}
(\byear{2012}).
\btitle{Introduction to the non-asymptotic analysis of random matrices}.
\bjournal{Compressed Sensing}
\bpages{210--268}.
\end{barticle}
\endbibitem

\bibitem[\protect\citeauthoryear{Woodroofe}{1978}]{Woodroofe}
\begin{barticle}[author]
\bauthor{\bsnm{Woodroofe},~\bfnm{Michael}\binits{M.}}
(\byear{1978}).
\btitle{Large deviations of likelihood ratio statistics with applications to
  sequential testing}.
\bjournal{Ann. Statist.}
\bpages{72--84}.
\end{barticle}
\endbibitem

\bibitem[\protect\citeauthoryear{Yousefi et~al.}{2010}]{Yousefi}
\begin{barticle}[author]
\bauthor{\bsnm{Yousefi},~\bfnm{Mohammadmahdi~R}\binits{M.~R.}},
  \bauthor{\bsnm{Hua},~\bfnm{Jianping}\binits{J.}},
  \bauthor{\bsnm{Sima},~\bfnm{Chao}\binits{C.}} \AND
  \bauthor{\bsnm{Dougherty},~\bfnm{Edward~R}\binits{E.~R.}}
(\byear{2010}).
\btitle{Reporting bias when using real data sets to analyze classification
  performance}.
\bjournal{Bioinformatics}
\bvolume{26}
\bpages{68--76}.
\end{barticle}
\endbibitem

\bibitem[\protect\citeauthoryear{Zou, Hastie and Tibshirani}{2006}]{ZHT}
\begin{barticle}[author]
\bauthor{\bsnm{Zou},~\bfnm{Hui}\binits{H.}},
  \bauthor{\bsnm{Hastie},~\bfnm{Trevor}\binits{T.}} \AND
  \bauthor{\bsnm{Tibshirani},~\bfnm{Robert}\binits{R.}}
(\byear{2006}).
\btitle{Sparse principal component analysis}.
\bjournal{J. Comp. Graph. Stat.}
\bvolume{15}
\bpages{265--286}.
\end{barticle}
\endbibitem

\end{thebibliography}
\end{document}